\documentstyle[12pt]{article}
\font\twlgot =eufm10 scaled \magstep1
\font\egtgot =eufm8
\font\sevgot =eufm7
\font\twlmsb =msbm10 scaled \magstep1
\font\egtmsb =msbm8
\font\sevmsb =msbm7
\newfam\gotfam
\def\pgot{\fam\gotfam\twlgot}
\textfont\gotfam\twlgot
\scriptfont\gotfam\egtgot
\scriptscriptfont\gotfam\sevgot
\def\got{\protect\pgot}
\newfam\msbfam
\textfont\msbfam\twlmsb
\scriptfont\msbfam\egtmsb
\scriptscriptfont\msbfam\sevmsb
\def\Bbb{\protect\pBbb}
\def\pBbb{\relax\ifmmode\expandafter\Bb\else\typeout{You cann't use
Bbb in text mode}\fi}
\def\Bb #1{{\fam\msbfam\relax#1}}

\def\thebibliography#1{\bigskip\section*{\centering
Bibliography\\}\bigskip\list
{\arabic{enumi}.}{\settowidth\labelwidth{#1}\leftmargin\labelwidth
\advance\leftmargin\labelsep
\usecounter{enumi}}
\def\newblock{\hskip .11em plus .33em minus .07em}
\sloppy\clubpenalty4000\widowpenalty4000
\sfcode`\.=1000\relax}
\newcommand{\Si}{\Sigma}

\def\op#1{\mathop{\fam0 #1}\limits}

\newcommand{\pr}{{\rm pr\,}}

\newcommand{\Id}{{\rm Id\,}}
\def\Ker{{\rm Ker\,}}
\newcommand{\ben}{\begin{eqnarray}}
\newcommand{\een}{\end{eqnarray}}
\newcommand{\be}{\begin{eqnarray*}}
\newcommand{\ee}{\end{eqnarray*}}
\newcommand{\bea}{\begin{eqalph}}
\newcommand{\eea}{\end{eqalph}}

\newcommand{\cT}{{\cal T}}

\newcommand{\cL}{{\cal L}}

\newcommand{\cE}{{\cal E}}
\newcommand{\cH}{{\cal H}}
\newcommand{\cF}{{\cal F}}
\newcommand{\bL}{{\bf L}}

\newcommand{\al}{\alpha}
\newcommand{\bt}{\beta}
\newcommand{\dl}{\delta}
\newcommand{\la}{\lambda}
\newcommand{\La}{\Lambda}
\newcommand{\f}{\phi}
\newcommand{\F}{\Phi}
\newcommand{\p}{\pi}

\newcommand{\om}{\omega}
\newcommand{\Om}{\Omega}
\newcommand{\m}{\mu}
\newcommand{\n}{\nu}
\newcommand{\g}{\gamma}
\newcommand{\G}{\Gamma}

\newcommand{\ve}{\varepsilon}
\newcommand{\th}{\theta}

\newcommand{\si}{\sigma}

\newcommand{\pdr}{\partial}

\newcommand{\w}{\wedge}

\newcommand{\wt}{\widetilde}
\newcommand{\wh}{\widehat}
\newcommand{\ol}{\overline}

\newcommand{\dr}{\partial}

\newcounter{eqalph}
\newcounter{equationa}

\newenvironment{proposition}[1]{\bigskip{\bf Proposition #1.}}{$\Box$
\bigskip}
\newenvironment{corollary}[1]{\bigskip{\bf Corollary #1.}}{$\Box$\bigskip}
\newenvironment{definition}[1]{\bigskip{\bf Definition #1.}}{$\Box$\bigskip}
\newenvironment{remark}{\bigskip{\bf Remark.}}{$\Box$\bigskip}
\newenvironment{theorem}[1]{\bigskip{\bf Theorem #1.}}{$\Box$\bigskip}
\newenvironment{lemma}[1]{\bigskip{\bf Lemma #1.}}{$\Box$\bigskip}

\newenvironment{example}[1]{\bigskip{\bf Example #1.}}{$\Box$\bigskip}

\newenvironment{eqalph}{\stepcounter{equation}
\setcounter{equationa}{\value{equation}}
\setcounter{equation}{0}

\begin{eqnarray}}{\end{eqnarray}
\setcounter{equation}{\value{equationa}}}

\begin{document}
\hbox{}

\centerline{\large\bf FIVE LECTURES ON THE JET MANIFOLD}
\medskip

\centerline{\large\bf METHODS IN FIELD THEORY}
\bigskip

\centerline{\bf Gennadi A Sardanashvily}
\medskip

\centerline{Department of Theoretical Physics, Physics Faculty,}

\centerline{Moscow State University, 117234 Moscow, Russia}

\centerline{E-mail: sard@grav.phys.msu.su}

\begin{abstract}
The fibre bundle formulation of gauge theory is generally accepted. The
jet manifold machinery completes this formulation and provides the adequate
mathematical description of dynamics of fields represented by sections of
fibre bundles. Theory of differential operators, Lagrangian and Hamiltonian
formalisms on bundles have utilized widely the language of jet manifolds.
Moreover, when not restricted to principal connections, differential geometry
also is phrased in jet terms. However, this powerful tool remains almost
unknown to physicists. These Lectures give introduction to jet manifolds,
Lagrangian and Hamiltonian formalisms in jet manifolds and their application
to a number of fundamental field models.
\end{abstract}
\bigskip\bigskip

\centerline{\large \bf Contents}
\medskip

\noindent
Introduction\newline
Bibliography\newline
LECTURE 1. Jet Manifolds\newline
LECTURE 2. General Connections\newline
LECTURE 3. Lagrangian Formalism\newline
LECTURE 4. Hamiltonian Formalism\newline
LECTURE 5. Field Theory
\bigskip\bigskip

\centerline{\large \bf Introduction}\bigskip\bigskip

We follow the generally accepted geometric description of classical
fields by sections of fibred manifolds
\[
\pi: Y\to X
\]
whose base $X$ is regarded generally as an $n$-dimensional parameter
space, e.g., a world manifold.

Lagrangian and Hamiltonian formalisms on fibred manifolds utilize the
language of jet spaces. Given a fibred manifold
$Y\to X$, the $k$-jet space $J^kY$ of $Y$ comprises the equivalence classes
$j^k_xs$, $x\in X$, of sections $s$ of $Y$ identified by the first $(k+1)$
terms of their Taylor series at points $x\in X$. One exploits
the well-known facts
that: (i) the $k$-jet space of sections of a fibred
manifold $Y$ is a finite-dimensional smooth manifold and (ii)
a $k$-order differential operator on sections of a fibred manifold $Y$
can be described as a morphism of $J^kY$ to a vector bundle over $X$.
As a consequence, the dynamics of field systems is played out
on finite-dimensional configuration and phase spaces. Moreover, this
dynamics is phrased in the geometric terms due to
the 1:1 correspondence between sections of the jet bundle
$J^1Y\to Y$ and connections on the fibred manifold $Y\to X$.

In the Lectures, we are concerned only with
first order Lagrangian and Hamiltonian systems, for
the most of contemporary field models are described by
first order Lagrangian densities.

In the framework of the first order Lagrangian formalism,
the finite-dimensional configuration space of sections of a fibred manifold
$Y\to X$ is the jet manifold $J^1Y$ coordinatized by
\[
(x^\m, y^i, y^i_\m)
\]
where $(x^\la, y^i)$ are fibred coordinates of $Y$
and $y^i_\mu$ are so-called derivative coordinates or velocities.
A first order Lagrangian density on the configuration space $J^1Y$ is
represented by an exterior form
\be
&&L=\cL(x^\m, y^i, y^i_\m)\om, \\
&& \om=dx^1\w ...\w dx^n.
\ee

 Given a Lagarngian density $L$, we have the associated
Legendre morphism $\wh L$
of the configuration space $J^1Y$ to the Legendre bundle
\[
\Pi=\op\w^n T^*X\op\otimes_Y TX\op\otimes_Y V^*Y
\]
over $Y$. This bundle is provided with the corresponding
induced coordinates
\[
( x^\la ,y^i,p^\la_i)
\]
such that
\[
 p^\m_i\circ\wh L=\pi^\m_i.
\]
The Legendre bundle $\Pi$ carries the multisymplectic form
\[
\Om =dp^\la_i\w dy^i\w\om\otimes\dr_\la.
\]
In case of $X=\Bbb R$, the form $\Om$ recovers
the standard symplectic form in analytical mechanics.

Building on the multisymplectic form $\Om$, one can develop
the so-called multimomentum Hamiltonian formalism of field theory where
canonical momenta correspond to derivatives of fields with respect to all world
coordinates, not only the temporal one. On the mathematical level, this is
the straightforward multisymplectic
generalization of the standard Hamiltonian formalism in analytical mechanics
to fibred manifolds over an $n$-dimensional base $X$, not only $\Bbb R$.

We shall say that a connection $\g$
on the fibred Legendre manifold $\Pi\to X$
is a Hamiltonian connection if the
exterior form $\g\rfloor\Om$ is closed.
Then, a multimomentum Hamiltonian form $H$ on $\Pi$ is defined to be an
exterior form  such that
 \begin{equation}
dH=\g\rfloor\Om \label{w2}
\end{equation}
for some Hamiltonian connection $\g$.
The crucial point consists in the fact that every multimomentum Hamiltonian
form admits splitting
\begin{equation}
H =p^\la_idy^i\w\om_\la
-p^\la_i\G^i_\la\om -\wt{\cH}_\G\om=p^\la_idy^i\w\om_\la-\cH\om \label{w1}
\end{equation}
where $\G$ is some connection on the fibred manifold $Y$.
Moreover, every multimomentum
Hamiltonian form itself sets up the
associated connection on $Y$ and, so meets the canonical splitting
(\ref{w1}). One can think of the splitting (\ref{w1}) as being a
workable definition of multimomentum Hamiltonian forms.
Given a multimomentum Hamiltonian form $H$ (\ref{w1}), the equality
(\ref{w2}) leads to the first order partial differential equations
\be
&&\dr_\la r^i(x) =\dr^i_\la\cH, \\
&&\dr_\la r^\la_i(x) =-\dr_i\cH
\ee
for sections $r$ of the fibred Legendre manifold $\Pi\to X$. We call them
the Hamilton equations.

If a Lagrangian density is regular,
the Lagrangian formalism on fibred manifolds and the
multimomentum Hamiltonian formalism are naturally equivalent to each
other. In this case, there
exists the unique  multimomentum Hamiltonian form such that the
first order Euler-Lagrange equations  and the corresponding
Hamilton equations are equivalent, otherwise in case of degenerate
Lagrangian densities.

In field theory, if a Lagrangian density is not regular,
the system of the Euler-Lagrange equations becomes
underdetermined and requires a supplementary gauge-type condition.
In gauge theory, these supplementary conditions are the
familiar gauge conditions. Such a gauge condition is introduced
by hand and
singles out a representative from each gauge coset. In general case, the
gauge-type conditions however remain elusive. In the framework of
the multimomentum Hamiltonian formalism, they appear automatically
as a part of the Hamilton equations.
The key point consists in the fact that, given a degenerate
Lagrangian density, one must consider a family of different
associated multimomentum
Hamiltonian forms in order that solutions of the corresponding Hamilton
equations exaust solutions of the Euler-Lagrange
equations. For a wide class of degenerate Lagrangian densities, such
complete families exist at least locally, and the adequate relations
between Lagrangian and multimomentum Hamiltonian formalisms can be given.
In particular, we spell out the models with degenerate quadratic and affine
Lagrangian densities. The most of contemporary field
models belong to these types. As a consequence, the tools are at hand
to canonically analize constraint field systems on covariant and
finite-dimensional levels.

In order to illustrate the jet manifold methods, we apply them to
a number of standard field models: gauge theory, electromagnetic
fields, Proca fields, matter fields and Einstein's gravitation theory.
At the same time, it is the multimomentum Hamiltonian formalism which
enables us to describe field systems on composite manifolds, e.g., the
gauge gravitation theory (see E-prints: gr-qc/9405013,9407032 and
hep-th/9409159). We refer also to E-prints: hep-th/9403172,9405040 as
the brief of this work.

\newpage

\centerline{\large \bf Lecture 1. JET MANIFOLDS}
\bigskip\bigskip

The most of differential geometric methods applied to field theory are based
on principal bundles. The Lectures 1,2
brief the basics of general theory of fibred manifolds, jet manifolds
and connections.

In application to field theory, one handles first and secon order jet
manifolds as a rule.

All morphisms throughout are differentiable mappings of
class $C^\infty$. Manifolds are real, Hausdorff,
finite-dimensional and second-countable (hence paracompact).
Unless otherwise stated, manifolds are proposed to be connected manifolds,
without boundary.

We use the conventional symbols
$\otimes$, $\vee$ and $\wedge$ for the tensor, symmetric and
exterior products respectively.
By $\rfloor$ is meant the interior product (i.e., contraction) of
multivectors on the right and differential forms on the left.

The symbols $\pdr^A_B$ denote the partial derivatives with respect to
coordinates possessing multi-indices $^B_A$.

Throughout this Lecture, by $M$ is meant an $m$-dimensional manifold. Let
\be
&&\Psi_M=\{U_\xi, \phi_\xi\}, \\
&&\phi_\xi (z)=z^\la e_\la, \qquad z\in U_\xi,
\ee
be an atlas of local coordinates $(z^\la)$ of $M$
where $\{e_\la\}$ is a fixed basis of $\Bbb R^m$.
The tangent bundle $TM$ of the manifold $M$ and the cotangent bundle
$T^*M$ of  $M$ are provided with atlases of the induced coordinates
$(z^\la, \dot z^\la)$ and $(z^\la,\dot z_\la)$  relative to the holonomic
fibre bases $\{\dr_\la\}$ and $\{dz^\la\}$ respectively. We have
\be
&& TM\ni t=\dot z^\la\dr_\la, \qquad T\phi_\xi(\dr_\la)=e_\la, \\
&& T^*M\ni t^*=\dot z_\la dz^\la, \qquad \dr_\la\rfloor dz^\al=
\delta^\al_\la.
\ee
If $f:M\to M'$ is a manifold mapping, by
\be
&& Tf: TM\to TM', \\
&&\dot z'^\la\circ Tf= \frac{\dr f^\la}{\dr z^\al}\dot z^\al,
\ee
is meant the morphism tangent to $f$.

Given a manifold product $A\times B$, we shall denote by $\pr_1$ and $\pr_2$
the canonical surjections
\be
&&\pr_1:A\times B\to A, \\
&& \pr_2:A\times B\to B.
\ee

\section{Manifold mappings}

Throughout the Lectures, we handle the following types of
manifold mappings: injection, surjection,
immersion, imbedding  and submersion. In this Section, we shall
recall these notions.

There exist two types of manifold mappings $f:M\to M'$
when the tangent morphism $Tf$ to
$f$ meets the maximal rank ${\rm rg}_zf$ at a point $z\in M$. These are an
immersion when $\dim M\leq \dim M'$ and a submersion
when $\dim M\geq \dim M'$. If a manifold mapping $f$ is both an
immersion and a submersion when $\dim M=\dim M'$, it is a
local diffeomorphism.

Let $z$ be a point of $M$ and $T_zM$ the tangent space to $M$
at $z$. A mapping $f:M\to M'$ is called the immersion  at a point
$z\in M$ when the tangent morphism $Tf$ to $f$ is an injection  of
the tangent space $T_zM$ into the tangent space $T_{f(z)}M'$ to $M'$ at
$f(z)\in M'$.

The following assertions are equivalent.
\begin{itemize}
\item A manifold mapping $f$ is an immersion at a point $z\in M$.
\item There exist: (i) an open neighborhood $U$ of $z$, (ii) an open
neighborhood $U'\supset f(U)$ of $f(z)$, (iii) a manifold $V$, (iv)
 a  diffeomorphism
\[
 \psi :U\times V\to U'
\]
 and (v) a point $a\in V$ so that
\[
f(b)=\psi(b,a)
\]
for each point $b\in U$.
\item  Given the above-mentioned neighborhoods $U$ of $z$ and $U'$
of $f(z)$, there exist coordinates $({z'}^\la)$ of $U'$ so that
\[
{z'}^\la\circ f=0,\qquad \la =\dim M+1,...,\dim M',
\]
whereas $({z'}^\la\circ f)$, $\la =1,...,\dim M$, set up
coordinates of $U$.
\end{itemize}

One can refer to these statements as the equivalent definitions of an
immersion at a point.

Let us recall that, given a manifold mapping $f:M\to M'$,
the function
\[
z\to {\rm rg}_zf
\]
is a lower semicontinuous function on $M$. It follows that, if $f$
is an immersion  at a point $z$,
then there exists an open neighborhood $U$ of $z$ so that $f$ is an
immersion at all points of $U$. Moreover, in accordance
with the rank theorem,  $f$ is an injection of some open neighborhood of $z$
into $M'$.

A manifold mapping $f$ of $M$ is termed the immersion if it is an
immersion at all points of $M$.

A triple $ f:M\to M'$
is called the submanifold if $f$ is both an immersion and an injection of $M$
into $M'$. For the sake of simplicity, we denote a submanifold by
the symbols $f$ or $f(M)$.
In particular, if a mapping $f$ is an immersion at a point $z\in M$, it is
locally a submanifold around $z$.

\begin{example}{1.1}
Let us consider the manifold $M=\Bbb R^1$ coordinatized by $z$ and the
manifold $M'=\Bbb R^2$
coordinatized by $({z'}^1, {z'}^2)$. The mapping
\[
{z'}^1=0, \qquad {z'}^2=z^2,
\]
of $\Bbb R^1$ into $\Bbb R^2$ exemplifies the injection which is not
a submanifold.
\end{example}

It should be noticed that a submanifold fails to be a topological
subspace of a manifold in general. The topology of a submanifold is finer
than the relative topology.

\begin{example}{1.2}
The picture \newline
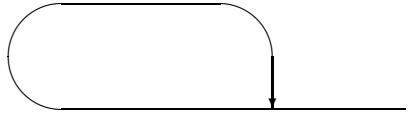
\[
\put(0,0){\oval(100,40)[t]}
\put(0,0){\oval(100,40)[bl]}
\put(0,-20){\line(1,0){100}}
\put(50,0){\vector(0,-1){20}}
\]
\newline
illustrates a submanifold $\Bbb R^1\to\Bbb R^2$ which is not a topological
subspace.
\end{example}

A submanifold which also is a
topological subspace is called the imbedded submanifold. A
diffeomorphism onto an imbedded submanifold is termed imbedding.
There is the following criterion of imbedding.

\begin{proposition}{1.3} A mapping $f:M\to M'$
is an imbedding if and only if, for each point $z\in M$,
there exists an open neighborhood $U'$ of $f(z)$ with coordinates
$({z'}^\la)$ such that the intersection $f(M)\cap U'$ consists of all points
$z'\in U'$ whose coordinates meet the condition
\[
{z'}^\la (z')=0, \qquad \la =\dim M+1,...,\dim M'.
\]
\end{proposition}

For instance, every open subset $U$ of a manifold $M$ is endowed with the
manifold structure such that the canonical injection
$i_U: U\hookrightarrow  M$ is imbedding. Hencerforth, we shall refer to an
open subset as being an open imbedded submanifold.

Let $z$ be a point of $M$ and $T_zM$ be the tangent space to $M$
at $z$. A mapping $f:M\to M'$ is called  the submersion at a point
$z\in M$ when the tangent morphism $Tf$ to $f$ is a surjection
of $T_zM$ onto the tangent space $T_{f(z)}M'$ to $M'$ at $f(z)$.
The following assertions are equivalent.
\begin{itemize}
\item A manifold mapping $f$ is a submersion at a point $z\in M$.
\item There exist: (i) an open neighborhood $U$ of $z$, (ii) an open
neighborhood $U'\supset f(U)$ of $f(z)$, (iii) a manifold $V$ and
(iv) a mapping $\psi:U\to V$ so that the mapping
\be
&&\psi:U\to U'\times V, \\
&&\pr_1\circ\psi = f\mid_U,
\ee
is a diffeomorphism. In other words, there exists a local splitting of $M$.
This splitting is called the local bundle splitting if
$U=f^{-1}(U')$ is the preimage of an open subset $U'\subset X$ by $f$.
\item Given the above-mentioned neigborhoods $U$ of $z$ and $U'$ of $f(z)$,
there exist  coordinates
$({z'}^\la)$ of $U'$ and fibred  coordinates  $(z^\la)$  of $U$
 so that
\[
z^\la ={z'}^\la\circ f, \qquad \la =1,...,\dim M'.
\]
\item There exist an open neighborhood $U'$ of $f(z)$ and imbedding
$s$ of $U'$ into $M$ so that
\[
(s\circ f)(z)=z.
\]
In other words, there is a section of the manifold $M$ over $U'$.
\end{itemize}

One can refer to these statements as the equivalent definitions of a
submersion at a point.

If a mapping $f$ of $M$ is a submersion at a point $z\in M$,
then there exists an open neighborhood $U$ of $z$ so that  $f$ is
a submersion at all points of $U$. Moreover, $f$ is a surjection
of some open neighborhood of $z$ onto an open subset of $M'$.

A manifold mapping $f$ of $M$ is termed the submersion if it is a submersion
at all points of $M$.

A triple $f:M\to M'$ is termed the fibred manifold if a mapping $f$ is
both a submersion and a surjection of $M$ onto $M'$.

\section{Fibred manifolds}

Throughout the Lectures, by $Y$ is meant a fibred manifold
\begin{equation}
\pi :Y\to X. \label{1.1}
\end{equation}
We use symbols $y$ and $x$ for points of its total space
$Y$ and its base $X$ respectively. Unless otherwise stated, $X$ is an
$n$-dimensional manifold. It is endowed with an atlas of local coordinates
$(x^\lambda)$.

We shall refer to the notations
\be
&& \om=dx^1\w ...\w dx^n,\\
&& \om_\la =\dr_\la\rfloor\om.
\ee

A fibred manifold $Y\to X$, by definition, is provided with an atlas of
fibred coordinates
\ben
&& (x^\la, y^i), \label{1.2}\\
&& x^\la \to {x'}^\la(x^\m), \nonumber\\
&& y^i \to {y'}^i(x^\m,y^j),\nonumber
\een
compatible with the fibration (\ref{1.1}).

A fibred manifold $Y\to X$ is called
the locally trivial fibred manifold if there exists
a fibred coordinate atlas of $Y$ over  an open covering
$\{\pi^{-1}(U_\xi)\}$ of $Y$ where $\{U_\xi\}$ is an
open covering of the base $X$. In other words, all points of the same fibre
$\pi^{-1}(x)$ of a bundle $Y$ can be
covered by the same fibred coordinate chart.
In this case, we have the standard
fibre manifold $V$  for all local bundle splittings
\[
\psi_\xi :\pi^{-1}(U_\xi)\to U_\xi\times V.
\]

By a differentiable fibre bundle (or simply a bundle) is
meant the locally trivial fibred manifold (\ref{1.1}) provided with a
family of equivalent bundle atlases
\be
&&\Psi = \{U_\xi, \psi_\xi, \rho_{\xi\zeta}\}, \\
&&\psi_\xi (y)=(\rho_{\xi\zeta}\circ\psi_\zeta)(y),
\qquad y\in U_\xi\cap U_\zeta,
\ee
where $\rho$ are transition functions. Let us recall that two bundle
atlases are equivalent if their union also is a bundle atlas.

If $Y\to X$ is a bundle, the fibred coordinates
(\ref{1.2}) of $Y$ are assumed to
be bundle coordinates associated with a bundle atlas  $\Psi$
of $Y$, that is,
\begin{equation}
y^i(y)=(v^i\circ\pr_2\circ \psi_\xi)(y), \qquad \pi (y)\in U_\xi,
\label{1.4}
\end{equation}
where $v^i$ are coordinates of the standard fibre $V$ of $Y$.

Note that a bundle $Y\to X$ is
uniquely defined by a triple $(X, V,\Psi)$ of its base $X$, its standard fibre
$V$ and a bundle atlas $\Psi$.

Given fibred manifolds $Y\to X$ and $Y'\to X'$, let a pair
\begin{equation}
\Phi : Y\to Y', \qquad f: X\to X', \label{1.3}
\end{equation}
be a fibre-to-fibre manifold mapping such that the diagram
\[
\begin{array}{rcccl}
& {Y} &  \op\longrightarrow^{\Phi} & {Y'} &  \\
{} &\put(0,10){\vector(0,-1){20}} & & \put(0,10){\vector(0,-1){20}} & {} \\
& {X} & \op\longrightarrow_{f} & {X'} &
\end{array}
\]
is commutative. Briefly,
one says that  (\ref{1.3}) is the fibred morphism $\Phi$ over $f$.
If $f=\Id_X$ is the identity mapping of $X$,
the fibred morphism (\ref{1.3}) is termed the fibred
morphism
\[
Y\op\to_XY'
\]
over $X$.

If the fibred morphism $\Phi$ [Eq.(\ref{1.3})] over $X$ is a submanifold,
$Y\to X$ is called the fibred submanifold of $Y'\to X$.

In particular, let $X_X$ denotes the fibred manifold
\[
\Id_X: X\hookrightarrow X.
\]
Given a fibred manifold $Y\to X$, a fibred morphism
$X_X\to Y$ over $X$, by definition, is a global section of $Y$.
It is a closed imbedded submanifold. Recall that, for each point
$x\in X$, a fibred manifold over $X$ has a local section (or simply a
section) over an open neighborhood of $x$.

More generally, let $N$ be
an imbedded submanifold of $X$. A fibred morphism $N_N\to Y$ over
$N\hookrightarrow X$ is called a local section of $Y$ over $N$.
Due to the properties which are required of a manifold, we can refer to the
well-known theorem.

\begin{theorem}{2.1} If a fibred manifold $Y\to X$ has a global
section, every local section of $Y$ over a closed imbedded
submanifold of $X$ is extended to a global section of $Y$.
\end{theorem}

A fibred morphism of a bundle $Y$ to a bundle $Y'$ is called a
bundle morphism. A subbundle is the image of a bundle by
a bundle morphism which is a submanifold.
A bundle imbedding and a bundle diffeomorphism are usually callled a bundle
monomorphism and a bundle isomorphism respectively.

Given a fibred manifold $Y\to X$, every manifold mapping
$f: X' \to X $ yields a fibred manifold over $X'$ as follows.
The pullback of the fibred manifold $Y$ by
the above-mentioned mapping $f$ is defined to be the
fibred manifold $f^*Y$ comprising the pairs
\[
\{(y,x')\in Y\times X' \mid \quad \pi(y) =f(x')\}
\]
together with the surjection $(y,x')\to x'$. Roughly speaking, the
fibre of $f^*X$ at a point $x'\in X'$ is the same as that of $Y$ at the
point $f(x)\in X$. In particular, every global section
$s$ of the fibred manifold $Y$ defines the corresponding global section
\[
(f^*s )(x') = ((s\circ f)(x'),x'),\qquad x'\in X',
\]
of the pullback $f^*Y\to X'$.

\begin{example}{2.2}
Let $Y\to X$ be a fibred manifold. If a mapping $f:X\to X'$
is a submanifold, the pullback
$f^*Y$ is called the restriction $Y\mid_{f(X')}$ of the fibred manifold
$Y$ to the submanifold $f(X')\subset X$ (or the portion of $Y$ over $f(X')$).
\end{example}

\begin{example}{2.3}
The product of fibred manifolds
$\pi:Y\to X$ and $\pi':Y'\to X$ over $X$, by definition, is the total space
of the pullbacks
\[
\pi^*Y'={\pi'}^*Y=Y\op\times_X Y'.
\]
\end{example}

Now, we consider bundles which possess algebraic structure.

A group bundle $Y\to X$ is defined to be a bundle together with the triple
\ben
&&\wh m: Y\op\times_X Y \to Y,\nonumber \\
&& \wh i: Y\to Y, \nonumber \\
&& e:X\to Y \label{1.5}
\een
of bundle morphisms $\wh m$, $\wh i$ and a global
section $e$ which make each fibre  of $Y$ into the group:
\be
&&  \wh m((e\circ \pi)(y) ,y)=\wh m(y,(e\circ \pi)(y)) =y, \\
&&\wh m(\wh i(y) ,y) =\wh m(y, \wh i(y))=(e\circ\pi)(y).
\ee
Note that a standard fibre of a group bundle is not a group in general.

A general affine bundle is defined to be the triple $(Y,\overline
Y,r)$ of a bundle $Y\to X$ and a group bundle $\overline Y\to X$
together with a bundle morphism
\begin{equation}
r : Y\op\times_X\overline Y\to Y\label{1.5'}
\end{equation}
over $X$ which makes each fibre $Y_x$, $x\in X$,
of $Y$ into a general affine space modelled on the left [resp. on the right]
on the group $\overline Y_x$ acting freely and transitively on $Y_x$
on the left [resp. on the right].

\begin{example}{2.4}
A vector bundle is an
additive group bundle provided with atlases of linear bundle coordinates.
Its standard fibre, by definition, is a vector space.
\end{example}

\begin{example}{2.5}
An affine bundle $Y$ is a general affine
bundle modelled on a vector bundle $\overline Y$.
It is endowed with atlases of
affine bundle coordinates $(x^\lambda, y^i)$ such that
\[
r : (x^\la,
y^i)\times(x^\la, \overline y^i)\mapsto (x^\la, y^i +\overline y^i)
\]
where $(x^\lambda,
\overline y^i)$ are linear bundle coordinates of the vector bundle $\overline
Y$. In particular,
every vector bundle $Y$ has the canonical structure of an affine bundle
modelled on $Y$ itself by the morphism
\[
r :(y,y')\mapsto y+y'.
\]
\end{example}

\begin{example}{2.6}
A principal  bundle
\[
\pi_P :P\to X
\]
with a structure
Lie group $G$ is a general affine bundle modelled on the right
on the trivial group
bundle $X\times G$ where the group $G$ acts freely and
transitively on $P$ on the right:
\ben
&&r_g : p\mapsto pg, \label{1}\\
&& p\in P,\quad g\in G. \nonumber
\een
A principal bundle $P$ is also the general affine bundle modelled on the
left on the associated group bundle $\wt P$ with the standard fibre
$G$ on which the structure group $G$ acts by the adjoint representation.
The corresponding bundle morphism reads
\[
\wt P\times P\ni (\wt p,p)\mapsto \wt pp\in P.
\]
Note that the standard fibre of the group bundle $\wt P$ is the group
$G$ , while that of the principal bundle $P$ is the group
space of $G$ on which the structure group $G$ acts on the left.
\end{example}

\section{Vertical tangent bundles}

In this Section, tangent, cotangent, vertical tangent and vertical
cotangent bundles of fibred manifolds are considered.

Let $TY\to Y$ be the tangent bundle
of a fibred manifold $Y\to X$. The following diagram commutes
\[
\begin{array}{rcccl}
& {TY} &  \op\longrightarrow^{T\pi} & {TX} &  \\
{_{\pi_Y}} &\put(0,10){\vector(0,-1){20}} & & \put(0,10){\vector(0,-1){20}} &
{_{\pi_X}} \\
& {Y} & \op\longrightarrow_{\pi} & {X} &
\end{array}
\]
where
\begin{equation}
T\pi :TY\to TX \label{1.6}
\end{equation}
is a fibred manifold. Note that $T\pi$ is still the bundle morphism of
the bundle $TY\to Y$ to $TX$ over $\pi$ and the fibred morphism of
the fibred manifold $TY\to X$
to $TX$ over $X$. There is also the canonical surjection
\[
\pi_T: TY\op\to_Y TX\op\times_XY.
\]

Given the fibred coordinates (\ref{1.2}) of a fibred manifold $Y$, the
corresponding induced coordinates of $TY$ are
\be
&&(x^\la,y^i,\dot x^\la, \dot y^i), \\
&& \dot y'^i=\frac{\dr y'^i}{\dr y^j}\dot y^j.
\ee
A glance at this expression shows that
the tangent bundle $TY\to Y$ of a fibred manifold $Y$ has the vector
subbundle
\[
VY = \Ker T\pi
\]
where $T\pi$ is regarded as the fibred morphism of $TY\to X$ to $TX$
over $X$. The subbundle $VY$ consists of tangent vectors to fibres of
$Y$. It is called the vertical tangent bundle of $Y$ and provided
with the induced coordinates $(x^\la,y^i,\dot y^i)$ with respect to
the fibre bases $\{\dr_i\}$.

\begin{remark}
Let $\Phi$ be a fibred morphism of a fibred manifold $Y\to X$ to a
vector bundle $Y'\to X$ over $X$. By $\Ker\Phi$ is meant the preimage
$\Phi^{-1}(\wh 0(X))$ of the global zero section $\wh 0(X)$ of the vector
bundle $Y'$ by $\Phi$.
\end{remark}

The vertical cotangent bundle $V^*Y\to Y$ of $Y$, by definition, is the
vector bundle dual to the vertical tangent bundle $VY\to Y$.
It is not a subbundle of the cotangent bundle $T^*Y$ of $Y$, but there
is the canonical surjection
\[
T^*Y\op\to_YV^*Y.
\]
We shall denote by $\{\ol dy^i\}$ the
fibre bases for $V^*Y$ which are dual to the fibre bases $\{\dr_i\}$ for
$VY$. Compare the transformation laws
\[
\ol dy'^i=\frac{\dr y'^i}{\dr y^j}\ol dy^j
\]
and
\[
dy'^i=\frac{\dr y'^i}{\dr y^j}dy^j +\frac{\dr y'^i}{\dr x^\la}dx^\la
\]
where $\{dx^\la, dy^i\}$ are holonomic fibre bases for $T^*Y$.

With $VY$ and $V^*Y$, we have the following exact sequences of bundles over a
fibred manifold $Y\to X$:
\bea
&& 0\to VY\hookrightarrow TY\to Y\op\times_X TX\to 0,
\label{1.8a} \\
&& 0\to Y\op\times_X T^*X\hookrightarrow T^*Y\to V^*Y\to 0
\label{1.8b}
\eea
For the sake of simplicity, we shall denote by $TX$ and $T^*X$
the pullbacks
\[
Y\op\times_X TX, \qquad Y\op\times_X T^*X
\]
and other pullbacks involving tangent and cotangent bundles of $X$.

Every splitting
\be
&&Y\op\times_X TX\op\hookrightarrow_Y TY, \\
&&\dr_\la\mapsto\dr_\la +\G^i_\la (y)\dr_i,
\ee
of the exact sequence (\ref{1.8a}) and
\be
&& V^*Y\to T^*Y,\\
&&\ol d^i\mapsto dy^i -\G^i_\la (y)dx^\la,
\ee
of the exact sequence (\ref{1.8b}), by
definition, corresponds to a certain connection on the fibred manifold
$Y\to X$, and {\it vice versa}.

Let $\Phi$ be a fibred morphism  of a fibred manifold
$Y\to X$ to a fibred manifold
$Y'\to X'$ over $f:X\to X'$. The tangent morphism
\[
T\Phi :TY\to TY'
\]
to $\F$ reads
\begin{equation}
(\dot{x'}^\la,\dot{y'}^i)\circ T\F =(\dr_\m f^\la\dot
x^\m,\dr_\m\Phi^i\dot x^\m +\dr_j\Phi ^i\dot y^j). \label{1.7}
\end{equation}
It is both the linear bundle morphism
\[
\begin{array}{rcccl}
& {TY} &  \op\longrightarrow^{T\Phi} & {TY'} &  \\
{} & \put(0,10){\vector(0,-1){20}} & & \put(0,10){\vector(0,-1){20}} & {} \\
& {Y} & \op\longrightarrow_{\Phi} & {Y'} &
\end{array}
\]
over $\Phi$ and the fibred morphism
\[
\begin{array}{rcccl}
& {TY} &  \op\longrightarrow^{T\Phi} & {TY'} &  \\
{} &\put(0,10){\vector(0,-1){20}} & & \put(0,10){\vector(0,-1){20}} & {} \\
& {TX} & \op\longrightarrow_{Tf} & {TX'} &
\end{array}
\]
over the tangent morphism $Tf$ to $f$.

In particular,
if $Y\to X$ is a bundle with an algebraic structure, also the
bundle $TY\to TX$ [Eq.(\ref{1.6})] possesses an algebraic structure as a
consequence of compatibility with the corresponding tangent morphisms to the
morphisms (\ref{1.5}) and (\ref{1.5'}).

\begin{example}{3.1}
If $Y$ is a vector bundle, so is the bundle (\ref{1.6}).
\end{example}

\begin{example}{3.2}
If $Y$ is an affine bundle modelled on a vector bundle $\overline
Y$, so is the bundle (\ref{1.6}) modelled on the vector
bundle $T\overline Y\to TX$.
\end{example}

\begin{example}{3.3}
Given a principal bundle $P\to X$, the bundle
$TP\to TX$ is a principal bundle
\[
TP\op\times_{TX} T(X\times G)\to TP
\]
with the structure group $TG=G\times {\got g}_l$
where ${\got g}_l$ is the left Lie algebra of left-invariant vector fields
on the group $G$.
\end{example}

The restriction of the tangent morphism $T\Phi$ to
the vertical tangent subbundle $VY$ of $TY$
yields  the vertical tangent morphism
\[
\begin{array}{rcccl}
 & {VY} &  \op\longrightarrow^{V\Phi} & {VY'} &  \\
{} &\put(0,10){\vector(0,-1){20}} & & \put(0,10){\vector(0,-1){20}} & {} \\
 & {Y} & \op\longrightarrow_{\Phi} & {Y'} &
\end{array}
\]
\begin{equation}
\dot y'^i\circ V\Phi =\dr_j\Phi^i\dot y^j.\label{2}
\end{equation}
This also is a linear bundle morphism over $\Phi$.

Vertical tangent bundles of many fibred manifolds utilized in field
theory meet the following simple structure.

\begin{definition}{3.4}
One says that a fibred manifold $Y\to X$ has vertical splitting if
there exists the linear bundle isomorphism
\begin{equation}
\al : VY\op\to_Y Y\op\times_X \overline Y \label{1.9}
\end{equation}
where $\overline Y\to X$ is a vector bundle.
\end{definition}

\begin{example}{3.5}
A vector bundle $Y\to X$ has the canonical vertical splitting
\begin{equation}
VY=Y\op\times_X Y. \label{1.10}
\end{equation}
\end{example}

\begin{example}{3.6} An
affine bundle $Y$ modelled on a vector bundle $\overline Y$ has the canonical
vertical splitting
\begin{equation}
VY=Y\op\times_X\overline Y.\label{48}
\end{equation}
\end{example}

\begin{example}{3.7}
A principal bundle $P\to X$ with a structure Lie group $G$ possesses the
canonical trivial vertical splitting
\ben
&&\al:VP\to P\times{\got g}_l, \nonumber \\
&& \pr_2\circ\al\circ e_m=J_m ,\label{1.28}
\een
where $\{J_m\}$ is a basis for the left Lie algebra
${\got g}_l$ and $e_m$ denotes the
corresponding fundamental vector fields on $P$.
\end{example}

The fibred coordinates (\ref{1.2}) of a fibred manifold $Y$ are called
adapted to the vertical splitting (\ref{1.9}) if the
 induced coordinates of the vertical tangent bundle $VY$ take the form
\[
\dot y^i = \overline y^i\circ\al
\]
where $(x^\m,\overline y^i)$ are bundle coordinates
of $\overline Y$. In this case, coordinate transformations
$\dot y^i\to\dot y'^i$ are independent of the coordinates $y^i$.

For instance, linear
bundle coordinates of a vector bundle and affine bundle coordinates of an
affine bundle are always adapted to their canonical vertical splittings
(\ref{1.10}) where $\dot y^i=y^i$ and (\ref{48}) where $\dot y^i=\ol y^i$.

We shall refer to the following fact.

\begin{lemma}{3.8}
Let $Y\to X$ and $Y'\to X$ be fibred manifolds
and $\Phi: Y\to Y'$  a fibred morphism over $X$. Let $V\Phi$ be the vertical
tangent morphism to $\Phi$. If $Y'$ admits vertical splitting
\[
VY'=Y'\op\times_X\overline Y,
\]
then there exists the linear bundle morphism
\begin{equation}
\ol V\Phi:VY\op\to_Y Y\op\times_X \ol Y'\label{64}
\end{equation}
over $Y$ given by the coordinate expression
\[
{\ol y'}^i\circ\ol V\Phi =\dr_j\Phi^i\dot y^j.
\]
\end{lemma}

\section{Tangent-valued forms}

Tangent-valued forms play the prominent role in the mathematical technique
which we utilize. Connections and
the major ingredients in the jet manifold machinery
are phrased in terms of tangent-valued forms.

Recall that an exterior r-form on a manifold $M$  with
local coordinates $(z^\la)$, by definition, is a section
\[
\f =\f_{\la_1\dots\la_r}
dz^{\la_1}\wedge\dots\wedge dz^{\la_r}
\]
of the skew-symmetric tensor bundle $\op\w^r T^*M$.
Its contraction with a vector field
\[
u = u^\m\dr_\m
\]
on $M$ is given by the expression
\be
&& u\rfloor\f =\sum_{p=1}^r (-1)^{p-1} u^\m
\f_{\la_1\dots\la_{p-1}\m\la_{p+1}\dots\la_r} \\
&&\qquad
dz^{\la_1}\wedge\dots\wedge dz^{\la_{p-1}}\wedge
dz^{\la_{p+1}}\wedge\dots \wedge dz^{\la_r}.
\ee
We denote by $\op\La^r\cT^*(M)$ the sheaf of exterior
r-forms on a manifold $M$.

A tangent-valued $r$-form on a manifold $M$ is defined to be a section
\[
\phi = \phi_{\la_1\dots\la_r}^\m
dz^{\la_1}\wedge\dots\wedge dz^{\la_r}\otimes\dr_\m
\]
of the bundle
\[
\op\w^r T^*M\otimes TM\to M.
\]
In particular, tangent-valued 0-forms are vector fields on
$M$.

\begin{example}{4.1}
There is the 1:1 correspondence between the tangent-valued
1-forms
\[
\th:M\to T^*M\otimes TM
\]
on $M$ and the linear bundle morphisms
\ben
&&\th:TM\to TM,\nonumber\\
&&\th: T_zM\ni t\mapsto t\rfloor\th(z)\in T_zM,\label{29}
\een
and
\be
&&\th:T^*M\to T^*M,\\
&&\th: T^*_zM\ni t^*\mapsto \th(z)\rfloor t^*\in T^*_zM,
\ee
over $M$.
For instance, $\Id_{TM}$ corresponds to the canonical tangent-valued 1-form
\be
&&\th_M = dz^\la\otimes \dr_\la, \\
&& \dr_\la \rfloor \th_M = \dr_\la,
\ee
on the manifold $M$.
\end{example}

Tangent-valued $r$-forms on a manifold $M$ constitute the sheaf
\[
\op\La^r\cT^*(M)\otimes\cT(M)
\]
where $\cT(M)$ is the sheaf of vector fields on $M$.
The sheaf $\cT(M)$ is brought into the
sheaf of Lie algebras with respect to the commutation bracket. This algebraic
structure is generalized to tangent-valued forms on $M$ by means of
the Fr\"olicher-Nijenhuis (F-N) bracket.

The F-N bracket is defined to be the sheaf morphism
\[
\op\La^r \cT^*(M)\otimes\cT(M)\times\op\La^s
\cT^*(M)\otimes\cT(M) \to \op\La^{r+s} \cT^*(M)\otimes\cT(M),
\]
\be
&& [\phi ,\si] = [\al\otimes u,\bt\otimes v] \\
&&\qquad= \al\wedge\bt\otimes [u,v] +
\al\wedge \bL_u\bt\otimes v - (-1)^{rs}\bt\wedge \bL_v\al\otimes u \\
&&\qquad +(-1)^r(v\rfloor \al)\wedge d\bt\otimes u-
(-1)^{rs+s}(u\rfloor\bt)\wedge d\al\otimes v,
\ee
\[
\al\in\op\La^r\cT^*(M),\quad \bt\in
\op\La^s\cT^*(M), \quad u,v\in \cT(M),
\]
where $\bL_u$ and $\bL_v$ are the Lie derivatives of exterior forms.
Its coordinate expression reads
\be
&& [\phi,\si] =
(\phi_{\la_1\dots\la_r}^\nu\dr_\n\si_{\la_{r+1}\dots\la_{r+s}}^\m \\
&& \qquad -(-1)^{rs}\si_{\la_1\dots\la_s}^\nu\dr_\nu\phi_{\la_{s+1}\dots
\la_{r+s}}^\m -r\phi_{\la_1\dots\la_{r-1}\nu}^\m
\dr_{\la_r}\si_{\la_{r+1}\dots\la_{r+s}}^\nu \\
&& \qquad +(-1)^{rs}s\si_{\la_1\dots\la_{s-1}\nu}^\m\dr_{\la_s}
\phi_{\la_{s+1}\dots\la_{r+s}}^\nu)
dz^{\la_1}\wedge\dots\wedge dz^{\la_{r+s}}\otimes\dr_\m.
\ee
The F-N bracket brings the sheaf of
tangent-valued forms into the sheaf of graded Lie algebras:
\[
[\phi,\si] = -(-1)^{\mid\phi\mid\mid\si\mid}[\si,\phi],
\]
\[
[\th,[\phi,\si]] =[[\th,\phi],\si] +
(-1)^{\mid\th\mid\mid\phi\mid} [\phi,[\th,\si]],
\]
 \[
\phi,\si,\th \in\La\cT^*(M)\otimes\cT(M),
\]
where by $\mid\phi\mid$ is meant the degree of the form $\phi$.

Given a tangent-valued form  $\th$, the Nijenhuis
differential is defined to be the sheaf morphism
\ben
&& d_\th : \si\mapsto d_\th\phi = [\th,\si], \label{33}\\
&& d_\th[\phi,\si] = [d_\th\phi,\si]
+ (-1)^{\mid\th\mid\mid\phi\mid}[\phi,d_\th\si],\nonumber\\
&& d_\th d_\eta - (-1)^{\mid\eta\mid\mid\th\mid}d_\eta d_\th= d_{[\th,\eta]}.
\nonumber
\een

\begin{example}{4.2}
If $\th=u$ is a vector field on the manifold $M$, we get the
Lie derivative of tangent-valued forms
\be
&& \bL_u\si=[u,\si] =(u^\n\dr_\n\si_{\la_1\dots\la_s}^\m -
\si_{\la_1\dots\la_s}^\n\dr_\n u^\m \\
&& \qquad +s\si^\m_{\la_1\dots\la_{s-1}\nu}\dr_{\la_s}u^\nu)
dx^{\la_1}\wedge\dots\wedge dx^{\la_s}\otimes\dr_\m.
\ee
\end{example}

The Nijenhuis differential
(\ref{33}) can be extended to exterior forms by the rule
\begin{equation}
d_\th\si=\al\w\bL_u\si +(-1)^rd\al\w(u\rfloor\si) =\th\rfloor d\si
+(-1)^rd(\th\rfloor\si), \label{32}
\end{equation}
\[
\th=\al\otimes u, \qquad \al\in\op\La^r\cT^*(M), \qquad u\in\cT(M), \qquad
\si\in\op\La^s\cT^*(M).
\]
Its coordinate expression is
\be
&& d_\th\si =
(\th_{\la_1\dots\la_r}^\nu\dr_\nu\si_{\la_{r+1}\dots\la_{r+s}} \\
&& \qquad +(-1)^{rs}s\si_{\la_1\dots\la_{s-1}\nu}\dr_{\la_s}
\th_{\la_{s+1}\dots\la_{r+s}}^\nu)
dz^{\la_1}\wedge\dots\wedge dz^{\la_{r+s}}.
\ee

\begin{example}{4.3}
If $\th=\th_M$, the Nijenhuis differential (\ref{32}) recovers
the familiar exterior differential
\[
d_{\th_M}\si=d\si.
\]
\end{example}

 Let  $Y\to X$ be a fibred manifold provided with the fibred coordinates
(\ref{1.2}). Hencerforth, we shall mean by differential forms (or simply
forms) on fibred manifolds exterior forms, tangent-valued forms
and pullback-valued forms.

We consider the following particular subsheafs of
exterior and tangent-valued forms on a fibred manifold $Y\to X$:
\begin{itemize}
\item exterior horizontal forms
\be
 && \phi : Y\to\op\w^r T^*X, \\
&&\phi =\phi_{\la_1\dots\la_r}(y)dx^{\la_1}\wedge\dots\wedge dx^{\la_r};
\ee
\item tangent-valued horizontal forms
\be
 && \phi : Y\to\op\w^r T^*X\op\otimes_Y TY,\\
&& \phi =dx^{\la_1}\wedge\dots\wedge dx^{\la_r}\otimes
[\phi_{\la_1\dots\la_r}^\m (y)\dr_\m +
\phi_{\la_1\dots\la_r}^i(y) \dr_i];
\ee
\item tangent-valued projectable horizontal forms
\[
\phi =dx^{\la_1}\wedge\dots\wedge dx^{\la_r}\otimes
[\phi_{\la_1\dots\la_r}^\m(x)\dr_\m +
\phi_{\la_1\dots\la_r}^i(y) \dr_i];
\]
\item vertical-valued horizontal forms
\be
&&\phi : Y\to\op\w^r T^*X\op\otimes_Y VY,\\
&&\phi =\phi_{\la_1\dots\la_r}^i(y)
dx^{\la_1}\wedge\dots\wedge dx^{\la_r}\otimes\dr_i.
\ee
\end{itemize}

\begin{example}{4.4}
Recall that a vector field $u$

on a fibred manifold $Y\to X$ is termed projectable when it covers a
vector field $\tau_u$ on $X$ such that the following diagram commutes:
\[
\begin{array}{rcccl}
 & {Y} &  \op\longrightarrow^{u} & {TY} &  \\
{_\pi} &\put(0,10){\vector(0,-1){20}} & & \put(0,10){\vector(0,-1){20}} &
{_{T\pi}} \\
 & {X} & \op\longrightarrow_{\tau_u} & {TX} &
\end{array}
\]
\[
u=u^\m (x)\dr_\m +u^i(y)\dr_i,
\]
\[
\tau_u=u^\mu(x)\dr_\mu.
\]
 In particular, a vertical vector field
\[
u= u^i\dr_i: Y\to VY
\]
on $Y$ is a projectable vector field which covers $\tau_u=0$.
\end{example}

\begin{example}{4.5}
Vertical-valued horizontal 1-forms
\be
&&\sigma : Y\to T^*X\op\otimes_Y VY, \\
&& \si = \si_\la^i dx^\la\otimes\dr_i,
\ee
on $Y$ are termed the soldering forms.
\end{example}

By pullback-valued forms on a fibred manifold $Y\to X$ are meant
the morphisms
\ben
&&Y\to \op\w^r T^*Y\op\otimes_Y TX, \label{1.11} \\
&& \phi =\phi_{\la_1\dots\la_r}^\m (y)
dx^{\la_1}\wedge\dots\wedge dx^{\la_r}\otimes \dr_\m,\nonumber
\een
and
\ben
&&Y\to \op\w^r T^*Y\op\otimes_Y V^*X, \label{87}\\
&& \phi =\phi_{\la_1\dots\la_ri}(y)
dx^{\la_1}\wedge\dots\wedge dx^{\la_r}\otimes \ol dy^i.\nonumber
\een

Note that the forms (\ref{1.11}) are not tangent-valued forms.
The pullbacks
\[
\phi =\phi_{\la_1\dots\la_r}^\mu (x)
dx^{\la_1}\wedge\dots\wedge dx^{\la_r}\otimes\dr_\mu
\]
of tangent-valued forms on $X$ onto $Y$ by $\pi$ exemplify the pullback-valued
forms (\ref{1.11}). In
particular, we shall refer to the pullback $\pi^*\th_X$ of the canonical
form $\th_X$ on
the base $X$ onto $Y$ by $\pi$. This is a pullback-valued horizontal 1-form on
$Y$
which we denote by the same symbol
\ben
&&\th_X:Y\to T^*X\op\otimes_Y TX, \nonumber\\
&& \th_X =dx^\la\otimes\dr_\la. \label{12}
\een

The pullback-valued forms (\ref{87}) are exemplified by
the canonical bundle monomorphism
\ben
&&\op\w^nT^*X\op\otimes_YV^*Y\op\hookrightarrow_Y\op\w^{n+1}T^*Y
\nonumber\\
&& \om\otimes\ol dy^i\mapsto\om\w dy^i.\label{86}
\een

All horizontal $n$-forms on a fibred
manifold $Y\to X$ are called horizontal densities.

\section{First order jet manifolds}

The major ingredients in the jet machinery are the following.
\begin{itemize}
\item Jet spaces of sections of fibred manifolds are finite-dimensional
manifolds.
\item Jets are expressed into familiar tangent-valued forms.
\item There exists the canonical horizontal splitting of the exact
sequences (\ref{1.8a}) and (\ref{1.8b}) over the first order jet manifolds.
\end{itemize}

Given a fibred manifold $Y\to X$, let us consider the equivalence
classes $j^1_xs$, $x\in X$, of sections $s$ of $Y$ so that different sections
$s$ and $s'$ belong to the same class $j^1_xs$ if and only if
\[
Ts\mid _{T_xX} =Ts'\mid_{T_xX}.
\]
Roughly speaking, sections $s\in j^1_xs$  are identified by their values
\[
s^i(x)={s'}^i(x)
\]
and values of their first order partial derivatives
\[
\dr_\mu s^i(x)=\dr_\mu{s'}^i(x)
\]
at the point $x$ of $X$. One terms  $j^1_xs$
the jet of $s$ at the point $x\in X$. We denote by $J^1Y$ the set of
all jets of sections of a fibred manifold $Y$.

There are the natural surjections
\begin{equation}
\pi_1:J^1Y\ni j^1_xs\mapsto x\in X, \label{1.14}
\end{equation}
and
\begin{equation}
\pi_{01}:J^1Y\ni j^1_xs\mapsto s(x)\in Y, \label{1.15}
\end{equation}
which form the commutative diagram
\[
\begin{array}{rcl}
{J^1Y}  & \op\longrightarrow^{\pi_{01}} &  {Y}  \\
{_{\pi_1}} & \searrow  \swarrow & {_\pi}\\
& {X} &
\end{array}
\]

There exist several  equivalent ways in order to
provide the jet set $J^1Y$ with the
manifold structure. The result is the following.

\begin{proposition}{5.1}
Let $Y\to X$ be a fibred manifold endowed with fibred coordinate atlases
(\ref{1.2}). The jet set $J^1Y$ of $Y$, being provided with the adapted
coordinate atlases
\ben
&&(x^\la,y^i,y_\la^i),\label{49}\\
&&(x^\la,y^i,y_\la^i)(j^1_xs)=(x^\la,s^i(x),\dr_\la s^i(x)),\nonumber\\
&&{y'}^i_\la = (\frac{\dr{y'}^i}{\dr y^j}y_\m^j +
\frac{\dr{y'}^i}{\dr x^\m})\frac{\dr x^\m}{\dr{x'}^\la},\label{50}
\een
satisfies the conditions which we require of a manifold. It is called
the first order jet manifold (or simply the jet manifold) of $Y$.
The surjection (\ref{1.14}) is a fibred manifold. The
 surjection (\ref{1.15}) is a bundle. If $Y\to X$ is a bundle, so is
the surjection (\ref{1.14}).
\end{proposition}

A glance at the transformation law (\ref{50}) shows
that the bundle $J^1Y\to Y$
is an affine bundle. We call it the jet bundle.
It is  modelled on the vector bundle
\begin{equation}
T^*X \op\otimes_Y VY\to Y.\label{23}
\end{equation}

\begin{example}{5.2}
Let $Y$ be the trivial bundle
\[
\pr_2: V\times{\Bbb R}^m\to {\Bbb R}^m.
\]
The corresponding jet manifold
$J^1Y\to {\Bbb R}^m$ [Eq.(\ref{1.14})]
is a trivial bundle. Note that, if a fibred manifold
$Y\to X$ is trivial, the fibred jet manifold $J^1Y\to X$ fails to
be trivial in general. For instance, the jet manifold of the bundle
\[
\pr_2: \Bbb R\times X\to X
\]
is the affine bundle modelled on the pullback
\[
T^*X\op\times_X\Bbb R
\]
which is not a trivial bundle over $X$ in general.
\end{example}

\begin{example}{5.3}
Given fibred manifolds $Y\to X$ and $Y'\to X$, there is the natural
diffeomorphism
\begin{equation}
J^1(Y\op\times_X Y')=J^1Y\op\times_X J^1Y'. \label{42}
\end{equation}
\end{example}

\begin{proposition}{5.4}
There exist the following canonical affine bundle monomorphisms over $Y$:
\begin{itemize}\begin{enumerate}
\item the contact map
\ben
&&\la:J^1Y\hookrightarrow
T^*X \op\otimes_Y TY,\label{18}\\
&&\la=dx^\la\otimes\wh{\dr}_\la=dx^\la
\otimes (\dr_\la + y^i_\la \dr_i),\nonumber
\een
\item the complementary morphism
\ben
&&\th_1:J^1Y \hookrightarrow T^*Y\op\otimes_Y VY,\label{24}\\
&&\th_1=\wh{d}y^i \otimes \dr_i=(dy^i- y^i_\la dx^\la)\otimes
\dr_i.\nonumber
\een
\end{enumerate}\end{itemize}
\end{proposition}

These monomorphisms enable us to handle jets as
familiar tangent-valued forms.

Let $\Phi$ be a fibred morphism  of a fibred manifold
$Y\to X$ to a fibred manifold $Y'\to X$ over a diffeomorphism $f$ of $X$.
There exists the first order jet prolongation (or simply the jet prolongation)
of $\Phi$ to the morphism
\[
J^1\Phi : J^1Y  \to J^1Y',
\]
\begin{equation}
J^1\Phi :j_x^1s\mapsto j_{f(x)}^1(\Phi\circ
s\circ f^{-1}),\label{26}
\end{equation}
\[
{y'}^i_\la\circ J^1\Phi=\dr_\la(\Phi^i\circ f^{-1}) +\dr_j(\Phi^iy^j_\la
\circ f^{-1}).
\]
It is both
an affine bundle morphism over $\Phi$ and a fibred morphism
over the diffeomorphism $f$:
\[
\begin{array}{rcccl}
& {J^1Y} &  \op\longrightarrow^{J^1\Phi} & {J^1Y'} &  \\
{_{\pi_{01}}} &\put(0,10){\vector(0,-1){20}} & & \put(0,10){\vector(0,-1){20}}
& {_{{\pi'}_{01}}} \\
& {Y} & \op\longrightarrow_{\Phi} & {Y'} & \\
{} &\put(0,10){\vector(0,-1){20}} & & \put(0,10){\vector(0,-1){20}} &
{} \\
& {X} & \op\longrightarrow_{f} & {X'} &
\end{array}
\]

The first order jet prolongations (\ref{26}) of
fibred morphisms satisfy the chain rules
\[
J^1(\Phi\circ\Phi')=J^1\Phi\circ J^1\Phi',
\]
\[
J^1(\Id_Y)=\Id_{J^1Y}.
\]
If $\Phi$ is a surjection [resp. an injection], so is $J^1\Phi$.

In particular, every section $s$ of a fibred manifold $Y\to X$ admits the
first order jet prolongation to the section $\ol s=J^1s$ of the
fibred jet manifold $J^1Y\to X$:
\begin{equation}
(J^1s)(x)\op =^{\rm def} j_x^1s, \label{27}
\end{equation}
\[
(y^i,y_\la^i)\circ J^1s= (s^i(x),\dr_\la s^i(x)).
\]
We have
\[
\la\circ J^1s=Ts
\]
where $\la$ is the contact map (\ref{18}).

Every  vector field
\[
u = u^\la\dr_\la + u^i\dr_i
\]
on a fibred manifold $Y\to X$ has the jet lift to the projectable
vector field
\ben
&&\overline u =r_1\circ J^1u: J^1Y\to J^1TY\to TJ^1Y,\nonumber \\
&& \overline u =
u^\la\dr_\la + u^i\dr_i + (\dr_\la u^i+y^j_\la\dr_ju^i
 - y_\m^i\dr_\la u^\m)\dr_i^\la, \label{1.21}
\een
on the fibred jet manifold $J^1Y\to X$ where $J^1TY$ is the jet
manifold of the fibred manifold $TY\to X$.
To construct $\ol u$, we use the canonical fibred morphism
\[
r_1: J^1TY\to TJ^1Y,
\]
\[
\dot y^i_\la\circ r_1 = (\dot y^i)_\la-y^i_\m\dot x^\m_\la,
\]
over
\[
J^1Y\op\times_YTY.
\]
In particular, there exists the canonical isomorphism
\ben
&&VJ^1Y=J^1VY, \label{1.22}\\
&& \dot y^i_\la=(\dot y^i)_\la,\nonumber
\een
where $J^1VY$
is the jet manifold of the fibred manifold $VY\to X$ and $VJ^1Y$ is the
vertical tangent bundle of the fibred manifold $J^1Y\to X$.
As a consequence, the jet lift (\ref{1.21}) of a vertical vector field $u$
on the fibred manifold $Y\to X$ consists with its first order jet
prolongation to the vertical vector field
\[
\overline u=J^1u=u^i\dr_i +(\dr_\la u^i+y^j_\la\dr_ju^i)\dr^\la_i
\]
on the fibred jet manifold $J^1Y\to X$.

If $Y\to X$ is a bundle endowed with an algebraic structure, this
algebraic structure also has the jet
prolongation to the jet bundle $J^1Y\to X$ due to the jet
prolongations of the corresponding morphisms.

\begin{example}{5.5}
If $Y$ is a vector bundle, $J^1Y\to X$
does as well. In particular, let $Y$ be a vector bundle and
$\langle\rangle$ the linear fibred morphism
\be
&&\langle\rangle: Y\op\times_X Y^*\op\to_X X\times\Bbb R,\\
&& r\circ\langle\rangle = y^iy_i,
\ee
where $(y^i)$ and $(\ol y^i)$ are the dual bundle coordinates of $Y$ and
$Y^*$ respectively.
The jet prolongation of $\langle\rangle$ is the linear fibred morphism
\be
&& J^1\langle\rangle : J^1Y\op\times_X J^1Y^* \op\to_X T^*X\times\Bbb R,\\
&& \dot x_\m \circ J^1\langle\rangle = y^i_\m y_i +y^iy_{i\m}.
\ee
\end{example}

\begin{example}{5.6}
Let $Y\to X$ and $Y'\to X$ be vector bundles and $\otimes$ the bilinear
bundle morphism
\be
&& \otimes :Y\op\times_X Y' \op\to_X Y\op\otimes_X Y', \\
&& y^{ik}\circ\otimes = y^iy^k.
\ee
The jet prolongation of $\otimes$ is the bilinear bundle
morphism
\be
 && J^1\otimes : J^1Y\op\times_X J^1Y'\op\to_X J^1(Y\op\otimes_X Y'), \\
&& y^{ik}_\m\circ J^1\otimes =y^i_\m y^k + y^i y^k_\m.
\ee
\end{example}

\begin{example}{5.7}
If $Y$ is an affine
bundle modelled on a vector bundle $\overline Y$, then
$J^1Y\to X$ is an affine bundle modelled on the vector bundle $J^1\overline
Y\to X$.
\end{example}

\begin{example}{5.8}
Let $J^1P$ be the first order jet manifold of a principal bundle $P\to
X$ with a structure Lie group $G$. The jet prolongation
\[
J^1P\times J^1(X\times G)\to J^1P
\]
of the canonical action (\ref{1}) brings the jet bundle $J^1P\to X$ into a
general affine bundle modelled on the group bundle
\begin{equation}
J^1(X\times G) = G\times (T^*X\op\otimes_X {\got g}_l) \label{1.30}
\end{equation}
 over
$X$. The jet bundle $J^1P\to X$ however fails to be a principal bundle since
the group bundle (\ref{1.30}) is not a trivial bundle over $X$ in general.
At the same time, $J^1P$ is the $G$ principal bundle
\[
C\op\times_X P\to C
\]
over the quotient
\begin{equation}
C=J^1 P/G\label{68}
\end{equation}
 of the jet bundle $J^1 P\to P$ by the first
order jet prolongations of the canonical morphisms (\ref{1}).
\end{example}

Application of jet manifolds to differential geometry is based on the
the canonical morphism
\[
J^1Y\op\times_Y TX\to J^1Y\op\times_Y TY
\]
over $J^1Y$ what means the canonical horizontal splitting of
the tangent bundle $TY$ determined over $J^1Y$ as follows.

The canonical morphisms (\ref{18}) and (\ref{24}) give rise to the bundle
monomorphisms
\ben
&& \wh\la: J^1Y\op\times_X TX\ni\dr_\la\mapsto\wh{\dr}_\la = \dr_\la\rfloor
\la\in J^1Y\op\times_Y TY, \label{30} \\
 &&\wh\th_1: J^1Y\op\times_Y V^*Y\ni
dy^i\mapsto\wh d y^i=\th_1 \rfloor dy^i\in J^1Y\op\times_Y T^*Y,
\label{1.18}
\een
\[
\wh{\dr}_\la=\dr_\la + y^i_\la \dr_i,\qquad
\wh{d}y^i =dy^i-y^i_\la dx^\la.
\]
The morphism (\ref{30}) determines the canonical horizontal
splitting of the pullback
\begin{equation}
J^1Y\op\times_Y TY=\wh\la(TX)\op\oplus_{J^1Y} VY,\label{1.20}
\end{equation}
\[
\dot x^\la\dr_\la
+\dot y^i\dr_i =\dot x^\la(\dr_\la +y^i_\la\dr_i) + (\dot y^i-\dot x^\la
y^i_\la)\dr_i.
\]
Similarly, the morphism (\ref{1.18})
yields the dual canonical horizontal splitting of
the pullback
\begin{equation}
J^1Y\op\times_Y T^*Y=T^*X\op\oplus_{J^1Y} \wh\th_1(V^*Y),\label{34}
 \end{equation}
\[
\dot x_\la dx^\la
+\dot y_i dy^i =(\dot x_\la + \dot y_iy^i_\la)dx^\la + \dot y_i(dy^i-
y^i_\la dx^\la).
\]
In other words, over $J^1Y$, we have the canonical horizontal splittings of
the tangent and cotangent bundles $TY$ and $T^*Y$ and the corresponding
splittings of the exact sequences (\ref{1.8a}) and (\ref{1.8b}).

\begin{example}{5.9}
Building on the canonical splittings (\ref{1.20}) and (\ref{34}),
one gets the following canonical horizontal splittings of
\begin{itemize}
\item a projectable vector field
\begin{equation}
u =u^\la\dr_\la +u^i\dr_i=u_H +u_V =u^\la (\dr_\la +y^i_\la
\dr_i)+(u^i - u^\la y^i_\la)\dr_i \label{31}
\end{equation}
on a fibred manifold $Y\to X$,
\item an exterior 1-form
\[
\si =\si_\la dx^\la + \si^idy^i=(\si_\la + y^i_\la\si_i)dx^\la+\si_i(dy^i-
y^i_\la dx^\la),
\]
\item a tangent-valued projectable horizontal form
\be
&&\phi = dx^{\la_1}\wedge\dots\wedge dx^{\la_r}\otimes
(\phi_{\la_1\dots\la_r}^\m\dr_\m +
\phi_{\la_1\dots\la_r}^i\dr_i)\\
&&\quad = dx^{\la_1}\wedge\dots\wedge dx^{\la_r}\otimes
[\phi_{\la_1\dots\la_r}^\m (\dr_\m  +y^i_\m \dr_i)
+(\phi_{\la_1\dots\la_r}^i - \phi_{\la_1\dots\la_r}^\m y^i_\m)\dr_i]
\ee
and, e.g., the canonical 1-form
\ben
&&\th_Y=dx^\la\otimes\dr_\la + dy^i\otimes\dr_i
=\la + \th_1=dx^\la\otimes\wh{\dr}_\la+\wh d
y^i\otimes\dr_i\nonumber \\
&&\qquad = dx^\la\otimes(\dr_\la + y^i_\la \dr_i)
+ (dy^i-y^i_\la dx^\la)\otimes\dr_i.\label{35}
\een
\end{itemize}
\end{example}

The splitting (\ref{35}) implies the
canonical horizontal splitting of the exterior differential
\begin{equation}
d=d_{\th_Y}=d_H+d_V=d_\la + d_{\th_1}. \label{1.19}
\end{equation}
Its components $d_H$ and $d_V$ act on the pullbacks
\[
\phi_{\la_1\dots\la_r}(y)dx^{\la_1}\wedge\dots\wedge dx^{\la_r}
\]
of horizontal exterior forms on a fibred manifold $Y\to X$ onto $J^1Y$ by
$\pi_{01}$. In this case, $d_H$ makes the sense of the total differential
\be
&& d_H\phi_{\la_1\dots\la_r}(y)dx^{\la_1}\wedge\dots\wedge dx^{\la_r}\\
&&\qquad =(\dr_\m + y^i_\m\dr_i)
\phi_{\la_1\dots\la_r}(y)dx^\m\w dx^{\la_1}\wedge\dots\wedge dx^{\la_r},
\ee
whereas $d_V$ is the vertical differential
\be
&& d_V\phi_{\la_1\dots\la_r}(y)dx^{\la_1}\wedge\dots\wedge dx^{\la_r}\\
&&\qquad =\dr_i
\phi_{\la_1\dots\la_r}(y)(dy^i-y^i_\m dx^\m)\w dx^{\la_1}\wedge\dots\wedge
dx^{\la_r}.
 \ee
If $\phi=\wt\phi\om$ is an exterior horizontal density on $Y\to X$, we have
\[
d\phi=d_V\phi=\dr_i\wt\phi dy^i\w\om.
\]

\newpage


\centerline{\large \bf Lecture 2. GENERAL CONNECTIONS}
\bigskip\bigskip

The most of differential geometric methods in field theory are based on
principal connections which model mediators of fundamental
interactions. We follow the general concept of connections as sections
of jet bundles, without appealing to transformation groups.

\section{First order connections}

One can introduce the notion of connections in several equivalent ways.
We start from the traditional definition of a connection as  horizontal
splitting of the tangent space to $Y$ at every point $y\in Y$. Roughly
speaking, a connection on a fibred manifold $Y\to X$ sends every
tangent vector $\dot x^\la\dr_\la$ to the base $X$ at $x=\pi (y)$ to the
tangent vector
\[
\dot x^\la (\dr_\la +\G^i_\la (y)\dr_i)
\]
to $Y$ at $y$.

\begin{definition}{6.1}
A connection on a fibred manifold $Y\to X$ is
defined to be  horizontal splitting of the exact sequences (\ref{1.8a}) and
(\ref{1.8b}). It means that there exists an imbedding
\begin{equation}
\G_T:TX\op\hookrightarrow_Y TY, \label{3}
\end{equation}
such that
\begin{equation}
TY=\G_T(TX)\op\oplus_Y VY.\label{11}
\end{equation}
\end{definition}

The imbedding (\ref{3}) implies the linear bundle morphism
\ben
&& \G=\G_T\circ\pi_T: TY\to TY, \nonumber \\
&& \dot x^\la\dr_\la +\dot y^i\dr_i\mapsto \dot x^\la (\dr_\la
+\G^i_\la (y)\dr_i), \label{4}
\een
which meets the conditions
\[
\G^2=\G, \qquad TV=\Ker\G
\]
and is given by the corresponding tangent-valued 1-form $\G$ [Eq.(\ref{29})]
associated with the splitting (\ref{3}). This form, in turn, yields the
linear bundle morphism
\ben
&&\G:T^*Y\to T^*Y, \nonumber \\
&& \dot x_\la dx^\la +\dot y_idy^i\mapsto (\dot x_\la +\G^i_\la (y)\dot y_i
)dx^\la, \label{5}
\een
which meets the conditions
\[
\G^2=\G, \qquad \G\mid_{T^*X}=\Id_{T^*X}
\]
and determines the horizontal splitting
\[
\G_T:V^*Y\op\to_Y (\th_Y-\G)(T^*Y)\subset T^*Y,
\]
\begin{equation}
T^*Y=T^*X\op\oplus_Y \G_T(V^*X), \label{6}
\end{equation}
of the exact sequence (\ref{1.8b}). The relations (\ref{4}) and (\ref{5})
lead to the equivalent definition of connections as tangent-valued forms.

\begin{definition}{6.2}
A connection on a fibred manifold $Y\to X$ is defined to be a
 tangent-valued projectable  horizontal 1-form $\G$ on $Y$ such that
\[
\G\rfloor\phi=\phi
\]
for all exterior horizontal 1-forms $\phi$ on $Y$. It is given by
the coordinate expression
\begin{equation}
\G =dx^\la\otimes(\dr_\la +\G^i_\la (y)\dr_i), \label{37}
\end{equation}
\[
{\G'}^i_\la = (\frac{\dr{y'}^i}{\dr y^j}\G_\m^j +
\frac{\dr{y'}^i}{\dr x^\m})\frac{\dr x^\m}{\dr{x'}^\la}.
\]
\end{definition}

It is readily observed that
the canonical horizontal splitting (\ref{1.20}) and (\ref{34}) of the
tangent and cotangent bundles $TY$ and $T^*Y$ of $Y$ over the jet bundle
$J^1Y\to Y$ enable us to recover the horizontal splittings (\ref{11}) and
(\ref{6}) by means of a section of this jet bundle.

\begin{definition}{6.3}
A first order jet field (or simply a jet field) on a
fibred manifold $Y$ is defined to be a section $\G$ of the affine jet
bundle $J^1Y\to Y$.
A first order connection $\G$ on a fibred manifold
$Y$ is defined to be a global jet field
\ben
&&\G :Y\to J^1Y,\nonumber\\
&&y^i_\la\circ\G =\G^i_\la (y).\label{61}
\een
\end{definition}

By means of the contact map $\la$ [Eq.(\ref{18})],
every connection $\G$ [Eq.(\ref{61})]
on a fibred manifold $Y\to X$ can be represented by a
tangent-valued projectable  horizontal 1-form $\la\circ\G$ [Eq.(\ref{37})]
on $Y$ which we denote by the same symbol $\G$.

Substituting a the
connection $\G$ [Eq.(\ref{37})] into the canonical horizontal
splittings (\ref{1.20}) and (\ref{34}), we obtain the familiar horizontal
splitting
\ben
&&\dot x^\la\dr_\la +\dot y^i\dr_i = \dot x^\la (\dr_\la
+\G^i_\la\dr_i) + (\dot y^i-\dot x^\la\G^i_\la)\dr_i, \nonumber\\
&&\dot x_\la dx^\la +\dot y_idy^i = (\dot x_\la +\G^i_\la\dot
y_i)dx^\la + \dot y_i(dy^i-\G^i_\la dx^\la) \label{9}
\een
of the tangent and cotangent bundles $TY$ and $T^*Y$ with
respect to the connection $\G$ on $Y$. Conversely, every horizontal splitting
(\ref{9}) determines a tangent-valued form (\ref{37}) and, consequently, a
global jet field on $Y\to X$.

\begin{example}{6.4}
Let $Y\to X$ be a vector bundle. A linear connection on $Y$ reads
\begin{equation}
\G=dx^\la\otimes[\dr_\la-\G^i{}_{j\la}(x)y^j\dr_i]. \label{8}
\end{equation}
\end{example}

\begin{example}{6.5}
Let $Y\to X$ be an affine bundle modelled on a vector bundle
$\ol Y\to X$. An affine connection on $Y$ reads
\[
\G=dx^\la\otimes[\dr_\la+(-\G^i{}_{j\la}(x)y^j+\G^i{}_\la (x))
\dr_i]
\]
where
\[
\G=dx^\la\otimes[\dr_\la-\G^i{}_{j\la}(x)\ol y^j\dr_i]
\]
is a linear connection on $\ol Y$.
\end{example}

\begin{example}{6.6}
Let $\G$ be a connection
on a fibred manifold $Y\to X$ and $\G'$ a connection on a fibred
manifold $Y'\to X$. In accordance with the isomorphism (\ref{42}),
there exists the
unique connection $\G\times\G'$ on the product $Y\op\times_X Y$ such that
the diagram
\[
\begin{array}{rcl}
{J^1Y\op\times_XJ^1Y'}  & \op\longrightarrow^{} &  {J^1(Y\op\times_XY')}  \\
{_{(\G,\G')}} & \nwarrow  \nearrow & {_{\G\times\G'}}\\
& {_{Y\op\times_XY'}} &
\end{array}
\]
commutes. It is called the product connection.
\end{example}

Let $\G$ be a connection on a fibred manifold $Y\to X$ and $\Phi$ a
fibred diffeomorphism of $Y$ to a fibred manifold $Y'\to X$ over a
diffeomorphism of $X$. Then, the composition
\[
\G'=J^1\Phi\circ\G\circ\Phi^{-1}
\]
>from the commutative diagram
\[
\begin{array}{rcccl}
& {J^1Y} &  \op\longrightarrow^{J^1\Phi} & {J^1Y'} &  \\
{_\G} &\put(0,-10){\vector(0,1){20}} & & \put(0,-10){\vector(0,1){20}} &
{_{\G'}} \\
& {Y} & \op\longrightarrow_{\Phi} & {Y'} &
\end{array}
\]
is a connection on the fibred manifold
$Y'$. In particular, if $\Phi$ is a fibred diffeomorphism over $X$, we have
\[
y'^i_\la\circ\G'=(\dr_\la\Phi^i+\G^j_\la\dr_j\Phi^i)\circ\Phi^{-1}.
\]

Since the affine jet bundle $J^1Y\to Y$ is modelled on the vector bundle
(\ref{23}), connections on a fibred manifold $Y$ constitute the affine space
modelled on the linear space of soldering forms on $Y$. It follows that, if
$\G$ is a connection and
\[
\si=\si^i_\la dx^\la\otimes\dr_i
\]
is a soldering form on a fibred manifold $Y$, its sum
\[
\G+\si=dx^\la\otimes[\dr_\la+(\G^i_\la +\si^i_\la)\dr_i]
\]
is a connection on $Y$. Conversely, if $\G$ and $\G'$ are
connections on a fibred manifold $Y$, then
\[
\G-\G'=(\G^i_\la -{\G'}^i_\la)dx^\la\otimes\dr_i
\]
is a soldering form on $Y$.

One introduces the following basic forms involving a connection $\G$ and a
soldering form $\si$:
\begin{itemize}
\item the curvature of $\G$:
\ben
&&R = \frac12 d_\G\G =\frac12 R^i_{\la\m} dx^\la\w dx^\m\otimes\dr_i=
\nonumber \\
&&\quad \frac12 (\dr_\la\G^i_\m -\dr_\m\G^i_\la +\G^j_\la\dr_j\G^i_\m
-\G^j_\m\dr_j\G^i_\la) dx^\la\w dx^\m\otimes\dr_i; \label{13}
\een
\item the torsion of $\G$ with respect to $\si$:
\ben
&&\Om = d_\si\G =d_\G\si =\frac 12 \Om^i_{\la\m} dx^\la \w
dx^\m\otimes\dr_i= \nonumber\\
&&\quad (\dr_\la\si^i_\m +\G^j_\la\dr_j\si^i_\m -\dr_j\G^i_\la\si^j_\m)
dx^\la\w dx^\m\otimes \dr_i; \label{14}
\een
\item the soldering curvature of $\si$:
\ben
&&\ve =\frac12 d_\si\si =\frac12 \ve^i_{\la\m} dx^\la \w
dx^\m\otimes\dr_i= \nonumber\\
&&\quad  \frac12 (\si^j_\la\dr_j \si^i_\m - \si^j_\m\dr_j \si^i_\l)
dx^\la\w dx^\m\otimes \dr_i. \label{15}
\een
\end{itemize}

They satisfy the following relations:
\be
&&\G'=\G +\si,\\
&& R'=R+\ve + \Om,\\
&&\Om'=\Om +2\ve.
\ee

\begin{example}{6.7} The curvature (\ref{13}) of the linear
connection (\ref{8}) reads
 \ben
&& R^i_{\la\m}(y)=-R^i{}_{j\la\m}(x)y^j, \nonumber\\
&&R^i{}_{j\la\m}=\dr_\la\G^i{}_{j\m} -\dr_\m\G^i{}_{j\la}
+\G^k{}_{j\mu}\G^i{}_{k\la} -\G^k{}_{j\la}\G^i{}_{k\mu}.\label{25}
\een
\end{example}

A connection $\G$ on a fibred manifold $Y\to X$ yields the affine
bundle morphism
\ben
&&D_\G:J^1Y\ni z\mapsto z-\G(\pi_{01}(z))\in T^*X\op\otimes_Y VY,\label{38}
\\
&&D_\G =(y^i_\la -\G^i_\la)dx^\la\otimes\dr_i. \nonumber
\een
It is  called the covariant differential relative to the connection $\G$.
The corresponding covariant
derivative of  sections $s$ of  $Y$ reads
 \begin{equation}
\nabla_\G s=D_\G\circ J^1s=[\dr_\la s^i-
(\G\circ s)^i_\la]dx^\la\otimes\dr_i. \label{39}
\end{equation}
In particular, a section $s$ of a fibred manifold $Y\to X$
is called an integral section for a
connection $\G$ on $Y$ if $ \nabla_\G s=0$, that is,
\begin{equation}
\G\circ s=J^1s.\label{40}
\end{equation}

Now, we  consider some particular  properties of linear
connections on vector bundles.

Let $Y\to X$ be a vector bundle and
$\G$ a linear connection on $Y$. Then, there
is the unique linear connection  $\G^*$ on the dual vector bundle $Y^*\to X$
such that the following diagram is commutative:
\[
\begin{array}{rcccl}
 & {J^1Y\op\times_X J^1Y^*} & \op\longrightarrow^{J^1\langle\rangle} &
{T^*X\times\Bbb R} & \\
{_{\G\times\G^*}} &\put(0,-10){\vector(0,1){20}} & &
\put(0,-10){\vector(0,1){20}} & {_{\wh 0\times\Id}} \\
 & {Y\op\times_X Y^*} &  \op\longrightarrow_{\langle\rangle} & {X\times\Bbb R}
&
 \end{array}
\]
where $\wh 0$ is the global zero section of the cotangent bundle $T^*X$.
The connection $\G^*$ is called the dual connection to $\G$. It has the
coordinate expression
\[
\G^*_{i\la}=-\G^j{}_{i\la}(x)y_j.
\]

\begin{example}{6.8} A linear connection $K$ on the tangent bundle $TX$ of a
manifold $X$ and the dual connection $K^*$ to $K$ on the cotangent bundle
$T^*X$ are given by the coordinate expressions
\ben
&& K^\al_\la=-K^\al{}_{\nu\la}(x)\dot x^\nu,\nonumber\\
&&K^*_{\al\la}=K^\nu{}_{\al\la}(x)\dot x_\nu, \label{408}
\een
\end{example}

 Let $Y$ and $Y'$ be vector bundles over $X$. Given linear connections
$\G$ and $\G'$ on $Y$ and $Y'$ respectively, there is the unique linear
connection $\G\otimes\G'$ on the tensor product
\[
Y\op\otimes_X Y'\to X
\]
such that the diagram
\[
\begin{array}{rcccl}
 & {J^1Y\op\times_X J^1Y'} & \op\longrightarrow^{J^1\otimes} &
{J^1(Y\op\otimes_X Y')} &  \\
{_{\G\times\G'}} &\put(0,-10){\vector(0,1){20}} & &
\put(0,-10){\vector(0,1){20}} & {_{\G\otimes\G'}} \\
 & {Y\op\times_X Y'} &  \op\longrightarrow_{\otimes} & {Y\op\otimes_X Y'} &
\end{array}
\]
commutes.
It is called the tensor product connection and has the coordinate expression
\[
(\G\otimes\G')^{ik}_\la=\G^i{}_{j\la}y^{jk}+{\G'}^k{}_{j\la}y^{ij}.
\]

\begin{example}{6.9}
The vertical tangent
morphism $V\G$ to $\G$ defines the connection
\ben
&&  V\G :VY\to VJ^1Y=J^1VY,\nonumber \\
&& V\G =dx^\la\otimes(\dr_\la
+\G^i_\la\frac{\dr}{\dr y^i}+\dr_j\G^i_\la\dot y^j \frac{\dr}{\dr \dot y^i}),
\label{43}
 \een
on the composite manifold $VY\to Y\to X$ due to the canonical isomorphism
(\ref{1.22}). The connection $V\G$ is projectable to the connection $\G$ on
$Y$, and it is a linear bundle morphism over $\G$.
The connection (\ref{43}) yields the  connection
\begin{equation}
V^*\G =dx^\la\otimes(\dr_\la +\G^i_\la\frac{\dr}{\dr
y^i}-\dr_j\G^i_\la \dot y_i \frac{\dr}{\dr \dot y_j}) \label{44}
\end{equation}
 on the composite manifold $ V^*Y\to Y\to X$ which is the dual connection to
$V\G$ over $\G$.
\end{example}

\section{Second order jet manifolds}

Considering the first order jet manifold of the
fibred jet manifold $J^1Y\to X$,
we come to several second order jet manifolds.
They are
(i) the repeated jet manifold $J^1J^1Y$,
(ii) the sesquiholonomic jet manifold $\wh J^2Y$
and (iii) the second order jet manifold $J^2Y$ of $Y$. A first order
connection on the fibred jet manifold $J^1Y\to X$, by definition,
is the second order connection on the fibred manifold $Y\to X$.

\begin{definition}{7.1} The repeated jet manifold
$J^1J^1Y$  is defined to be the first order jet manifold of the fibred jet
manifold $J^1Y\to X$.
\end{definition}

 Given the coordinates (\ref{50}) of $J^1Y$, the repeated jet
manifold $J^1J^1Y$ is provided with the adapted coordinates
\begin{equation}
(x^\la ,y^i,y^i_\la ,y_{(\m)}^i,y^i_{\la\m}).\label{51}
\end{equation}

There are the following two repeated jet bundles:
\ben
(i) \quad&&\pi_{11}:J^1J^1Y\to J^1Y, \label{S1}\\
&&y_\la^i\circ\p_{11} = y_\la^i,\nonumber\\
(ii) \quad&&J^1\pi_{01}:J^1J^1Y\to J^1Y,\label{S'1}\\
&& y_\la^i\circ J^1\pi_{01} = y_{(\la)}^i.\nonumber
\een
Their affine
difference over $Y$ yields the Spencer bundle morphism
\[
\dl=J^1\pi_{01} - \pi_{11} :J^1J^1Y\op\to_Y T^*X\op\otimes_Y VY,
\]
\[
\dot x_\la\otimes \dot y^i\circ\dl ={(\la)}^i-y^i_\la.
\]
 The kernel of this morphism is the sesquiholonomic affine subbundle
\ben
&&\wh J^2Y\to J^1Y,\label{52} \\
&&y^i_{(\la)}= y^i_\la,\nonumber
\een
of the repeated jet bundles (\ref{S1}) and (\ref{S'1}).
It is modelled on the vector bundle
\[
\op\otimes^2 T^*X\op\otimes_{J^1Y} VY.
\]
Given the coordinates (\ref{51}) of $J^1J^1Y$, the sesquiholonomic jet manifold
$\wh J^2Y$ is coordinatized by
\[
(x^\la ,y^i, y^i_\la,y^i_{\la\m}).
\]

\begin{definition}{7.2} The
second order jet manifold $J^2Y$
of a fibred manifold $Y\to X$ is defined to be the affine subbundle
\ben
&&\pi_{12}: J^2Y\to J^1Y, \label{53}\\
&&y^i_{\la\m}=y^i_{\m\la},\nonumber
\een
of the bundle (\ref{52}) which is modelled on the vector bundle
\[
\op\vee^2 T^*X\op\otimes_{J^1Y} VY.
\]
It is the union of
all equivalence classes  $j_x^2s$ of sections $s$ of $Y\to X$ such that
\[
y^i_\la (j_x^2s)=\dr_\la s^i(x),\qquad
y^i_{\la\m}(j_x^2s)=\dr_\m\dr_\la s^i(x).
\]
\end{definition}

In other words, sections $s\in j^2_xs$ are identified by their values
and values of their first and second order derivatives at the point $x\in X$.
Given the fibred coordinates (\ref{51}) of the repeated jet manifold
$J^1J^1Y$, the second order jet
manifold $J^2Y$ is coordinatized by
\[
(x^\la ,y^i, y^i_\la,y^i_{\la\m}=y^i_{\m\la}).
\]

We have the following affine bundle monomorphisms
\[
J^2Y\hookrightarrow \wh J^2Y\hookrightarrow J^1J^1Y
\]
 over $J^1Y$
and the canonical splitting
\[
\wh J^2Y =J^2Y\op\oplus_{J^1Y} (\op\w^2 T^*X \op\otimes_Y VY),
\]
\[
y^i_{\la\m} = \frac12(y^i_{\la\m}+y^i_{\m\la}) +
(\frac12(y^i_{\la\m}-y^i_{\m\la}).
 \]

Let $\Phi$ be a fibred morphism of a fibred manifold $Y\to X$ to
a fibred manifold $Y'\to X$ over a
diffeomorphism of $X$. Let $J^1\Phi$ be its first order jet prolongation
(\ref{26}). One can consider the first order jet prolongation $J^1J^1\Phi$
of the fibred morphism $J^1\Phi$.
The restriction of the morphism $J^1J^1\Phi$ to the second order jet manifold
$J^2Y$ of $Y$ is called the second order jet prolongation $J^2\Phi$ of a fibred
morphism $\Phi$.

In particular, the repeated jet prolongation
$J^1J^1s$ of a section $s$ of $Y\to X$ is a  section of the fibred manifold
$J^1J^1Y\to X$. It takes its values into $J^2Y$ and
consists  with the second order jet prolongation $J^2s$ of $s$:
\[
(J^1J^1s)(x)=(J^2s)(x)=j^2_xs.
\]

Given a second order jet manifold $J^2Y$ of $Y$, we have (i)
 the fibred morphism
\[
r_2: J^2TY\to TJ^2Y,
\]
\[
(\dot y^i_\la, \dot y^i_{\la\al})\circ r_2 =
((\dot y^i)_\la-y^i_\m\dot x^\m_\la, (\dot y^i)_{\la\al} -y^i_\m\dot
x^\m_{\la\al} - y^i_{\la\m}\dot x^\m_\al),
\]
and (ii) the canonical isomorphism
\[
VJ^2Y=J^2VY
\]
 where $J^2VY$
is the second order jet manifold of the  fibred manifold $VY\to X$ and $VJ^2Y$
is the vertical tangent bundle of the fibred manifold $J^2Y\to X$.

As a consequence, every  vector field $u$
on a fibred manifold $Y\to X$ admits the second order jet lift  to the
projectable  vector field
\[
\ol u_2 =r_2\circ J^2u: J^2Y\to TJ^2Y.
\]
In particular, if
\[
u = u^\la\dr_\la + u^i\dr_i
\]
is a projectable vector field on $Y$, its second order jet lift reads
\ben
&& \ol u_2 =
u^\la\dr_\la + u^i\dr_i + (\dr_\la u^i+y^j_\la\dr_ju^i
 - y_\m^i\dr_\la u^\m)\dr_i^\la+ \nonumber \\
&& \qquad [(\dr_\al +y^j_\al\dr_j +y^j_{\beta\al}\dr^\beta_j) (\dr_\la
+y^k_\la\dr_k)u^i  -y^i_\m\dot
x^\m_{\la\al} - y^i_{\la\m}\dot x^\m_\al]\dr_i^{\la\al}. \label{80}
 \een

There exist the following generalizations of the contact map (\ref{18}) and the
complementary morphism (\ref{24}) to the second order jet manifold
$J^2Y$:
\ben
(i)\quad &&\la:J^2Y\op\to_{J^1Y}
T^*X \op\otimes_{J^1Y} TJ^1Y,\nonumber\\
 &&\la=dx^\la\otimes\wh{\dr}_\la=dx^\la
\otimes (\dr_\la + y^i_\la \dr_i + y^i_{\m\la}\dr_i^\m),\label{54}\\
(ii)\quad &&\th_1:J^2Y \op\to_{J^1Y}T^*J^1Y\op\otimes_{J^1Y} VJ^1Y,\nonumber\\
  &&\th_1=(dy^i- y^i_\la dx^\la)\otimes\dr_i +
(dy^i_\m- y^i_{\m\la} dx^\la)\otimes\dr_i^\m.\label{55}
\een

The contact map (\ref{54}) defines the canonical horizontal splitting of the
exact sequence
\[
0\to VJ^1Y\op\hookrightarrow_{J^1Y} TJ^1Y\op\to_{J^1Y}
 J^1Y\op\times_X TX\to 0.
\]
Hence, we get the canonical horizontal splitting of a projectable
vector field $\ol u$ on $J^1Y$ over $J^2Y$:
\ben
&&\ol u=u_H+u_V = u^\la[\dr_\la+ y^i_\la+y^i_{\m\la}] \nonumber \\
&& \qquad + [(u^i-y^i_\la u^\la)\dr_i + (u^i_\m- y^i_{\m\la}u^\la)\dr^\m_i] .
\label{79}
\een

Building on the morphisms (\ref{54}) and (\ref{55}), one can obtain the
horizontal splittings  of the canonical tangent-valued 1-form
\[
\th_{J^1Y}=dx^\la\otimes\dr_\la + dy^i\otimes\dr_i +
dy^i_\m\otimes\dr_i^\m=\la +\th_1
\]
on $J^1Y$ and the exterior differential
\begin{equation}
 d= d_{\th_{J^1Y}}=d_\la +d_{\th_1}= d_H +d_V. \label{56}
\end{equation}
They are similar to the horizontal splittings (\ref{35}) and (\ref{1.19}).

Now, we consider connections on jet manifolds.

\begin{definition}{7.3} A second order jet field [resp. a second order
connection]  $\ol\G$ on a fibred manifold $Y\to X$ is defined to be
a first order jet field [resp. a first order connection]
on the fibred jet manifold $J^1Y\to X$, i.e.
this is a section [resp. a global section] of the  bundle (\ref{S1}).
\end{definition}

In the coordinates (\ref{51}) of the repeated jet manifold $J^1J^1Y$, a second
order jet field $\ol\G$ is given by the expression
 \[
 (y^i_\la,y^i_{(\m)},y^i_{\la\m})\circ\ol\G=
(y^i_\la,\ol\G^i_{(\m)},\ol\G^i_{\la\m}).
\]
Using the contact map (\ref{54}), one can represent it by the
tangen-valued horizontal 1-form
\begin{equation}
\ol\G=dx^\m\otimes (\dr_\m
+\ol\G^i_{(\m)}\dr_i+\ol\G^i_{\la\m}\dr^\la_i)\label{58}
\end{equation}
 on the fibred jet manifold $J^1Y\to X.$

A second order jet field $\ol\G$ on $Y$ is termed a sesquiholonomic [resp.
holonomic] second order jet field if it takes its values into the subbundle
$\wh J^2Y$ [resp. $J^2Y$] of $J^1J^1Y$. We have the coordinate equality
$\ol\G^i_{(\m)}=y^i_\m$
 for a sesquiholonomic second order jet field and the additional equality
$\ol\G^i_{\la\m}=\ol\G^i_{\m\la}$
 for a holonomic second order jet field.

Given a first order connection $\G$ on a fibred manifold $Y\to X$,
one can construct a second order connection on $Y$, that is,
a connection on the fibred jet  manifold $J^1Y\to X$ as follows.

The first
order jet prolongation $J^1\G$ of the connection $\G$ on $Y$ is
 a section of the repeated jet bundle (\ref{S'1}), but not the
bundle $\pi_{11}$ (\ref{S1}).
Let $K^*$ be a linear symmetric connection (\ref{408}) on the cotangent bundle
$T^*X$ of $X$:
\[
K_{\la\m}^*=K^\al{}_{\la\m}\dot x_\al, \qquad K^\al{}_{\la\m}=K^\al{}_{\m\la}.
\]
 There exists the  affine fibred morphism
\[
 r_K: J^1J^1Y\to J^1J^1Y, \qquad  r_K\circ r_K=\Id_{J^1J^1Y},
\]
\[
(y^i_\la ,y_{(\m)}^i,y^i_{\la\m})\circ r_K=
(y^i_{(\la)} ,y_\m^i,y^i_{\m\la}+ K^\al{}_{\la\m}(y^i_\al
- y^i_{(\al)})).
\]
 Using the coordinate transformation law of coordinates
(\ref{51}), one can verify the following relations
\be
&&{y'}^i_\m\circ r_k={y'}^i_{(\m)}, \qquad
{y'}^i_{(\m)}\circ r_k={y'}^i_\m, \\
&&{y'}^i_{\la\m}\circ r_k={y'}^i_{\m\la} +{K'}^\al{}_{\la\m}({y'}^i_\al
- {y'}^i_{(\al)}).
\ee
Hence, given a first order connection $\G$ on a fibred
manifold $Y\to X$, we have the second order connection
\[
J\G\op =^{\rm def}r_K\circ J^1\G,
\]
\begin{equation}
J\G=dx^\m\otimes
[\dr_\mu+\Gamma^i_\mu\dr_i +(\dr_\la\Gamma^i_\m+
\dr_j\Gamma^i_\mu y^j_\la -
K^\alpha{}_{\la\mu} (y^i_\alpha-\Gamma^i_\alpha))
\dr_i^\la],\label{59}
\end{equation}
on $Y$. This is an affine morphism
\[
\begin{array}{rcccl}
 & {J^1Y} &  \op\longrightarrow^{J\G} & {J^1J^1Y} &  \\
{_{\pi_{01}}} &\put(0,10){\vector(0,-1){20}} & & \put(0,10){\vector(0,-1){20}}
& {_{\pi_{11}}} \\
 & {Y} & \op\longrightarrow_{\G} & {J^1Y} &
\end{array}
\]
over the first order connection $\G$.

Note that the curvature $R$ (\ref{13}) of a first order connection $\G$ on a
fibred manifold $Y\to X$ induces the soldering form
\begin{equation}
\ol\si_R=R^i_{\la\m}dx^\m\otimes\dr^\la_i \label{60}
\end{equation}
 on the  fibred jet
manifold $J^1Y\to X$.  Also the
torsion (\ref{14}) of a first order connection $\G$ with respect to a soldering
form $\si$ on $Y\to X$ and the soldering curvature (\ref{15}) of $\si$
determine soldering forms on $J^1Y\to X$.

\section{Principal connections}

The general approach to connections as jet fields is suitable to formulate the
classical concept of principal connections.

Unless otherwise stated, a structure group $G$ of a principal bundle
is assumed to be a real finite-dimensional Lie group, and $\dim G>0$.

If $P\to X$
is a principal bundle with a structure group $G$,
the exact sequence (\ref{1.8a}) can be reduced to the exact sequence
\begin{equation}
0\to V^GP\op\hookrightarrow_X T^GP\to TX\to 0\label{1.33}
\end{equation}
 where
\[
 T^GP=TP/G,\qquad V^GP=VP/G
\]
are  the quotients of the tangent bundle $TP$ of $P$ and the
vertical tangent bundle $VP$ of $P$ respectively
by the canonical action (\ref{1}) of $G$
on $P$ on the right. The bundle $V^GP\to X$ is called the adjoint bundle.
Its standard fibre is the right
Lie algebra ${\got g}_r$ of the right-invariant vector fields on the group
$G$. The group $G$ acts on this standard fibre by the adjoint representation.

A principal connection $A$ on
a principal bundle $P\to X$ is defined to be a
 $G$-equivariant global jet field on $P$ such that
\[
J^1r_g\circ A= A\circ r_g
\]
for each canonical morphism (\ref{1}). We have
\be
&&A\circ r_g=J^1r_g\circ A, \qquad g\in G,\\
&& A=dx^\la\otimes(\dr_\la +A^m_\la(p)e_m), \qquad p\in P,\\
&& A^m_\la(qg)= A^m_\la(p) {\rm ad}g^{-1}(e_m).
\ee
A principal connection $A$ determines  splitting
$TX\hookrightarrow T^GP$ of the exact sequence (\ref{1.33}).

Let $A:P\to J^1P$ be a principal connection. Building on the complementary
morphism (\ref{24}) and the canonical horizontal splitting (\ref{1.28}),
we recover the familiar connection form $\ol A=\al\circ\theta_1
\circ A$ on the principal bundle $P$. Given an atlas
\[
\Psi^P=\{U_\xi, \psi^P_\xi, \rho_{\xi\zeta}\}
\]
of $P$, we have the local connection 1-forms
\begin{equation}
A_\xi=z^*_\xi\ol A=-A^m_\la(x)dx^\la\otimes J_m \label{1.32}
\end{equation}
where $\{z_\xi\}$ is the family of local sections of $P$ associated with
the atlas $\Psi^P$:
\[
(\pr_2\circ\psi^P_\xi\circ z_\xi)(x)=1_G, \qquad x\in U_\xi.
\]

There is the 1:1 correspondence between the principal connections on a
principal bundle $P\to X$  and the global sections of the bundle (\ref{68}).
 We shall call
$C$ the principal connection bundle. It is an affine bundle modelled on
the vector bundle
 \begin{equation}
\ol C =T^*X \otimes V^GP,\label{3.1}
\end{equation}
and there is the canonical vertical
splitting
\[
VC=C\times\ol C.
\]

\begin{remark} The bundle $C$ [Eq.(\ref{68})] fails to be a bundle
with a structure group. The jet prolongation
\[
J^1P\times J^1(X\times G)\to J^1P
\]
of the canonical action (\ref{1}) brings the fibred jet manifold
$J^1P\to X$ into the general affine bundle modelled on the right on
the group bundle
\[
J^1(X\times G)=G\times (T^*X\times{\got g}_l)
\]
over $X$. However, the latter fails to be generally a trivial bundle
because of $T^*X$. Therefore,
$J^1P\to X$ is not a principal bundle. At the same time, the bundle
\[
J^1P=C\op\times_XP\to C
\]
is the $G$-principal bundle over $C$.
\end{remark}

Given a bundle atlas $\Psi^P$ of $P$, the principal connection bundle $C$
is provided with  the fibred coordinates $(x^\mu,k^m_\mu)$ so that
\[
(k^m_\mu\circ A)(x)=A^m_\mu(x)
\]
 are coefficients of
the local connection 1-form (\ref{1.32}).
The first order jet manifold $J^1C$ of  $C$ is endowed
 with the adapted coordinates
\begin{equation}
(x^\mu, k^m_\mu, k^m_{\mu\la}).\label{S2}
\end{equation}

The affine jet bundle $J^1C\to C$
is modelled on the vector bundle
\[
T^*X\op\otimes_C (C\times T^*X\otimes V^GP).
\]
There exists the canonical splitting
\begin{equation}
J^1C=C_+\op\oplus_C C_-=(J^2P/G)\op\oplus_C
(\op\w^2 T^*X\op\otimes_C V^GP) \label{N31}
\end{equation}
over $C$ where
\[
C_-=C\times\op\w^2 T^*X\op\otimes_X V^GP
\]
and $C_+\to C$ is the affine bundle modelled on the vector bundle
\[
\ol C_+=\op\vee^2 T^*X\op\otimes_C V^GP.
\]
In the coordinates (\ref{S2}), the splitting (\ref{N31}) reads
\[
 k^m_{\mu\la}=\frac12(
k^m_{\mu\la}+k^m_{\la\mu} +c^m_{nl}k^n_\la k^l_\mu)
 +\frac12( k^m_{\mu\la}-k^m_{\la\mu}
 -c^m_{nl}k^n_\la k^l_\mu)
\]
where $c^k_{mn}$ are structure constants of the Lie algebra ${\got g}_r$
with respect to its basis $\{I_m\}$.

There are  the corresponding canonical
projections
\be
&&{\cal S}:={\rm pr}_1: J^1 C\to C_+,\\
&& {\cal S}^m_{\la\mu}=
k^m_{\mu\la}+k^m_{\la\mu} +c^m_{nl}k^n_\la k^l_\mu,
\ee
and
\be
&& \cF:={\rm pr}_2: J^1 C\to C_-,\\
&&\cF=\frac{1}{2}\cF^m_{\la\m}dx^\la\w dx^\m\otimes I_m,\\
&& \cF^m_{\la\mu}=
k^m_{\mu\la}-k^m_{\la\mu} -c^m_{nl}k^n_\la k^l_\mu.
\ee

For every principal connection $A$, we observe that
\be
&&\cF\circ J^1A=F,\\
&& F=\frac{1}{2}F^m_{\la\m}dx^\la\w dx^\m\otimes I_m,\\
&& F^m_{\la\m}=\dr_\la A^m_\m
-\dr_\m A^m_\la -c^m_{nk}A^n_\la A^k_\m,
\ee
is the strength of $A$.

Given a symmetric linear connection $K^*$
on the cotangent bundle $T^*X$ of $X$,  every principal connection $A$ on
a principal bundle $P$ gives rise to the connection
\[
S_A: C\to C_+, \qquad S_A\circ A={\cal S}\circ J^1A,
\]
on the principal connection bundle $C$. In the coordinates
(\ref{S2}), the connection $S_A$ reads
\ben
&& S_A{}^m_{\mu\la}=\frac{1}{2} [c^m_{nl}k^n_\la
k^l_\mu  +\dr_\mu A^m_\la+\dr_\la A^m_\mu \nonumber \\
&&\qquad -c^m_{nl}
(k^n_\mu A^l_\la+k^n_\la A^l_\mu)] -
K^\beta{}_{\mu\la}(A^m_\beta-k^m_\beta). \label{3.7}
\een

Let $Y\to X$ be a bundle associated with a principal bundle $P\to X$.
The structure group $G$ of $P$ acts freely on the standard fibre $V$ of
$Y$ on the left. The total space of the $P$-associated bundle $Y$,
by definition, is the quotient
\[
Y=(P\times V)/G
\]
of the product $P\times V$
by identification of its elements $(pg\times gv)$
for all $g\in G$. For each $P$ associated
bundle $Y$, there exists the representation morphism
\[
\wt P\times Y\op\to_XY.
\]

The $P$-associated bundle $Y$ is provided with atlases
$\Psi=\{U_\xi, \psi_\xi\}$ associated with atlases
$\Psi^P=\{U_\xi, z_\xi\}$ of the principal bundle $P$ as follows:
\[
\psi^{-1}_\xi (x\times V)= [z_\xi (x)]_V (V), \qquad x\in U_\xi,
\]
where by $[p]_V$ is meant the restriction of the canonical map
$P\times V\to Y$ to the subset $p\times V$.

Every principal connection $A$ on a principal bundle $P$ yields the
associated connection $\G$ on a $P$-associated bundle $Y$ such that
the diagram
\[
\begin{array}{rcccl}
 & {J^1P\times V} &  \longrightarrow & {J^1Y} &  \\
{_{A\times{\Id}_V}} &\put(0,-10){\vector(0,1){20}} & & \put(0,-10)
{\vector(0,1){20}} & {_{\G}} \\
 & {P\times V} & \longrightarrow & {Y} &
\end{array}
\]
is commutative. We call it the associated principal connection. With
respect to the associated atlases $\Psi$ of $Y$ and $\Psi^P$ of $P$,
this connection is written
\begin{equation}
\G=dx^\la\otimes [\dr_\la +A^m_\mu (x)I_m{}^i{}_jy^j\dr_i] \label{S4}
\end{equation}
where $A^m_\mu (x)$ are coefficients of the local connection 1-form
(\ref{1.32}) and $I_m$ are generators of the structure group $G$
on the standard fibre $V$ of the bundle $Y$. The curvature of the
connection (\ref{S4}) reads
\[
R^i_{\la\mu}= F^m_{\la\mu}I_m{}^i{}_jy^i.
\]

\section{Higher order jet manifolds}

In this Section, we brief some notions of the
higher order jet formalism in order to apply
them to differential operators.

We use the multi-index $\La$,
$\mid\La\mid=k$ for symmetrized collections $(\la_1...\la_k)$. By $\La + \la$
is meant the symmetrized collection $(\la_1...\la_k\la)$.

\begin{definition}{9.1} The $k$-order jet manifold $J^kY$ of a
fibred manifold $Y\to
X$ is defined to comprise all equivalence classes $j^k_xs$, $x\in X$,
of sections $s$ of $Y$ so that sections $s$ and $s'$ belong to the
same class $j^k_xs$ if and only if
\[
\dr_\La s^i(x)=\dr_\La {s'}^i(x), \qquad 0\leq \mid\La\mid \leq k.
\]
It is  provided with atlases
of the adapted coordinates $(x^\la, y^i_\La)$, $0\leq \mid\La\mid \leq k$,
which are compatible with the natural surjections
\[
\pi_{ik}: J^kY\to J^iY
\]
and the composite fibration
\[
\pi_{0k}: J^kY\to J^{k-1}Y\to ... \to Y\to X.
\]
\end{definition}

\begin{proposition}{9.2} There exists the following fibred
monomorphism
\ben
&&\la: J^{k+1}Y\to T^*X\op\otimes_{J^kY} TJ^kY,\nonumber\\
&&\la =dx^\la\otimes(\dr_\la + y^i_{\La+\la}\dr_i^\La), \qquad
0\leq \mid\La\mid \leq k ,\label{82}
\een
whose particular cases are the contact maps (\ref{18}) and (\ref{54}).
\end{proposition}

The $k$-order contact map (\ref{82})
sets up the canonical horizontal splitting of the exact sequence
\[
0\to VJ^kY\hookrightarrow TJ^kY\to J^kY\op\times_X TX\to 0.
\]
In particular, we get the  canonical horizontal splitting of a
projectable vector field $\ol u$ on $J^kY\to X$ over $J^{k+1}Y$:
\[
\ol u=u_H+u_V= u^\la (\dr_\la + y^i_{\La+\la}\dr_i^\La) + (u^i_\La -
 y^i_{\La+\la})\dr_i^\La, \qquad 0\leq \mid\La\mid \leq k .
\]
This splitting  is the $k$-order generalization of the splittings
(\ref{31}) and (\ref{80}).

Given a $k$-order jet manifold $J^kY$ of $Y$, we have the
fibred morphism
\[
r_k: J^kTY\to TJ^kY
\]
and the canonical isomorphism
\[
VJ^kY=J^kVY
\]
where $J^kVY$
is the $k$-order jet manifold of the fibred manifold $VY\to X$ and $VJ^kY$ is
the vertical tangent bundle of the fibred manifold $J^kY\to X$.
As a consequence, every  vector field $u$
on a fibred manifold $Y\to X$ has the $k$-order jet lift to the projectable
vector field
\ben
&&\ol u_k :=r_k\circ J^ku: J^kY\to TJ^kY, \nonumber \\
&& \ol u_k =
u^\la\dr_\la + u^i\dr_i + u_\La^i\dr_i^\La, \nonumber\\
&& u_{\La+\la}^i = \wh\dr_\la u_\La^i - y_{\La+\m}^iu^\m, \label{84}
\een
on $J^kY$ where
\begin{equation}
\wh\dr_\la = (\dr_\la + y^i_{\Si+\la}\dr_i^\Si), \qquad
0\leq \mid\Si\mid \leq k. \label{85}
\end{equation}
The expression (\ref{84}) is the $k$-order
generalization of the expressions (\ref{1.21}) and (\ref{80}).

Given the fibred jet manifold $J^kY\to X$, let us consider the repeated jet
manifold  $J^1J^kY$ provided with the adapted coordinates
\[
(x^\m, y^i_\La, y^i_{\La\la}), \qquad |\Lambda|\leq k.
\]
Just as in the case of $k=1$,
there exist two fibred morphisms of $J^1J^kY$
to $J^1J^{k-1}Y$ over $X$. Their difference over $J^{k-1}Y$ is
the $k$-order Spencer morphism
\[
J^1J^kY\to T^*X\op\otimes_{J^{k-1}Y} VJ^{k-1}Y
\]
where $ VJ^{k-1}Y$ is the vertical tangent bundle of the fibred manifold
$J^{k-1}Y\to X$. Its kernel is the $k$-order
sesquiholonomic  subbundle $\wh J^{k+1}Y$ of the bundle
$J^{k+1}Y\to Y$ which is coordinatized by
\[
(x^\m,y^i_\La, y^i_{\Si\m}), \qquad |\La|\leq k, \qquad |\Si|=k.
\]

\begin{proposition}{9.3} There exist the fibred monomorphisms
\begin{equation}
J^kY\hookrightarrow\wh J^kY\hookrightarrow J^1J^{k-1}Y \label{76}
\end{equation}
and the canonical splitting
\begin{equation}
\wh J^{k+1}Y= J^{k+1}Y\op\oplus_{J^kY}(T^*X\w\op\vee^{k-1}T^*X\op\otimes_Y
VY),\label{75}
\end{equation}
\end{proposition}

Let $\Phi: Y\to Y'$
be a fibred morphism over a diffeomorphism $f$ of $X$.  It yields the
$k$-order jet prolongation
\[
J^k\Phi: J^kY\ni j^k_xs\mapsto j^k_{f(x)}(\Phi\circ s\circ f^{-1})
\in  J^kY'
 \]
of $\Phi$. In particular,
every section $s$ of a fibred manifold $Y\to X$ admits the $k$-order jet
prolongation to the section
\[
(J^ks)(x)\op=^{\rm def} j^k_xs
\]
of the fibred manifold $J^kY\to X$.

We have the following integrability condition.

\begin{lemma}{9.4} Let $\ol s$ be a section of the fibred manifold
$J^kY\to X$. Then, the
following conditions are equivalent:
\begin{itemize}\begin{enumerate}
\item $\ol s=J^ks$ where $s$ is a section of $Y\to X$,
\item $J^1\ol s:X\to \wh J^{k+1}Y$,
\item $J^1\ol s:X\to J^{k+1}Y$.
\end{enumerate}\end{itemize}
\end{lemma}

Let $Y\to X$ and $Y'\to X$ be fibred manifolds and $J^kY$ the $k$-order
jet manifold of $Y$.

\begin{definition}{9.5} Every fibred morphism
\begin{equation}
\cE: J^kY\op\to_X Y'  \label{70}
\end{equation}
is called the $k$-order differential operator (of class $C^\infty$) on
sections of $Y$ (or simply on $Y$). It
sends every section $s$ of the fibred manifold $Y$ to the section
$\cE\circ J^ks$ of the fibred manifold $Y'$.
\end{definition}

Building on Proposition 9.3 and Lemma 9.4,
we can describe reduction of higher order differential operators
to the first order ones.

\begin{proposition}{9.6} Given a fibred manifold $Y$, every
first order differential operator
\begin{equation}
\cE'': J^1J^{k-1}Y\op\to_X Y'\label{72}
\end{equation}
 on $J^{k-1}Y$ implies the $k$-order differential operator
$\cE=\cE''\mid_{J^kY}$ on $Y$.
Conversely, if a first order differential operator
 on $J^{k-1}Y$ exists, any $k$-order differential operator (\ref{70}) on
$Y$ can be represented by the restriction $\cE=\cE''\mid_{J^kY}$ of some first
order differential operator (\ref{72}) on $J^{k-1}Y$ to the $k$-jet
manifold $J^kY$.
\end{proposition}

In particular, every $k$-order differential operator (\ref{70}) yields the
morphism
\begin{equation}
\cE':=\cE\circ\pr_2: \wh J^kY\op\to_X Y' \label{77}
\end{equation}
where $\pr_2:\wh J^kY\to J^kY$ is the surjection corresponding to the
canonical splitting (\ref{75}).
It follows that, for every section $s$ of a fibred manifold $Y\to X$,
we have the equality
 \[
\cE'\circ J^1J^{k-1}s= \cE\circ J^ks.
\]
Moreover, let $\ol s$ be a section of the fibred $(k-1)$-order
jet manifold $J^{k-1}Y\to X$ such
that its first order jet prolongation $J^1\ol s$ takes its values into the
sesquiholonomic jet manifold $\wh J^kY$. In virtue of Lemma 9.4,
there exists a section $s$ of $Y\to X$ such that
$\ol s=J^{k-1}s$ and
\begin{equation}
\cE'\circ J^1\ol s= \cE\circ J^ks.\label{S5}
\end{equation}

We call $\cE'$ [Eq.(\ref{77})] the  sesquiholonomic differential operator
and consider extensions of a $k$-order differential operator $\cE$
[Eq.(\ref{70})]
to first order differential operators (\ref{72}) only through its extension
to the sequiholonomic differential operator (\ref{77}).

Reduction of $k$-order differential operators to the first order ones
implies reduction of the associated $k$-order differential equations to
the first order differential equations as follows.

Let $Y'\to X$ be a composite manifold $Y'\to Y\to X$ where $Y'\to Y$
is a vector bundle. Let the $k$-order differential operator
(\ref{70}) on a fibred manifold $Y$ be a fibred morphism over $Y$.
We say that a section $s$
of  $Y$ satisfies the corresponding system of $k$-order
differential equations if
\begin{equation}
J^ks(X)\subset \Ker\cE. \label{73}
\end{equation}
As a shorthand, we shall write
\[
\cE\circ J^ks=0.
\]

Let a $k$-order differential operator $\cE$ on $Y$ be extended to a first
order differential operator $\cE''$ on $J^{k-1}Y$. Let $\ol s$ be a section of
$J^{k-1}Y\to X$. We shall say that $\ol s$ is a
sesquiholonomic solution of the corresponding system of first order
differential equations if
\begin{equation}
J^1\ol s(X)\subset \Ker\cE''\cap \wh J^kY. \label{74}
\end{equation}

\begin{proposition}{9.7} The system of the $k$-order differential equations
(\ref{73}) and the system of the first order differential equations (\ref{74})
are equivalent to each other.
\end{proposition}
\newpage

\centerline{\large \bf Lecture 3. LAGRANGIAN FORMALISM}
\bigskip
\bigskip

This Lecture is devoted to the Lagrangian formalism on fibred manifolds
and its De Donder Hamiltonian derivation.

The jet approach to Lagrangian systems has been mainly
stimulated by  the calculus of variations where the Lepagean
equivalents of a Lagrangian density play the prominent role. The
literature on this subject is extensive.
Point out basic structural
ambiguities in the higher order Langrangian formalism when the Cartan
forms, Legendre morphisms {\it etc.} fail to be uniquely defined.
Bearing in mind
physical applications, we shall
restrict our consideration to the first order Lagrangian systems whose
configuration space is the jet manifold $J^1Y$ of $Y$.

Here, we are not concerned deeply
with the variational principle and the calculus
of variations, but aim to detail different types of field equations
which one handles in the first order Lagrangian formalism on a fibred
manifold $Y\to X$.
They are the Cartan equations, the De Donder-Hamilton equations
and three types of Euler-Lagrange equations:
(i) the algebraic
Euler-Lagrange equations for sections of the repeated jet bundle
$J^1J^1Y \to J^1Y$,
(ii) the
first order differential Euler-Lagrange equations for sections of
the fibred jet manifold $J^1Y\to X$,
and (iii) the second order differential Euler-Lagrange
equations for sections of the fibred manifold $Y\to X$ itself.

To introduce the Euler-Lagrange equations, we start with the notion of
the Euler-Lagrange operator, for it is
the sesquiholonomic first order Euler-Lagrange operator which is the
Lagrangian counterpart of the Hamilton operator in the multimomentum
Hamiltonian formalism.

\section{}

Let
\[
\pi:Y\to X
\]
be a fibred manifold provided with fibred coordinates
$(x^\la, y^i)$ [Eq.(\ref{1.2})]. Unless otherwise stated, the dimension
of $X$ is $n>1$, for case of $n=1$
possesses essential percularities and
corresponds to the time-dependent mechanics.

In jet terms, a first order Lagrangian density is defined to be a
bundle morphism
 \[
 L:J^1Y\to\op\w^n T^*X
\]
over $Y$. It is viewed as an exterior horizontal density
\begin{equation}
 L=\cL(x^\mu, y^i, y^i_\mu)\om  \label{301}
\end{equation}
on the fibred jet manifold $J^1Y\to X$. The jet manifold $J^1Y$
 thus plays the role of a
finite-dimensional configuration  space of fields represented by
sections of $Y\to X$.

We further use the notation
\[
\pi^\la_i=\dr^\la_i\cL.
\]

\begin{remark}
In field theory, all Lagrangian densities are polynomial forms
relative to velocities $y^i_\la$.
Note that a polynomial form of degree $k$ on a vector space $\ol E$
is defined to be a linear form on the tensor space
\[
\Bbb R\oplus\ol E\oplus...\oplus(\op\otimes^k\ol E).
\]
Given an affine space $E$ modelled on a vector space $\ol E$, polynomial
forms on $E$ are factorized by morphisms $E\to\ol E$.
Since the jet bundle $J^1Y\to Y$ is affine,
every Lagrangian density of field theory factors as
\begin{equation}
L: J^1Y\op\to^D T^*X\op\otimes_Y VY\to\op\w^n T^*X \label{S6}
\end{equation}
where $D$ is the covariant differential (\ref{38}) relative to some
connection on the fibred manifold $Y\to X$.
\end{remark}

With a Lagrangian density $L$, the jet manifold $J^1Y$ carries the
the generalized Liouville form
\begin{equation}
\th_L=-\pi^\la_idy^i\w\om\otimes\dr_\la \label{302}
\end{equation}
and the Lagrangian multisymplectic form
\begin{equation}
\Om_L=d\pi^\la_idy^i\w\om\otimes\dr_\la.\label{322}
\end{equation}
They are pullback $TX$-valued forms on $J^1Y\to X$.

Given a first order Lagrangian density $L$ [Eq.(\ref{301})],
its Lepagean equivalent $\Xi_L$ is
an exterior $n$-form on the jet manifold $J^1Y$ such that
\begin{equation}
J^1s^*\Xi_L = L\circ J^1s \label{S7}
\end{equation}
for all sections $s$ of $Y$.
We shall follow the so-called De Donder-Weyl approach to the calculus of
variations which is based upon the Cartan
forms as Lepagean equivalents.
In first order theory, Cartan forms consist with the more particular
Poincar\'e-Cartan forms.

In the first order Lagrangian formalism, the Poincar\'e-Cartan form
associated with a
Lagrangian density $L$ always exists and is uniquely defined.
This is the exterior horizontal $n$-form
on the jet bundle $J^1Y\to Y$ which is
given by the coordinate expression
\ben
&&\Xi_L=\pi^\la_idy^i\w\om_\la +\pi\om, \label{303}\\
&& \pi=\cL-\pi^\la_iy^i_\la. \nonumber
\een

With $\Xi_L$, we have the following first order differential
Cartan equations for
sections $\ol s$ of the fibred jet manifold $J^1Y\to X$:
\begin{equation}
\ol s^*(u\rfloor d\Xi_L) = 0 \label{316}
\end{equation}
where $u$ is an arbitrary  vertical vector field on  $J^1Y\to X$.
In the coordinate form, these equations read
\bea
&&\dr^\mu_j\pi^\la_i (\dr_\la\ol s^i - \ol s^i_\la) =0, \label{S8a}\\
&& \dr_i\cL-(\dr_\la+\ol s^j_\la\dr_j
+\dr_\la\ol s^j_\m\dr^\m_j)\dr^\la_i\cL
+\dr_i\pi^\la_j(\dr_\la\ol s^j -\ol s^j_\la)=0. \label{S8b}
\eea
Solutions of the Cartan equations (\ref{316}) extremize the action
functional
\[
\op\int_X \ol s^*\Xi_L.
\]

If a section $\ol s$ of $J^1Y\to X$
is the jet prolongation of a section $s$ of $Y\to X$,
the form $\ol s^*\Xi_L$ comes to the familiar Lagrangian form
\[
L(s):=L\circ J^1s: X\to\op\w^n T^*X.
\]
The corresponding action functional reads
\[
\op\int_X L(s).
\]
This functional is proved to be stationary at a section $s$ iff
the jet prolongation
$J^1s$ of $s$ satisfies the Cartan equations (\ref{316}). It means that,
on holonomic sections $\ol s=J^1s$, the Cartan
equations are equivalent to the differential Euler-Lagrange equations.

In particular, if a Lagrangian density $L$ is
regular, the Cartan equations are equivalent to the Euler-Lagrange
equations since all solutions of the Cartan equations are holonomic.

We shall introduce the Euler-Lagrange equations as conditions for the
kernal of the Euler-Lagrange operator.

The Euler-Lagrange operator can be defined intrinsically as a
second order differential operator
\[
\cE_L:J^2Y\op\to_Y\op\w^{n+1}T^*Y
\]
of the variational type.
Here, the Euler-Lagrange operator associated with a
Lagrangian density $L$ will be reproduced as the restriction
of a certain exterior form $\La_L$ below to the repeated jet manifold
$J^1J^1Y$ to the second order jet manifold $J^2Y$.

Building on the pullbacks of the forms (\ref{322}) and (\ref{303})
onto the repeated jet manifold $J^1J^1Y$ by the bundle $\pi_{11}$
[Eq.(\ref{S1})], one can construct the exterior form
\begin{equation}
\La_L=d\Xi_L-\la\rfloor\Om_L
=(y^i_{(\la)}-y^i_\la)d\pi^\la_i\w\om +
(\dr_i-\wh\dr_\la\dr^\la_i)\cL dy^i\w\om,\label{304}
\end{equation}
\[
\la=dx^\la\otimes\wh\dr_\la,
\qquad
\wh\dr_\la =\dr_\la +y^i_{(\la)}\dr_i+y^i_{\m\la}\dr^\m_i,
\]
on $J^1J^1Y$.
Its restriction to the second order jet manifold $J^2Y$ of $Y$ recovers
the familiar variational Euler-Lagrange operator
\ben
&&\cE_L: J^2Y\op\to_Y\op\w^{n+1}T^*Y,\nonumber\\
&&\cE_L=\delta_i\cL dy^i\w\om=
 [\dr_i-(\dr_\la +y^i_\la\dr_i+y^i_{\m\la}\dr^\m_i)
\dr^\la_i]\cL dy^i\w\om ,\label{305}
\een
associated with the first order Lagrangian density $L$ [Eq.(\ref{301})].

The restriction of the form (\ref{304}) to the sesquiholonomic jet manifold
$\wh J^2Y$ of $Y$ yields the sesquiholonomic extension
\begin{equation}
\cE'_L:\wh J^2Y\op\to_Y\op\w^{n+1}T^*Y \label{2.26}
\end{equation}
 of the Euler-Lagrange operator (\ref{305}).
 It is given by the expression (\ref{305}), but with nonsymmetric
second order derivative coordinates $y^i_{\m\la}$.

\begin{definition}{10.1}
Given a Lagrangian density $L$, a sesquiholonomic jet field
\[
\ol\G =dx^\la\otimes(\dr_\la
+y^i_\la\dr_i+\ol\G^i_{\m\la}\dr^\m_i)
\]
 on the fibred jet manifold
$J^1Y\to X$ is termed a Lagrangian jet field
for $L$ if it takes its values into
$\Ker\cE'_L$:
\[
 \cE'_L\circ\ol\G=0,
\]
\begin{equation}
 \dr_i\cL-(\dr_\la+y^j_\la\dr_j
+\ol\G^j_{\m\la}\dr^\m_j)\dr^\la_i\cL=0. \label{2.27}
\end{equation}
\end{definition}

One can regard Eqs.(\ref{2.27}) as the system of linear algebraic
Euler-Lagrange equations for components $\ol\G^i_{\mu\la}$ of a Lagrangian
jet field associated with the Lagrangian density $L$.

\begin{proposition}{10.2} Let the determinant of the Hessian
\begin{equation}
\dr^\m_j\dr^\la_i\cL \label{313}
\end{equation}
of a Lagrangian density $L$ be different from zero at a point $z\in J^1Y$,
i.e., $L$ is regular at $z$.
Then, on an open neighborhood of $z$, the algebraic equations
(\ref{2.27}) admit an unique local solution.
\end{proposition}

In other words, there always exists an unique Lagrangian jet field for
a regular Lagrangian density $L$, otherwise for the degenerate ones.

It is readily observed that Lagrangian jet fields associated with
the same Lagrangian density $L$ differ from each other in soldering
forms
\[
\ol\si=\ol\si^j_{\mu\la}dx^\la\otimes\dr_j^\mu
\]
on the repeated jet bundle $J^1J^1Y\to J^1Y$ which satisfy the equations
\[
\ol\si^j_{\mu\la}\dr^\mu_j\dr^\la_i\cL=0.
\]
If a Lagrangian density is degenerate, these equations have
non-zero solutions. It follows that, in case of a degenerate Lagrangian
density, the corresponding differential Euler-Lagrange equations are
underdetermined, for higher order derivatives of field functions can not
be expressed uniquely in the less order ones.

 Let a Lagrangian jet field $\ol\G$ for the Lagrangian density $L$
has an integral section
$\ol s$  of the fibred jet manifold $J^1Y\to X$,
i.e., the first order jet prolongation $J^1\ol s$ of $\ol s$
takes its values into
$\Ker\cE'_L$. Then, $\ol s$ satisfies the system of first order
differential Euler-Lagrange equations
\[
\cE'_L\circ J^1\ol s=0
\]
which have the coordinate form
 \bea
&&\dr_\la\ol s^i=\ol s^i_\la, \label{306a}\\
&& \dr_i\cL-(\dr_\la+\ol s^j_\la\dr_j
+\dr_\la\ol s^j_\m\dr^\m_j)\dr^\la_i\cL=0. \label{306b}
\eea
Note that Eq.(\ref{306a}) selects sections $\ol s$ of
the fibred jet manifold $J^1Y\to X$
whose jet prolongations take their values into
the sesquiholonomic jet manifold $\wh J^2Y$.

The first order Euler-Lagrange equations (\ref{306a}) and
(\ref{306b}) are equivalent to the Cartan equations (\ref{S8a}) and
(\ref{S8b}) on sections $\ol s$ of $J^1Y\to X$ whose jet
prolongations $J^1\ol s$ take their values into the sesquiholonomic
jet manifold $\wh J^2Y$ of $Y$. In particular, this equivalence always
takes place when a Lagrangian density $L$ is regular.

 In virtue of Proposition 9.7,
the  system of the first order differential equations (\ref{306a}) and
(\ref{306b}) is equivalent to the system of the familiar second order
Euler-Lagrange equations
 \[
\cE_L\circ J^2s =0,
\]
\begin{equation}
 \dr_i\cL-(\dr_\la+\dr_\la s^j\dr_j
+\dr_\la\dr_\mu s^j \dr^\m_j)\dr^\la_i\cL=0,
\label{2.29}
\end{equation}
for sections $s$ of the fibred manifold $Y\to X$. In other words,
Eqs.(\ref{306a}) and (\ref{306b}) represent the familiar first order
reduction of the second order differential equations (\ref{2.29}).
At the same time, the Cauchy problems for Eqs.(\ref{2.29})
and the system of Eqs.(\ref{306a}) and (\ref{306b})
fail to be equivalent.

We have the following conservation laws on solutions of the first order
Euler-Lagrange equations.

Let
\[ u=u^\mu\dr_\mu + u^i\dr_i\]
be a vector field on a fibred manifold $Y$ and $\ol u$ its jet lift
(\ref{1.21}) onto the fibred jet manifold $J^1Y\to X$. Given a
Lagrangian density $L$ on $J^1Y$, let us computer the Lie derivative
${\bf L}_{\ol u}L$. We have
\begin{equation}
{\bf L}_{\ol u}L= [\wh \dr_\la(\pi^\la_i(u^i-u^\mu y^i_\mu ) +u^\la\cL
)+ (u^i-u^\mu y^i_\mu )(\dr_i-\wh\dr_\la\dr^\la_i)\cL]\om, \label{501}
\end{equation}
\[\wh\dr_\la =\dr_\la +y^i_\la\dr_i+y^i_{\m\la}\dr^\m_i.\]
On solutions $\ol s$ of the first order Euler-Lagrange equations, the
equality (\ref{501}) comes to the conservation law
\begin{equation}
\ol s^*{\bf L}_{\ol u}L= \frac{d}{dx^\la}[\pi^\mu_i(\ol s)(u^i-u^\mu \ol
s^i_\mu) +u^\la\cL (\ol s)]\om. \label{502}
\end{equation}

In particular, if $u$ is a vertical vector field such that
\[{\bf L}_{\ol u}L=0,\]
the conservation law (\ref{502}) takes the form of the current
conservation law
\begin{equation}
\frac{d}{dx^\la}[u^i\pi^\mu_i(\ol s)]=0. \label{503}
\end{equation}
In gauge theory, this conservation law is exemplified by the N\"oether
identities.

Now, let
\[\tau=\tau^\la\dr_\la\]
be a vector field on $X$ and
\[ u=\tau_\G=\tau^\mu (\dr_\mu+\G^i_\mu\dr_i)\]
its horizontal lift onto the fibred manifold $Y$ by a connection $\G$
on $Y$. In this case, the conservation law (\ref{502}) takes the form
\begin{equation}
{\bf L}_{\ol\tau_\G}L=-\frac{d}{dx^\la}[\tau^\mu T_\G{}^\la{}_\mu (\ol s)]
\om \label{504} \end{equation}
where $T_\G{}^\la{}_\mu (\ol s)$ are coefficients of the $T^*X$-valued
form
\begin{equation}
T_\G(\ol s)=-(\G\rfloor\Xi_L)\circ\ol s =[\pi^\la_i(\ol s^i_\mu-\G^i_\mu)
-\delta^\la_\mu\cL]dx^\mu\otimes\om_\la \label{S14}
\end{equation}
on $X$. One can think on this form as being the canonical energy-momentum
tensor of a field $s$ with respect to the connection $\G$ on $Y$. In
particular, when the fibration $Y\to X$ is trivial,
one can choose the trivial connection $\G=\theta_X$.
In this case, the form (\ref{S14}) is precisely the familiar canonical
energy-momentum tensor. If
\[{\bf L}_\tau\cL=0\]
for all vector fields $\tau$ on $X$, the conservation law (\ref{502})
comes to the conservation law
\[\frac{d}{dx^\la} T^\la{}_\mu (\ol s)=0\]
of the canonical energy-momentum tensor.

\section{}

In the framework of the first order Lagrangian formalism, there exist two
different morphisms called the
Legendre morphisms which lead to different candidates
for a phase space of fields.

(i)  Given a first order Lagrangian density $L$ [Eq.(\ref{301})],
let us consider the vertical tangent morphism  $VL$ to $L$.
Due to  the canonical vertical splitting
\[
VJ^1Y=J^1Y\op\times_Y (T^*X\op\otimes_Y VY)
\]
of the affine jet bundle $J^1Y\to Y$, this morphism yields
the linear morphism (\ref{64})
\[
\ol VL:J^1Y\op\times_Y(T^*X\op\otimes_Y VY)\op\to_{J^1Y}J^1Y\op\times_Y
\op\w^nT^*X
\]
over $J^1Y$. The corresponding morphism (\ref{29})
\[
\wh L:J^1Y\op\to_Y\Pi,
\]
\begin{equation}
\Pi=\op\w^nT^*X\op\otimes_YTX\op\otimes_YV^*Y, \label{00}
\end{equation}
is defined to be the Legendre morphism associated with the
Lagrangian density $L$ (or simply the Legendre morphism).
We call $\Pi\to Y$ [Eq.(\ref{00})] the Legendre bundle over the fibred
manifold $Y\to X$.
Given fibred coordinates (\ref{1.2}) of $Y$ and the corresponding
induced coordinates of the bundles $TX$, $T^*X$ and $V^*Y$, the
Legendre bundle $\Pi$ is provided with the fibred coordinates
\[
(x^\m,y^i,p^\la_i).
\]
With respect to these coordinates, the Legendre morphism
$\wh L$ reads
\[
 p^\m_i\circ\wh L=\pi^\m_i,
\]
The generalized Hamiltonian formalism
founded on the phase space (\ref{00}) is the multimomentum Hamiltonian
formalism developed in next Lecture.

(ii) The Poincar\'e-Cartan form $\Xi_L$ [Eq.(\ref{303})] yields the bundle
morphism $\wh\Xi_L$ of the jet bundle $J^1Y\to Y$ to the bundle
\begin{equation}
Z = \op\w^{n-1}T^*X\w T^*Y  \label{N41}
\end{equation}
over $Y$. It is termed the Legendre morphism associated with $\Xi_L$.
The bundle (\ref{N41}) is endowed with the coordinates
\be
&&(x^\la, y^i, p^\la_i, p),\\
&&{p'}^\la_i=J\frac{\dr y^j}{\dr {y'}^i}\frac{\dr {x'}^\la}{\dr x^\mu}
p^\mu_j,\\
&&p'=J(p-\frac{\dr y^j}{\dr {y'}^i}\frac{\dr {y'}^i}{\dr x^\mu}
p^\mu_j),\\
&&J^{-1}=\det (\frac{\dr {x'}^\la}{\dr x^\mu}).
\ee
Relative to these coordinates, the Legendre morphism
$\wh\Xi_L$ is written
\begin{equation}
(p^\m_i, p)\circ\wh\Xi_L =(\pi^\m_i,\pi ). \label{N42}
\end{equation}
The Hamiltonian formalism founded on the phase space (\ref{N41}) is the
De Donder Hamiltonian partner
of the Lagrangian formalism on fibred manifolds.

The bundle (\ref{N41}) carries the the canonical exterior $n$-form
\begin{equation}
\Xi= p\om + p^\la_i dy^i\w\om_\la \label{N43}
\end{equation}
defined by the relation
\[
u_1\rfloor...u_n\rfloor\Xi (w) = T\zeta (u_1)\rfloor...T\zeta(u_n)
\rfloor w,
\]
\[
  w\in Z, \qquad u_i\in T_wZ,
\]
where $\zeta$ is the canonical surjection $Z\to Y$.
It is readily observed
that the Poincar\'e-Cartan form $\Xi_L$ is the pullback of the
canonical form $\Xi$ [Eq.(\ref{N43})]
by the associated Legendre morphism (\ref{N42}).

The canonical form (\ref{N43}) provides the bundle
(\ref{N41}) with the multisymplectic structure characterized by
the  multisymplectic form
\begin{equation}
\Om_Z=d\Xi=dp^\la_i\w dy^i\w\om_\la+dp\w\om.\label{106}
\end{equation}

\begin{remark}  In case
of mechanics when $X=\Bbb R$, the form $\Xi$  reduces
to the Liouville form
\begin{equation}
\Xi = p_i dy^i-Edt \label{N0}
\end{equation}
for the homogeneous formalism
where $t$ is the temporal coordinate and $E$ is
the energy variable.$^{18}$
However, in field theory when $n>1$, the straightforward physical treatment
of the coordinate $p$ is not evident.
\end{remark}

 Given a Lagrangian density $L$,
let the image $Z_L$ of the configuration space $J^1Y$ by the
 morphism (\ref{N42}) be an imbedded subbundle
\[
i_L:Z_L\hookrightarrow Z
\]
of the bundle $Z$.
In the framework of the De Donder Hamiltonian
formalism, the pullback De Donder form
\[
H_L=i^*_L\Xi
\]
on $Z_L$ is introduced. By analogy with the Cartan equations
(\ref{316}), the corresponding  De Donder-Hamilton equations for
sections $\ol r$ of the bundle $Z_L\to X$ are written
\begin{equation}
\ol r^*(u\rfloor dH_L)=0 \label{N46}
\end{equation}
where $u$ is an arbitrary vertical vector field on
$Z_L\to X$. To bring the De Donder-Hamilton equations (\ref{N46})
into the explicit form, one should substitute solutions
$y^i_\la (x^\m,y^i,p_i^\mu)$ and $\cL (x^\m,y^i,p_i^\mu)$
 of the equations
\ben
&&p^\la_i=\pi^\la_i(x^\m,y^i,y^i_\mu), \nonumber\\
&& p=\cL -\pi^\m_i y^i_\m  \label{N60}
\een
into the Cartan equations (\ref{316}).

We thus observe that the De Donder Hamiltonian formalism is not formulated
instrinsically. It is the derivation of the Lagrangian formalism on fibred
manifolds.

If a Lagrangian density is regular, Eqs.(\ref{N60}) have
the unique solution and the De Donder-Hamilton equations take the
coordinate form
\be
&&\dr_\mu\ol r^i =- \dr^i_\mu\ol r, \\
&&\dr_\mu\ol r^\mu_i = \dr_i\ol r.
\ee
They are equivalent to the Cartan equations.

If a Lagrangian density is
degenerate, Eqs.(\ref{N60}) admit different solutions or no
solutions at all. At the same time,
when the Legendre morphism $\wh\Xi_L$ is a submersion
$J^1Y\to Z_L$, the De Donder-Hamilton equations (\ref{N46}) are
proved to be equivalent
to the Cartan equations (\ref{316}).

Lagrangian densities of field models are almost always of this type.
It follows that, in comparison with the Lagrangian machinery, its
De Donder Hamiltonian partner has no advantages of describing constraint
field systems, otherwise the multimomentum Hamiltonian formalism where
the phase space of fields is the Legendre bundle $\Pi$ [Eq.(\ref{00})]
over $Y$.

The relations between the De Donder-Hamiltonian formalism and the
multimomentum Hamiltonian one stemmed from the fact that
the total space of the bundle $Z$ [Eq.(\ref{N41})] represents also
the 1-dimensional affine bundle
\begin{equation}
\varsigma: Z\to\Pi, \label{100}
\end{equation}
\[
 (y^i,p_i^\mu)\circ\varsigma = (y^i,p_i^\mu),
\]
over the Legendre bundle $\Pi$ [Eq.(\ref{00})].
Furthermore, the exact sequence (\ref{1.8b}) gives rise to the exact
sequence
\begin{equation}
0\to\op\w^nT^*X\hookrightarrow Z\to\Pi\to 0 \label{101}
\end{equation}
over $Y$. In particular, every splitting of the exact sequence
(\ref{1.8b}) by a connection $\G$ on $Y\to X$ implies the corresponding
splitting
\ben
&& h_\G:\Pi\hookrightarrow Z, \label{102}\\
&& p\circ h_\G=-p^\la_i\G^i_\la, \nonumber
\een
of the exact sequence (\ref{101}). Moreover, there is the 1:1
correspondence between global sections of the bundle (\ref{100}) and
the multimomentum Hamiltonian forms on the Legendre bundle $\Pi$.

\newpage


\centerline{\large\bf Lecture 4. HAMILTONIAN FORMALISM}
\bigskip\bigskip

The Hamiltonian approach to field theory was called into
play mainly for canonical quantization of fields by analogy with quantum
mechanics. The major goal of this approach has consisted in establishing
simultaneous commutation relations of quantum fields in models with
degenerate Lagrangian densities, e.g., gauge theories.

In classical field theory, the conventional Hamiltonian formalism
fails to be so successful. In the straightforward manner, it takes the form
of the instantaneous Hamiltonian formalism when canonical variables are
field functions at a given instant  of time. The corresponding phase space
is infinite-dimensional. Hamiltonian dynamics played out on this phase
space is far from to be a partner
of the usual Lagrangian dynamics of
field systems. In particular, there are no Hamilton equations in the
bracket form which would be adequate to Euler-Lagrange field
equations.

This Lecture presents the covariant finite-dimensional Hamiltonian
machinery for field theory which has been intensively
developed from 70th as both
the De Donder Hamiltonian partner of the higher order Lagrangian
formalism in the framework of the calculus of variations and
the multisymplectic (or polysimplectic) generalization of the conventional
Hamiltonian formalism in analytical mechanics when canonical momenta
correspond to derivatives of fields with respect to all world coordinates,
not only time. Each approach goes hand-in-hand with the other. They
exemplify the generalized Hamiltonian dynamics which is not merely a
time evolution directed by the Poisson bracket, but it is governed by
partial differential equations where temporal and spatial coordinates
enter on equal footing. Maintaining covariance has the principal
advantages of describing field theories, for any preliminary
space-time splitting shades the covariant picture of field constraints.

\section{}

The multimomentum Hamiltonian formalism in field theory is formulated
intrinsically in the context of the multisymplectic Legendre bundles
over fibred manifolds just as the Hamiltonian formalism in analytical
mechanics is phrased in terms of symplectic manifolds.

By the multisymplectic Legendre bundle, we mean the Legendre bundle
$\Pi$ provided with the multisymplectic structure whose main ingredients
are (i) the generalized Liouville form,
(ii) the multisymplectic form,
(iii)  Hamiltonian connections,
and (iv) multisymplectic canonical transformations.

The total space of the Legendre bundle $\Pi\to Y$ [Eq.(\ref{00})]
over a fibred manifold $Y\to X$ represents the composite  manifold
\ben
&&\pi_{\Pi X}:=\pi\circ\pi_{\Pi Y}:\Pi\to Y\to X,\label{2.1'}\\
&&\Pi =\op\w^n T^*X\op\otimes_Y V^*Y\op\otimes_Y TX. \label{2.1}
\een
We call it the fibred Legendre manifold (or simply the Legendre
manifold). The symbol $q$ will be utilized for elements of $\Pi$.

Given fibred coordinates $(x^\la, y^i)$ of the fibred manifold $Y$ and
induced coordinates $(x^\la, \dot x^\la)$ and $(x^\la, \dot x_\la)$ of the
bundles $TX$ and $T^*X$, the Legendre manifold (\ref{2.1})
is provided with an atlas of fibred coordinates
\ben
&&( x^\la ,y^i,p^\la_i),\nonumber\\
&&x^\la \to {x'}^\la(x^\m), \nonumber\\
&& y^i \to {y'}^i(x^\m,y^j),\nonumber \\
&& {p'}^\la_i = J \frac{\dr y^j}{\dr{y'}^i} \frac{\dr
{x'}^\la}{\dr x^\m}p^\m_j, \label{2.3}\\
&& J^{-1}=\det (\frac{\dr {x'}^\la}{\dr x^\m}). \nonumber
\een
These coordinates are compatible with the composite fibration (\ref{2.1'})
of the Legendre bundle (\ref{2.1}).
We shall call them the canonical coordinates.

The coordinate transformations (\ref{2.3}) give the local expression for
isomorphisms of the Legendre manifold (\ref{2.1}). These isomorphisms,
by definition, are generated by
isomorphisms of the fibred manifold $Y\to X$ and the corresponding induced
isomorphisms of the bundles $V^*Y$, $TX$ and $T^*X$.
They preserve the composite fibration (\ref{2.1'}) of the
Legendre bundle $\Pi$.

There are the following noteworthy morphisms of the Legendre bundle:
\begin{itemize}
\item  the canonical isomorphism
\ben
&&i:\Pi \to\op\w^{n-1}T^*X\op\otimes_Y V^*Y, \nonumber\\
&&i=p^\la_i\om_\la\otimes\ol dy^i,\label{401}
\een
where by $\{\ol dy^i\}$ are meant the fibre bases for the vertical
cotangent bundle $V^*Y$ of $Y$;
\item  the canonical bundle monomorphism
\ben
&&\th :\Pi\op\hookrightarrow_Y\op\w^{n+1}T^*Y\op\otimes_Y TX, \label{402}\\
&& \th =-p^\la_idy^i\w\om\otimes\dr_\la, \label{2.4}
\een
which is the straigtforward carollary of the canonical bundle monomorphism
(\ref{86}).
\end{itemize}

\begin{definition}{12.1} The generalized Liouville
form on the Legendre bundle $\Pi\to Y$ is defined to be
the pullback-valued horizontal form $\th$ [Eq.(\ref{2.4})]
corresponding to the canonical bundle monomorphism (\ref{402}).
\end{definition}

Since $\th$ [Eq.(\ref{2.4})] is a pullback-valued form, one can not act
on it by the exterior differential in order to get the corresponding
multysimplectic form on the Legendre bundle $\Pi$. Under
these circumstances, we begin with
the following straightforword definition of the multisymplectic form.

\begin{definition}{12.2} The multisymplectic form on the Legendre bundle
$\Pi$ is defined to be the pullback-valued form given by the coordinate
expression
\begin{equation}
\Om =dp^\la_i\w dy^i\w\om\otimes\dr_\la \label{406}
\end{equation}
with respect to canonical coordinate atlases (\ref{2.3}) of $\Pi$.
\end{definition}

It is readily observed that this definition is
independent of choice of the
coordinate atlas (\ref{2.3}). The form (\ref{406}) is globally defined and
invariant under all isomorphisms of the  Legendre manifold.

The multisymplectic form (\ref{406}) is related to the
generalized Liouville form (\ref{2.4}) as follows.

Let $V\th$ be the vertical tangent morphism
to the monomorphism (\ref{402}) over $Y$. As a shorthand,
we can write
\[
\Om=-V\th.
\]

\begin{proposition}{12.3} For each exterior 1-form on $X$ and
its pullback $\phi$ by $\pi_{\Pi X}$ onto the Legendre manifold $\Pi$, the
multisymplectic form (\ref{406}) and the generalized Liouvolle form
(\ref{2.4}) satisfy the relations
\ben
&& d(\Om\rfloor\phi)=0,\nonumber\\
&&\Om\rfloor\phi = -d(\th\rfloor\phi).\label{202}
\een
\end{proposition}

Building on the generalized Liouville form (\ref{2.4})
and the multisymplectic form (\ref{406}), we develope
the multimomentum Hamiltonian formalism which is compatible with the
De Donder-Weyl variant of the Lagrangian formalism when the Lepagean
equivalent of a Lagrangian density is choosen to be the
Poincar\'e-Cartan form.

Let $J^1\Pi$ be the first order jet manifold of the fibred
Legendre manifold
$\Pi\to X$. It is  provided with the adapted fibred coordinates
\[
( x^\la ,y^i,p^\la_i,y^i_{(\m)},p^\la_{i\m}).
\]
In particular, the bundle surjection $J^1\pi_{\Pi Y}$ of $J^1\Pi$
onto $J^1Y$ reads
\[
y^i_\m\circ J^1\pi_{\Pi Y}=y^i_{(\m)}.
\]

\begin{definition}{12.4} We say that a jet field [resp. a connection]
\[
\g =dx^\la\otimes(\dr_\la +\g^i_{(\la)}\dr_i
+\g^\m_{i\la}\dr^i_\m)
\]
on the fibred Legendre manifold $\Pi\to X$ is a Hamiltonian jet field
[resp. a Hamiltonian connection] if the exterior form
\[
\g\rfloor\Om =dp^\la_i\w dy^i\w\om_\la
+\g^\la_{i\la}dy^i\w\om -\g^i_{(\la)} dp^\la_i\w\om
\]
on $\Pi$ is closed.
\end{definition}

It is easy to verify that a jet field $\g$ on $\Pi$ is a Hamiltonian
jet field iff it satisfies the conditions
\ben
&&\dr^i_\la\g^j_{(\m)}-\dr^j_\m\g^i_{(\la)}=0,\nonumber\\
&& \dr_i\g_{j\m}^\m- \dr_j\g_{i\m}^\m=0,\nonumber\\
&& \dr_j\g_{(\la)}^i+\dr_\la^i\g_{j\m}^\m=0.\label{422}
\een
Hamiltonian connections
constitute a subspace of the affine space of connections on the fibred
Legendre manifold $\Pi\to X$. This subspace is not empty as follows.

\begin{lemma}{12.5} Every connection $\G$ on a fibred manifold $Y\to X$
gives rise to the connection
\ben
&& \wt\G =dx^\la\otimes[\dr_\la +\G^i_\la (y)\dr_i\nonumber\\
&& \qquad +
(-\dr_j\G^i_\la (y)  p^\m_i-K^\m{}_{\nu\la}(x) p^\nu_j+K^\al{}_{\al\la}(x)
p^\m_j)\dr^j_\m]  \label{404}
\een
on the Legendre manifold $\Pi\to Y\to X$ where $K$
is a symmetric linear connection (\ref{408})
on the bundles $TX$ and $T^*X$.
\end{lemma}

Let $\G$ be a connection on the fibred manifold $Y\to X$ and
\ben
&&\wh\G:=\G\circ\pi_{\Pi Y}:\Pi\to Y\to J^1Y, \nonumber\\
&&\wh\G=dx^\la\otimes (\dr_\la +\G^i_\la\dr_i), \label{104}
\een
its pullback by $\pi_{\Pi Y}$ onto the Legendre bundle $\Pi$ over $Y$.
Then, the lift $\wt\G$ [Eq.(\ref{404})] of $\G$ onto
the fibred Legendre manifold $\Pi\to X$ obeys the identity
\begin{equation}
\wt\G\rfloor\Om =d(\wh\G\rfloor\th). \label{412}
\end{equation}
A glance at this identity shows that $\wt\G$ is a Hamiltonian connection.

Thus, Hamiltonian connections always exist on the Legendre bundle $\Pi$.

\begin{definition}{12.6}
By multisymplectic diffeomorphisms of the Legendre
manifold $\Pi$, we mean isomorphisms of the fibred manifold $\Pi\to X$
(but not {\it a priori} the composite manifold
$\Pi\to Y\to X$) which maintain
the multisymplectic form (\ref{406}).
\end{definition}

The following assertion shows that multisymplectic diffeomorphisms of
the Legendre manifold $\Pi$ preserve really
its composite fibration and, moreover,
they are almost exausted by isomorphisms of $\Pi$.

\begin{proposition}{12.7} Unless $n=1$, every multisymplectic
diffeomorphism $\Phi$ of the Legendre manifold $\Pi$ decomposes as
the semidirect product of an
isomorphism of $\Pi$ and an affine bundle morphism
\begin{equation}
q\to q+r(\pi_{\Pi Y}(q)), \qquad q\in\Pi,\label{105}
\end{equation}
of the Legendre bundle $\Pi$ over $Y$ where
\[
r=r^\la_i (y)\ol dy^i\w\om\otimes\dr_\la
\]
is a section of $\Pi\to Y$ such that, for each exterior 1-form
 on $X$ and its pullback $\phi$ by $\pi_{\Pi X}$ onto $\Pi$, the form
\[
r\rfloor\phi =r^\la_i (y)\phi_\la (x)\ol dy^i\w\om
\]
is closed.
\end{proposition}

In coordinate terms, multisymplectic diffeomorphisms represent
compositions of the coordinate transformations (\ref{2.3}) and
translations
\[
{p'}^\la_i = p^\la_i + r^\la_i(y)
\]
where
\[
 \dr_jr^\la_i(y)=\dr_ir^\la_j(y).
\]

Building on the relations (\ref{422}), it is easy to check
that every multisymplectic diffeomorphism sends a
Hamiltonian jet field to a Hamiltonian jet field.

It follows that the multisymplectic diffeomorphisms can be regarded as the
multisymplectic canonical transformations.
In view of Proposition 12.7, multisymplectic
canonical transformations have the
structure which is much simpler than that of canonical transformations
in the symplectic case of $n=1$.
Roughly speaking, the multisymplectic canonical transformations do not
admit exchange of the canonical variables $y^i$ for the
canonical momenta $p^\la_i$ since $y^i$ and $p^\la_i$ constitute
the linear spaces of different dimensions.

\section{}

In the multimomentum Hamiltonian formalism, Hamiltonian connections play
the same role as Hamiltonian vector fields in the symplectic geometry.
Accordingly, the definition of multimomentum Hamiltonian forms follows
the definition of Hamiltonians in symplectic geometry.

\begin{definition}{13.1} An exterior $n$-form $H$ on the multisymplectic
Legendre manifold $\Pi$ is called a multimomentum Hamiltonian form if, on an
open neighborhood of each point of $\Pi$,
there exists a Hamiltonian jet field satisfying the equation
\begin{equation}
\g\rfloor\Om  =dH. \label{4.1}
\end{equation}
This jet field $\g$ is termed the associated Hamiltonian jet field for
$H$.
\end{definition}

A glance at the relation (\ref{4.1}) shows that multimomentum
Hamiltonian forms may be considered modulo
closed forms, for closed forms do not make contribution into the Hamilton
equations.

\begin{proposition}{13.2} Let $H$ be a multimomentum Hamiltonian form. For
any exterior horizontal density
\begin{equation}
\wt H=\wt{\cH}\om \label{4.2}
\end{equation}
 on the fibred Legendre manifold $\Pi\to X$, the form $H-\wt H$
also is a multimomentum Hamiltonian form.
Conversely, if $H$ and $H'$ are multimomentum Hamiltonian forms,
their difference $H-H'$ is an exterior horizontal density on $\Pi\to X$.
\end{proposition}

In virtue of Proposition 13.2, multimomentum Hamiltonian
 forms constitute an affine space modelled
on the linear space of horizontal densities (\ref{4.2}). A glance
at Eq.(\ref{412}) shows that this affine space is not empty.

\begin{corollary}{13.3} Given a connection $\G$ on
a fibred manifold $Y\to X$, its lift (\ref{404}) onto the fibred Legendre
manifold $\Pi\to X$  is a Hamiltonian connection for the multimomentum
Hamiltonian form
\ben
&& H_\G =\wh\G\rfloor\th :\Pi\to\op\w^n T^*Y, \nonumber\\
&& H_\G =p^\la_i dy^i\w\om_\la -p^\la_i\G^i_\la (y)\om. \label{3.6}
\een
\end{corollary}

It follows that every multimomentum
Hamiltonian form on the Legendre bundle $\Pi$ over a fibred manifold
$Y\to X$ is expressed as
\begin{equation}
H=H_\G -\wt H_\G =p^\la_idy^i\w\om_\la
-p^\la_i\G^i_\la\om-\wt{\cH}_\G\om=p^\la_idy^i\w\om_\la-\cH\om\label{4.7}
\end{equation}
where $\G$ is some connection on $Y\to X$.
Given another connection
\[
\G'=\G+\si
\]
 where $\si$ is a soldering form on $Y$,  we have
\[
 \wt H_{\G'}=\wt H_\G-p^\la_i\si^i_\la\om.
\]

\begin{proposition}{13.4} Unless $n=1$, the multimomentum canonical
transformations maintain the splitting (\ref{4.7}) with
accuracy to a closed form.
\end{proposition}

In view of Proposition 13.4, one can think of Eq.(\ref{4.7})
as being a workable definition of
multimomentum Hamiltonian forms modulo closed forms.
One can start with this definition and prove afterwards that, for any
multimomentum Hamiltonian form (\ref{4.7}), there always exists an
associated Hamiltonian jet field since the algebraic
Hamilton equations (\ref{3.10a}) and (\ref{3.10b}) below
always have a local solution.

Note that the expression (\ref{4.7}) for multimomentum Hamiltonian
forms depends upon the arbitrary
specification of elements extrinsic to the multisymplectic structure such
as a connection $\G$ on a fibred manifold $Y\to X$. At the same time,
every multimomentum Hamiltonian form itself sets up the associated
connection $\G_H$ on $Y$ and thereby has the canonical
splitting (\ref{4.7}) as follows.

We call by a momentum morphism any bundle morphism
\ben
&&\Phi:\Pi\op\to_Y J^1Y,\nonumber\\
&& y^i_\la\circ\Phi= \Phi^i_\la(q), \qquad q\in\Pi, \label{2.6}
\een
over $Y$. Its composition with the
contact map (\ref{18}) yields the bundle morphism
\[
\la\circ\Phi:\Pi\op\to_Y T^*X\op\otimes_YTY
\]
represented by the horizontal pullback-valued 1-form
\begin{equation}
\Phi =dx^\la\otimes(\dr_\la +\Phi^i_\la (q)\dr_i)\label{2.7}
 \end{equation}
on the Legendre bundle $\Pi\to Y$.

For instance, let $\G$ be a connection on a fibred manifold $Y$. Then, the
composition $\wh\G$ [Eq.(\ref{104})] is a momentum morphism. Conversely,
every momentum morphism $\Phi$ of the Legendre bundle $\Pi$ over $Y$ defines
the associated connection
\[
 \G_\Phi =\Phi\circ\wh 0
\]
on $Y\to X$ where $\wh 0$ is the global zero section of the
Legendre bundle $\Pi\to Y$. In particular, we have
\[
\G_{\wh\G}=\G.
\]

\begin{proposition}{13.5}
 Given a multimomentum Hamiltonian form $H$ (\ref{4.7}) on the Legendre
bundle $\Pi\to Y$, the vertical tangent morphism $VH$ to
$H$ defines the associated momentum morphism
\ben
&&\wh H:\Pi\to J^1Y,\nonumber\\
&& y_\la^i\circ\wh H=\dr^i_\la\cH. \label{415}
\een
\end{proposition}

\begin{corollary}{13.6} Every multimomentum Hamiltonian form $H$ on the
Legendre bundle $\Pi$ over a fibred manifold $Y$ sets up
the associated connection
\[
\G_H =\wh H\circ\wh 0
\]
on this fibred manifold $Y\to X$.
\end{corollary}

In particular, we have
\[
\G_{H_\G}=\G
\]
where $H_\G$ is the multimomentum Hamiltonian form (\ref{3.6}) associated
with the connection $\G$ on $Y\to X$.

\begin{corollary}{13.7}
Every multimomentum Hamiltonian form (\ref{4.7}) admits the canonical
splitting
\begin{equation}
H=H_{\G_H}-\wt H.\label{3.8}
\end{equation}
\end{corollary}

 The horizontal density
\[
\wt H=\wt{\cH}\om
\]
in this canonical splitting meets the condition
\[
\dr^i_\la \wt{\cH}\circ\wh 0=0
\]
where $\wh 0$ is the canonical zero section of the Legendre bundle
$\Pi\to Y$. It will be termed the Hamiltonian density.

\begin{proposition}{13.8}
Every momentum morphism (\ref{2.6}) represented by
the pullback-valued form (\ref{2.7}) on $\Pi$ yields the associated
multimomentum Hamiltonian form
\begin{equation}
H_\Phi=\Phi\rfloor\th =p^\la_idy^i\w\om_\la -p^\la_i\Phi^i_\la\om.
\label{414}
\end{equation}
\end{proposition}

For instance, if a multimomentum Hamiltonian form $H$ satisfies the
condition
\[
 H_{\wh H}=H,
\]
then we have
\[
H=H_\G
\]
for some connection $\G$ on $Y$.

Note that the converse to Proposition 13.8 is not true.

To characterize completely the affine space of
multimomentum Hamiltonian forms
on the Legendre bundle, let us turn to the exact sequence (\ref{101})
\[
0\to \op\w^nT^*X\to Z\to\Pi \to 0.
\]
Every splitting of the exact sequence (\ref{1.8b}) by a connection $\G$ on
the fibred manifold $Y\to X$ yields the splitting
of the exact sequence (\ref{101}).
Let $h_\G$ be the corresponding section (\ref{102}) of the bundle
$Z\to\Pi$ and $\Xi$ the canonical form (\ref{N43}) on $Z$.
It is readily observed that the pullback $h_\G^*\Xi$ is precisely
the multimomentum Hamiltonian form $H_\G$ [Eq.(\ref{3.6})].
Building on the splitting (\ref{4.7}), one then can
justify that every section $h$ of the bundle $Z\to \Pi$
determines the corresponding multimomentum Hamiltonian form
$h^*\Xi$ on $\Pi$. Conversely, the splitting (\ref{4.7}) shows
that any multimomentum Hamiltonian
form defines a section of the bundle $Z\to\Pi$.

Thus, there is the 1:1 correspondence between the multimomentum
Hamiltonian forms on the Legendre bundle $\Pi$ and the global sections
of the bundle $Z\to\Pi$ [Eq.(\ref{100})].

\section{}

In this Section, we assign to every multimomentum Hamiltonian form its
Hamilton equations.

Given a multimomentum Hamiltonian form, the Hamilton equations provide the
conditions for a Hamiltonian jet field be associated with this Hamiltonian
form. In contrast with analytical mechanics, there always exists
a family of different Hamiltonian jet
fields for the same multimomentum Hamiltonian form. To characterize this
family, the Hamilton operator is introduced.
In particular, we aim to show that, for every multimomentum
Hamiltonian form, there exists the associated Hamiltonian connection.

\begin{definition}{14.1} The Hamilton operator $\cE_H$ for a
multimomentum Hamiltonian form $H$ on the Legendre bundle $\Pi$ is defined
to be the first order differential operator
\ben
&& \cE_H :J^1\Pi\to\op\w^{n+1} T^*\Pi,\nonumber \\
&& \cE_H=dH-\wh\Om=[(y^i_{(\la)}-\dr^i_\la\cH) dp^\la_i
-(p^\la_{i\la}+\dr_i\cH) dy^i]\w\om \label{3.9}
\een
where
\[
\wh\Om=dp^\la_i\w
dy^i\w\om_\la +p^\la_{i\la}dy^i\w\om -y^i_{(\la)} dp^\la_i\w\om
\]
is the pullback of the multisymplectic form (\ref{406}) onto $J^1\Pi$.
\end{definition}

The Hamilton operator (\ref{3.9}) is an affine morphism, and the
surjection
\begin{equation}
\Ker\cE_H\to\Pi \label{417}
\end{equation}
is an affine subbundle of the jet bundle $J^1\Pi\to\Pi$. It is
given by the coordinate conditions
\ben
&& y^i_{(\la)}=\dr^i_\la\cH,\nonumber\\
&&p^\la_{i\la}=-\dr_i\cH. \label{108}
\een

For any jet field
\[
\g =dx^\la\otimes(\dr_\la +\g^i_{(\la)}\dr_i+\g^\m_{i\la}\dr^i_\m)
\]
on the Legendre manifold, we have the equality
\[
\cE_H\circ\g =dH-\g\rfloor\Om.
\]
A glance at this equality shows that a jet field $\g$ is the Hamiltonian
jet field for a
multimomentum Hamiltonian form $H$ iff it takes its values into
$\Ker\cE_H$, that is, it is a section of the affine bundle (\ref{417}) and
obeys the algebraic Hamilton equations
\[
 \cE_H\circ\g=0
\]
which are written
\bea
&&\g^i_{(\la)} =\dr^i_\la\cH, \label{3.10a}\\
&& \g^\la_{i\la}=-\dr_i\cH. \label{3.10b}
\eea

Being affine, the bundle (\ref{417}) always has a global section due
to the properites required of manifolds. In other words,
a Hamiltonian connection for every multimomentum Hamiltonian form
always exists.

In virtue of Eq.(\ref{3.10a}), every  Hamiltonian connection $\g$
for a multimomentum Hamiltonian form $H$ satisfies the relation
\begin{equation}
J^1\pi_{\Pi Y}\circ \g= \wh H \label{109}
\end{equation}
and takes the form
\[
\g =dx^\la\otimes(\dr_\la +\dr^i_\la\cH\dr_i+\g^\m_{i\la}\dr^i_\m).
\]

A glance at Eq.(\ref{3.10b}) shows that there is a set of
Hamiltonian jet fields for the same multimomentum Hamiltonian form.
They differ from each other in local
soldering forms $\wt\si$ on $\Pi\to X$ which obey the equations
\[
\wt\si\rfloor\Om =0
\]
whcih are written
\ben
&&\wt\si^i_\la =0,\nonumber\\
&& \wt\si^\la_{i\la} =0. \label{2.9}
\een

Note that Eqs.(\ref{2.9}) for
all multimomentum Hamiltonian forms are the same. It follows that,
for  all Hamilton operators $\cE_H$, the affine bundle (\ref{417}) is
modelled on the same vector subbundle
\be
&&\ol y^i_\la=0, \\
&& \ol p^\la_{i\la}=0
\ee
of the vector bundle
\[
T^*X\op\otimes_\Pi V\Pi\to \Pi.
\]

\begin{remark}
If $n=1$, Eqs.(\ref{2.9}) have evidently only the zero solution
and consequently, there is always the unique Hamiltonian connection
associated with a given multimomentum
Hamiltonian form. This fact is well-known from analytical mechanics.
\end{remark}

\begin{example}{14.2}
As a test case, let us construct a Hamiltonian connection for a given
multimomentum Hamiltonian form $H$ if (i) a fibred manifold $Y\to X$ admits
the vertical splitting (\ref{1.9}) and (ii) relative the coordinates
adapted to this vertical splitting, the Hamiltonian density in
the canonical splitting (\ref{3.8}) of $H$ is independent of
coordinates $y^i$.
If a fibred manifold $Y$ admits vertical splitting,
every momentum morphism (\ref{2.6}) defines the connection $\g_\Phi$ on
the fibred Legendre manifold $\Pi\to X$ which is given by the expression
\ben
&& \g_\Phi =dx^\la\otimes[\dr_\la  +\Phi^i_\la(q)\dr_i \nonumber\\
&& \qquad+
(-\dr_j\Phi^i_\la(q) p^\m_i-K^\m{}_{\nu\la}(x)p^\nu_j+K^\al{}_{\al\la}
(x)p^\m_j)\dr^j_\m] \label{405}
\een
with respect to the coordinates adapted to this splitting. Here, $K$
denotes both a symmetric linear connection (\ref{408}) on $TX$ and
its dual on $T^*X$. In particular, we have
\[
\g_{\wh\G}=\wt\G.
\]
Let $H$ be the above-mentioned multimomentum Hamiltonian form and
$\wh H$ the associated momentum morphism (\ref{415}).
 Then, the connection $\g_{\wh H}$ [Eq.(\ref{405})] defined by $\wh H$
is a Hamiltonian connection
\[
\g_{\wh H} =dx^\la\otimes[\dr_\la  +\dr^i_\la\cH\dr_i+
(-\dr_j\dr^i_\la\cH p^\m_i-K^\m{}_{\nu\la} p^\nu_j+K^\al{}_{\al\la}
p^\m_j)\dr^j_\m]
\]
for $H$.
\end{example}

Let $r$ be a section of the fibred Legendre manifold $\Pi\to X$ such that
its jet prolongation $J^1r$ takes its values into the kernal $\Ker\cE_H$
of the Hamilton operator (\ref{3.9}).
Then, the algebraic Hamilton equations (\ref{3.10a}) and (\ref{3.10b})
come to the system of first order differential Hamilton equations
\bea
&&\dr_\la r^i =\dr^i_\la\cH, \label{3.11a}\\
 &&\dr_\la r^\la_i =-\dr_i\cH.\label{3.11b}
\eea
Conversely, if  $r$ is a
solution [resp. a global solution] of the Hamilton equations
(\ref{3.11a}) and (\ref{3.11b}) for a
multimomentum Hamiltonian form $H$, there exists an extension of this
solution to a Hamiltonian jet field $\g$ [resp. a Hamiltonian connection]
which has an integral section $r$, that is,
\[
\g\circ r=J^1r.
\]
Substituting $J^1r$ into Eq.(\ref{109}), we obtain the identity
\begin{equation}
J^1(\pi_{\Pi Y}\circ r)= \wh H\circ r \label{N10}
\end{equation}
for every solution of the Hamilton equations (\ref{3.11a}) and
(\ref{3.11b}).

Let us note that the Hamilton equations (\ref{3.11a}) and (\ref{3.11b}) can
be introduced without appealing to the Hamilton operator. Set
\begin{equation}
r^*(u\rfloor dH)= 0 \label{N7}
\end{equation}
for any vertical vector field $u$ on the
fibred Legendre manifold $\Pi\to X$. The
Hamilton equations (\ref{N7}) are similar to the Cartan equations
(\ref{316}), for the Poincar\'e-Cartan form appears to be the Lagrangian
counterpart of the multimomentum Hamiltonian forms.

\section{}

If the parameter space $X$ is $\Bbb R$, the
multimomentum Hamiltonian formalism reproduces the familiar time-dependent
Hamiltonian formalism of analytical mechanics.

In this case, a fibred manifold $Y\to X$ is trivial so that there are
the identifications
\ben
&&Y=\Bbb R\times F,\label{3.12'}\\
&& \Pi =\Bbb R\times T^*F \label{3.12}
\een
where $F$ is some manifold. The corresponding coordinates of $Y$ and $\Pi$
respectively are written
\ben
&& ( t,y^i), 	\nonumber\\
&& (t,y^i,p_i=\dot y_i) \label{4.5}
\een
where $t$ is the canonical parameter of $\Bbb R$.

The generalized Liouville form (\ref{2.4}) and the multisymplectic form
(\ref{406}) on the Legendre manifold $\Pi$ [Eq.(\ref{3.12})]
respectively appear the pullback of the canonical Liouville form
\[
\th=\dot y_i dy^i
\]
and the pullback of the canonical symplectic form
\[
 \Om =d\dot y_i\w dy^i
\]
on the cotangent bundle $T^*F$ by $\pr_2$ onto $\Pi$.

Let $\G_0$ be the trivial connection on $Y$ corresponding to splitting
(\ref{3.12'}) and $\wt\G_0$ its lift
(\ref{404}) onto $\Pi$. Relative to the coordinates (\ref{4.5}), they are
equal to the pullback-valued form (\ref{12})
\[
\th_X=dt\otimes\dr_t
\]
on $Y$ and $\Pi$
respectively. Hence, every  multimomentum Hamiltonian form on the
Legendre manifold $\Pi$ [Eq.(\ref{3.12})] is expressed as
\begin{equation}
 H=H_{\G_0}-\wt H=\dot
y_idy^i-\wt{\cH}_0dt. \label{N8}
\end{equation}
The corresponding algebraic Hamilton
equations (\ref{3.10a}) and (\ref{3.10b})
reduce to the well-known Hamilton equations
\be
&& u^i=\dr^i\wt{\cH}_0,\\
&& u_i=-\dr_i\wt{\cH}_0
\ee
for a Hamiltonian vector field
\[
u=u^i\dr_i +u_i\dr^i
\]
and the Hamiltonian $\wt\cH_0$ on $\Bbb R\times T^*F$. In particular,
there is the 1:1 correspondence between the Hamiltonian
vector fields $u$ and the Hamiltonian connections $\g$ on $\Bbb R\times
T^*F$. It is given by the relation
\begin{equation}
\g=dt\otimes u+\wt\G_0 \label{4.8}
\end{equation}
where
\[
\si=dt\otimes u
\]
means a soldering form on the fibred Legendre manifold
\[
\Bbb R\times T^*F\to \Bbb R.
\]
With respect to the canonical coordinates (\ref{4.5}), the relation
(\ref{4.8}) takes the form
\[
\g=dt\otimes (\dr_t+u).
\]

Conversely, let $T^*F$ be a standard symplectic manifold.
In the framework of the
1-dimensional multimomentum Hamiltonian formalism, every standard
time-dependent Hamiltonian
system with a Hamiltonian vector field $u$ and a Hamiltonian
$\wt{\cH}_0$ on $\Bbb R\times T^*F$
is described by the Hamiltonian connection (\ref{4.8})
associated with the multimomentum Hamiltonian form (\ref{N8})
on the Legendre manifold (\ref{3.12}).

There is the 1:1 correspondence between the
multimomentum Hamiltonian forms (\ref{N8}) and the global sections
of the bundle (\ref{100})
\[
Z=\Bbb R^2\times T^*F\to \Bbb R\times T^*F.
\]

Note that the expression (\ref{N8}) for a multimomentum Hamiltonian form
$H$ on $\Bbb R\times T^*F$ fails to consists with the canonical splitting
(\ref{3.8}) of $H$. Therefore, in the framework of the multimomentum
Hamiltonian formalism, there is no evident higher-dimensional
analogue of a Hamiltonian in mechanics.

Let us point at the following two percularities
of the multimomentum Hamiltonian
formalism when the parameter space $X$ is $\Bbb R$.
\begin{itemize}
\item In case of $n=1$, the canonical transformations, e.g.,
\[
{y'}^i=p_i, \qquad {p'}_i=-y^i,
\]
do not keep the fibration $\Pi\to Y$ of the Legendre
manifold and as a consequence, do not maintain the
splitting (\ref{4.7}) of multimomentum Hamiltonian forms in general.
Therefore, this splitting loses its canonical validity.
\item If $X=\Bbb R$, there exists the well-known bijection between the
multimomentum Hamiltonian forms (\ref{N8}) and the Hamiltonian connections
\[
\g=dt\otimes(\dr_t+\dr^i\wt{\cH}_0\dr_i -\dr_i\wt{\cH}_0\dr^i).
\]
Moreover, every Hamiltonian connection $\g$ has always an integral
section which is the integral curve
of the corresponding Hamiltonian vector field $u$.
\end{itemize}

In case of constraint Hamiltonian systems, this bijection makes necessary the
well-known Dirac-Bergman procedure of calculating the $k$-class
constraints.
In the framework of the multimomentum Hamiltonian formalism,
there exist different Hamiltonian connections associated with the same
multimomentum Hamiltonian form. Moreover, in field theory
when the primary constraint space $Q$ is the image of the
configuration space $J^1Y$
by the Legendre morphism $\wh L$, there is a family of
multimomentum Hamiltonian forms associated with the same
Lagrangian density as a rule. Therefore, one can choose either
the Hamilton equations or solutions of the Hamilton equations such
that these solutions live on the constraint space.

\section{}

This Section is devoted to the relations between the Lagrangian
formalism and the multimomentum Hamiltonian formalism in field theory when
a Lagrangian density is degenerate.
The major percularity of these relations consists in the fact that
one deals with a family of different multimomentum
Hamiltonian forms associated with the same degenerate Lagrangian density.

Given a fibred manifold $Y\to X$, let $H$ be a multimomentum
Hamiltonian form (\ref{4.7}) on the Legendre bundle $\Pi$ [Eq.(\ref{2.1})]
over $Y$ and $L$  a Lagrangian density (\ref{301})
on the jet manifold $J^1Y$ of $Y$. Let $\wh H$
be the corresponding momentum morphism (\ref{415}) and $\wh L$
the corresponding  Legendre morphism (\ref{00})
\ben
&&\wh L:J^1Y\op\to_Y\Pi,\nonumber\\
&& p^\m_i\circ\wh L=\pi^\m_i.\label{421}
\een
To define associated
Lagrangian and Hamiltonian systems, we consider the diagram
\[
\begin{array}{rcccl}
 & {\Pi} &  \op\longrightarrow^{\wh H} & {J^1Y} &  \\
{_{\wh L}} &\put(0,-10){\vector(0,1){20}} & & \put(0,10){\vector(0,-1){20}} &
{_{\wh L}} \\
 & {J^1Y} & \op\longleftarrow_{\wh H} & {\Pi} &
\end{array}
\]
 which does not commutative in general.
If the Legendre morphism $\wh L$
is a diffeomorphism, the corresponding Lagrangian
system meets naturally the unique equivalent Hamiltonian system such that
the associated momentum morphism
is the inverse diffeomorphism
\[
\wh H=\wh L^{-1}.
\]
 It follows that, when the Legendre morphism
 is regular at a point, the corresponding Lagrangian
system restrited to an open neighborhood of this point has the
equivalent local Hamiltonian system. In order to keep this local
equivalence in case of degenerate Lagrangian densities, we shall require of
associated Lagrangian and Hamiltonian systems that the image of the
configuration space $J^1Y$ by the Legendre morphism $\wh L$
contains all points where the momentum morphism $\wh H$ is regular.

\begin{definition}{16.1} Given a Lagrangian density $L$ on the jet
manifold $J^1Y$ of a fibred manifold $Y\to X$,
we shall say that a multimomentum
Hamiltonian form $H$ on the Legendre bundle $\Pi$ over $Y$ is associated
with $L$ if $H$ satisfies the relations
\bea
&&\wh L\circ\wh H\mid_Q=\Id_Q,\label{2.30a} \\
&& H=H_{\wh H}+L\circ\wh H,\label{2.30b}
\eea
 where
\begin{equation}
 Q=\wh L( J^1Y) \label{420}
\end{equation}
is the image of the configuration space $J^1Y$ by the Legendre morphism.
\end{definition}

In the terminology of constraint theory, we call $Q$ [Eq.(\ref{420})]
the Lagrangian constraint space (or simply the constraint space)
 by analogy with the first-class primary
constraint space in the familiar Hamiltonian formalism.
We shall observe that all Hamiltonian partners of solutions of the
Euler-Lagrange equations live on the Lagrangian constraint space.

In coordinates, the relations (\ref{2.30a}) and (\ref{2.30b}) are
expressed as
\[
 p^\m_i=\dr^\m_i\cL (x^\m,y^i,\dr^j_\la\cH)
\]
and
\[
\cH=p^\m_i\dr^i_\m\cH-\cL(x^\m,y^i,\dr^j_\la\cH).
\]
Acting on the both sides of the
second equality by the derivative $\dr^a_\al$,
one lifts the condition (\ref{2.30b}) to the condition
\begin{equation}
(p^\m_i-\dr^\m_i\cL\circ\wh H)\dr^i_\m\dr^a_\al\cH=0 \label{4.12}
\end{equation}
for the vertical
tangent morphisms to forms $H$, $H_{\wh H}$ and $L\circ\wh H$.
A glance at this condition  shows the following.
\begin{itemize}
 \item The condition (\ref{2.30a}) is the corollary of
Eq.(\ref{2.30b}) if the momentum morphism $\wh H$
is regular
\[
\det (\dr^i_\la\dr^j_\m\cH)\mid_Q \neq 0
\]
at all points of the constraint space $Q$.
\item The momentum morphism $\wh H$ is not regular outside
the Lagrangian constraint space $Q$ just as it has been
required of associated Lagrangian and Hamiltonian systems.
\end{itemize}

We shall refer to the following criterion in order to construct
multimomentum Hamiltonian forms associated with a given Lagrangian density.

\begin{proposition}{16.2}
Let $\Phi$ be a momentum morphism. Given a Lagrangian density $L$,
let us consider the multimomentum Hamiltonian form
\begin{equation}
 H_{L\Phi}=H_\Phi+L\circ\Phi\label{4.24}
\end{equation}
where $H_\Phi$ is the multimomentum Hamiltonian form (\ref{414})
associated with $\Phi$. If
the momentum morphism $\Phi$ obeys the relations
\ben
&&\wh L\circ\Phi\mid_Q=\Id_Q, \label{4.26}\\
&&\wh H_{L\Phi} = \Phi, \label{4.25}
\een
the multimomentum Hamiltonian form (\ref{4.24}), by very definition,
is associated with $L$.
 \end{proposition}

It is readily observed that Eq.(\ref{4.25}) is just Eq.(\ref{4.12})
and that the multimomentum Hamiltonian form (\ref{2.30b}) is of
the type (\ref{4.24}).

\begin{proposition}{16.3} If a multimomentum Hamiltonian form $H$ is
associated with a Lagrangian density $L$, the following relations hold:
 \ben
&& H\mid_Q=\Xi_L\circ\wh H\mid_Q, \label{4.9}\\
 && \dr_i\cH =-\dr_i\cL\circ\wh H \label{2.31}
\een
where $\Xi_L$ is the Poncar\'e-Cartan form (\ref{303}).
\end{proposition}

Let us emphasize that
different multimomentum Hamiltonian forms can be associated with the same
Lagrangian density.

\begin{example}{16.4}
All multimomentum Hamiltonian forms $H_\G$ [Eq.(\ref{3.6})] are
associated with the zero Lagrangian density
\[
L=0.
\]
In this case, the image of the Legendre morphism is
\[
Q=\wh 0(Y)
\]
where $\wh 0$ is the canonical zero
section of the Legendre bundle $\Pi\to Y$. The
condition (\ref{2.30a}) comes to the trivial identity $0=0$, and the
multimomentum Hamiltonian forms $H_\G$ obey Eq.(\ref{2.30b})
which takes the form
\[
\cH=p^\m_i\dr^i_\m\cH.
\]
\end{example}

In accordance with the relation (\ref{4.9}), the associated multimomentum
Hamiltonian forms restricted to the Lagrangian
constraint space $Q$ consist with
the pullbacks of the Poincar\'e-Cartan form, but generally they do
not coincide
with each other even on $Q$.

It should be noted that there exist Lagrangian densities which possess no
associated multimomentum Hamiltonian forms defined everywhere on the
Legendre bundle $\Pi$. The requirement to be of class
$C^\infty$ also is rather
rigid condition. But in field theories where
Lagrangian densities are almost
always quadratic and affine in velocities $y^i_\la$,
the associated multimomentum Hamiltonian forms always exist.

We aim to match solutions of the
Euler-Lagrange equations for a Lagrangian desnity $L$ and solutions
of the Hamilton equations for the associated
multimomentum Hamiltonian forms $H$ in accordance with the diagram
\[
\begin{array}{rcccccl}
 & {J^1\Pi} &  \op\longrightarrow^{J^1\wh H} & {J^1J^1Y} &
\op\longrightarrow^{J^1\wh L} & {J^1\Pi} & \\
 {}&\put(0,-10){\vector(0,1){20}} & & \put(0,-10){\vector(0,1){20}}
& &\put(0,-10){\vector(0,1){20}}  &{} \\
 & {\Pi} & \op\longrightarrow^{\wh H} & {J^1Y} &
\op\longrightarrow^{\wh L}& {\Pi}&
\end{array}
\]
where $J^1\wh L$ and $J^1\wh H$ are the
first order jet prolongations  of the Legendre morphism $\wh L$
[Eq.(\ref{421})] and the momentum morphism $\wh H$ [Eq.(\ref{415})]
respectively. They have the coordinate expressions
 \begin{equation}
(y^i_\m,y^i_{(\m)},y^i_{\m\la})\circ
J^1\wh H=(\dr^i_\m\cH,y^i_{(\m)},\wh\dr_\la\dr^i_\m\cH),\label{2.34}
\end{equation}
\[
\wh\dr_\la=\dr_\la+y^j_{(\la)}\dr_j+p^\nu_{j\la}\dr^j_\nu,
\]
and
\begin{equation}
(p^\la_i,y^i_{(\m)},p^\la_{i\m})\circ J^1\wh L=
(\pi^\la_i,y^i_{(\m)},\wh\dr_\m\pi^\la_i),\label{2.35}
\end{equation}
\[
\wh\dr_\la=\dr_\la+y^j_{(\la)}\dr_j+y^j_{\m\la}\dr_j^\m.
\]

It is essential that, if $\g$ is a Hamiltonian jet field for a
multimomentum Hamiltonian form $H$,
the composition $J^1\wh H\circ\g$ takes its
values into the sesquiholonomic subbundle $\wh J^2Y$ of the repeated jet
manifold $J^1J^1Y$:
\[
(y^i_\m, y^i_{(\m)})\circ J^1\wh H\circ\g=(\dr^i_\m\cH,\dr^i_\m\cH).
\]

Hereafter, we shall restrict our consideration to
Lagrangian densities of the following type:
\begin{itemize}\begin{enumerate}
\item a hyperregular Lagrangian density $L$ when the Legendre morphism
$\wh L$ is a diffeomorphism and a regular Lagrangian density $L$
when $\wh L$ is a local diffeomorphism;
\item a semiregular Lagrangian density
$L$ when the preimage $\wh L^{-1}(q)$ of any point $q$ of the constraint
space $\wh L(J^1Y)$ is a connected submanifold of $J^1Y$;
\item an almost regular Lagrangian density $L$ when the Legendre morphism
$\wh L$ is a  submersion and the Lagrangian constraint space
$Q$ is an imbedded submanifold of the Legendre manifold.
\end{enumerate}
\end{itemize}

The notions (ii) and (iii) of degeneracy are most appropriate since they
provide the workable relations between solutions of the Euler-Lagrange
equations and the Hamilton equations.

In case of hyperregular Lagrangian densities, the Lagrangian
formalism on fibred manifolds and the
multimomentum Hamiltonian formalism are equivalent.
If a Lagrangian density $L$ is hyperregular, there always exists
the unique associated multimomentum Hamiltonian form
\[
H=H_{\wh L^{-1}}+L\circ\wh L^{-1}.
\]
The corresponding momentum morphism (\ref{415}) is the diffeomorphism
\[
\wh H=\wh L^{-1}
\]
and, so is its first order jet prolongation $J^1\wh H$:
\begin{equation}
J^1\wh L\circ J^1\wh H =\Id_{J^1\Pi}. \label{4.15}
\end{equation}

\begin{proposition}{16.5} Let $L$ be a hyperregular Lagrangian density and
$H$ the associated multimomentum Hamiltonian form. The following relations
hold
\ben
&&\Xi_L=H\circ \wh L. \label{4.16}\\
&&\La_L=\cE_H\circ J^1\wh L, \label{4.18} \\
&& \cE_H=\La_L\circ J^1\wh H, \label{4.19}
\een
where $\cE$ is
the Hamilton operator  $\cE_H$ [Eq.(\ref{3.9})] for $H$ and
$\La_L$ is the exterior form (\ref{304}).
\end{proposition}

A glance at Eqs.(\ref{4.9}) and (\ref{4.16}) shows that just the
Poincar\'e-Cartan form is the Lagrangian counterpart of multimomentum
Hamiltonian forms (\ref{4.7}), whereas
the Lagrangian partner of the Hamilton operator is the exterior form
$\La_L$ [Eq.(\ref{304})]. This form fails
to define a differential operator in
the conventional sense, but its restriction to $\wh J^2J$
consists with the sesquiholonomic Euler-Lagrange operator $\cE'_L$
for the Lagrangian density $L$. It follows that, if $\g$ is a Hamiltonian
jet field for the associated multimomentum Hamiltonian form $H$, then the
composition $J^1\wh H\circ\g$ takes its values into the kernal of
$\cE'_L$. It follows that $J^1\wh H\circ\g\circ\wh L$
is a Lagrangian jet field for $L$.

\begin{proposition}{16.6} Let $L$ be a hyperregular Lagrangian density and
$H$ the associated multimomentum Hamiltonian form.

(i) Let $r: X\to\Pi$ be a solution of the Hamilton equations
(\ref{3.11a}) and (\ref{3.11b}) for the multimomentum Hamiltonian
form $H$. Then, the section
\begin{equation}
\ol s=\wh H\circ r \label{4.13}
\end{equation}
of the fibred jet manifold $J^1Y\to X$ satisfies the
first order Euler-Lagrange equations (\ref{306a}) and (\ref{306b}).

(ii) Conversely, if a section $\ol s$ of
the fibred jet manifold $J^1Y\to X$ is a solution of the first order
Euler-Lagrange equations (\ref{306a}) and (\ref{306b}),
the section
\[
r=\wh L\circ\ol s
\]
of the fibred Legendre manifold $\Pi\to X$ satisfies the
Hamilton equations (\ref{3.11a}) and (\ref{3.11b}).

(iii) In virtue of Eq.(\ref{N10}), we have
 \ben
&&\ol s=J^1s, \nonumber \\
&& s= \pi_{\Pi Y}\circ r \label{4.14}
\een
where $s$ is a solution of the second order Euler-Lagrange equations
(\ref{2.29}).
\end{proposition}

Propositions 16.6 shows that, given a hyperregular Lagrangian density and
the associated multimomentum Hamiltonian form,
the corresponding first order Euler-Lagrange equations (\ref{306a}) and
(\ref{306b}) are equivalent to the Hamilton
equations (\ref{3.11a}) and (\ref{3.11b}).

In case of a regular Lagrangian density $L$, the Lagrangian
constraint space $Q$
is an open subbundle of the Legendre bundle $\Pi\to Y$. If $Q\neq\Pi$,
the unique associated multimomentum Hamiltonian form fails to be
defined everywhere on $\Pi$ in general.
At the same time, an open constraint
subbundle $Q$ can be provided
with the induced multisymplectic structure, so that we may restrict our
consideration to multimomentum Hamiltonian forms on $Q$.
If a regular Lagrangian density is still
semiregular, the associated Legendre
morphism is a diffeomorphism of $J^1Y$ onto $Q$ and, on
$Q$, we	can recover
all results true for  hyperregular Lagrangian densities.
Otherwise, there exists an open neighborhood
of each point of the configuration space $J^1Y$ such that the
corresponding local Lagrangian system defined by $L$ is hyperregular.

\begin{example}{16.7}
Let $Y$ be the bundle $\Bbb R^3\to\Bbb R^2$ coordinatized
by $(x^1,x^2,y)$. The jet manifold $J^1Y$ and the Legendre bundle
$\Pi$ over $Y$ are coordinatized by
\[
(x^1,x^2,y,y_1,y_2)
\]
and
\[
(x^1,x^2,y,p^1,p^2)
\]
respectively. Set the Lagrangian density
\be
&&L=\cL\om,\\
&&\cL=\exp y_1 +\frac12(y_2)^2.
\ee
It is regular and semiregular, but not hyperregular. The corresponding
Legendre morphism reads
\be
&&p^1\circ\wh L=\exp y_1, \\
&& p^2\circ\wh L= y_2.
\ee
The image $Q$ of the configuration space under this morphism
is given by the coordinate relation $p^1>0$. It is an open
subbundle of the Legendre bundle.
On $Q$, we have the associated multimomentum Hamiltonian form
\be
&&H=p^\la dy\w\om_\la -\cH\om,\\
&&\cH=p^1(\ln p^1-1)+\frac12(p^2)^2,
\ee
which however is not extended to $\Pi$.
\end{example}

Contemporary field theories are almost never regular, but their Lagrangian
densities are nonetheless well-behaved. They are both semiregular and
almost regular as a rule.

\begin{proposition}{16.8} All multimomentum Hamiltonian forms associated
with a semiregular Lagrangian density $L$ coincide on the
Lagrangian constraint space $Q$:
\[
H\mid_Q=H'\mid_Q.
\]
Moreover, the Poincar\'e-Cartan form $\Xi_L$ [Eq.(\ref{303})]
for $L$ is the pullback
\[
\Xi_L=H\circ\wh L,
\]
\begin{equation}
\pi^\la_iy^i_\la-\cL=\cH(x^\m,y^i,\pi^\la_i), \label{2.32}
\end{equation}
of any associated multimomentum Hamiltonian form $H$ by the Legendre
morphism $\wh L$.
\end{proposition}

Coinciding on the Lagrangian constraint space $Q$,
the multimomentum Hamiltonian
forms associated with a semiregular Lagrangian density differ from each
other outside $Q$. The condition (\ref{2.30b}) restricts
rigidly the arbitrariness of these forms at points of $\Pi\setminus Q$.
Substituting this condition into Eq.(\ref{2.32}), we obtain
\[
(\pi^\la_i\circ\wh H-p^\la_i)\dr^i_\la\cH=\cH(x^\m, y^i,\pi^\la_i\circ\wh
H) -\cH(x^\m, y^i,p^\la_i)
\]
at each point of $\Pi\backslash Q$.
It follows that, at a point which does not belong the
constraint space, a multimomentum Hamiltonian form associated with a
semiregular Lagrangian density is written
\be
&&H=p^\la_idy^i\w\om_\la-\cH\om,\\
&&\cH=\ol p^\la_i\G_H{}^i_\la +\underline p^\mu_j\G_H{}^j_\mu
+\wt\cH (y^i,\ol p^\la_i),
\ee
where $\underline p^\mu_j$ are coordinates which obey the relation
\[
\underline p^\mu_j\neq \pi^\mu_j\circ\wh H
\]
and $\ol p^\la_i$ are the remaining ones. It is affine in momenta
$\underline p^\mu_j$. The
Hamilton equations (\ref{3.11a}) corresponding to the coordinates
$\underline p^\mu_j$ are reduced to the gauge-type conditions
\[
\dr_\mu r^j=\G_H{}^j_\mu
\]
independent of momenta.

\begin{example}{16.9}
Let $Y$ be the bundle $\Bbb R^3\to\Bbb R^2$ coordinatized
by $(x^1,x^2,y)$. The jet manifold $J^1Y$ and the Legendre bundle
$\Pi$ over $Y$ are coordinatized by
\[
(x^1,x^2,y,y_1,y_2)
\]
and
\[
(x^1,x^2,y,p^1,p^2)
\]
respectively. Set the Lagrangian density
\begin{equation}
L=\frac12(y_1)^2\om. \label{200}
\end{equation}
It is semiregular. The associated Legendre morphism reads
\be
&&p^1\circ\wh L=y_1, \\
&& p^2\circ\wh L= 0.
\ee
The corresponding constraint space $Q$ consists of points with
the coordinate $p^2=0$. Multimomentum Hamiltonian forms associated with
the Lagrangian density (\ref{200}) are given by the expression
\[
H=p^\la dy\w\om_\la - [\frac12 (p^1)^2 + c(x^1,x^2,y)p^2]\om
\]
where $c$ is arbitrary function of coordinates $x^1$, $x^2$ and $y$.
They are affine in the momentum coordinate $p^2$.
\end{example}

The relation (\ref{2.32}) implies the identity
\begin{equation}
(y^i_\la-\dr^i_\la\cH\circ\wh L)d\pi^\la_i\w\om-(\dr_i\cL+ \dr_i\cH\circ\wh
L)dy^i\w\om=0. \label{2.33}
\end{equation}
In particular, we have the equality
\begin{equation}
[y^i_\la-\dr^i_\la\cH (x^\m,y^j,\p^\m_j)]\dr^\la_i\dr^\alpha_k\cL=0
\label{4.11}
 \end{equation}
similar to Eq.(\ref{4.12}).

Building on Eq.(\ref{2.33}), one can reproduce Eq.(\ref{4.18})
\[
\La_L=\cE_H\circ J^1\wh L,
\]
but not Eq.(\ref{4.19}). It enables one to extend the
part (i) of Proposition 16.6 to multimomentum Hamiltonian forms
associated with semiregular Lagrangian densities.

\begin{proposition}{16.10}
Let a section $r$ of the Legendre manifold $\Pi\to X$
be a solution of the Hamilton equations (\ref{3.11a}) and (\ref{3.11b})
for some multimomentum Hamiltonian form $H$ associated with a semiregular
Lagrangian density $L$. If $r$ lives on the constraint space $Q$
[Eq.(\ref{420})], the section
\[
\ol s=\wh H\circ r
\]
of the fibred jet manifold $J^1Y\to X$ satisfies the first
order Euler-Lagrange equations (\ref{306a}) and (\ref{306b}).
\end{proposition}

The assertion (ii) from Proposition 16.6 however must be modified as
follows.

\begin{proposition}{16.11} Given a semiregular Lagrangian density $L$, let
a section $\ol s$ of the fibred jet manifold
$J^1Y\to X$ be a solution of the
first order Euler-Lagrange equations (\ref{306a}) and (\ref{306b}).
Let $H$ be a multimomentum Hamiltonian form associated with $L$ so that
its momentum morphism satisfies the condition
\begin{equation}
\wh H\circ \wh L\circ \ol s=\ol s.\label{2.36}
\end{equation}
 Then, the section
\[
r=\wh L\circ \ol s
\]
 of the fibred Legendre manifold $\Pi\to X$ is a solution of the
Hamilton equations (\ref{3.11a}) and (\ref{3.11b}) for $H$. It lives on the
constraint space $Q$.
\end{proposition}

\begin{lemma}{16.12} For every pair of sections $\ol s$ of $J^1Y\to X$
and $r$ of $\Pi\to X$ which satisfy either
Proposition 16.10 or Proposition 16.11, Eqs.(\ref{4.14})
where $s$ is a solution of the second order Euler-Lagrange equations
(\ref{2.29}) remain true.
\end{lemma}

Proposition 16.10 and Proposition 16.11 show that, if $H$ is a multimomentum
Hamiltonian form associated with a semiregular Lagrangian density $L$,
every solution of the corresponding Hamilton equations which  lives on the
constraint space $\wh L(J^1Y)$
yields a solution of the Euler-Lagrange equations for $L$.
At the same time, to exaust all solutions of the Euler-Lagrange
equations, one must consider a family of different
multimomentum Hamiltonian forms associated with $L$.

\begin{example}{16.13}
To illustrate these circumstances, let us consider the zero
Lagrangian density
\[
L=0.
\]
This Lagrangian density is semiregular.
Its Euler-Lagrange equations come to the identity $0=0$.
Every section $s$ of the fibred manifold $Y\to X$ is a solution
of these equations. Given a section $s$, let $\G$ be a connection on
$Y$ such that $s$ is its integrale section.
The multimomentum Hamiltonian form $H_\G$ [Eq.(\ref{3.6})] is associated
with $L$. The corresponding momentum morphism
satisfies Eq.(\ref{2.36}).
The Hamilton equations have the solution
\[
r=\wh L\circ s
\]
 given by the coordinate expression
\ben
&&r^i=s^i, \nonumber\\
&& r^\la_i=0. \label{4.28}
\een
\end{example}

 We shall say that a family of multimomentum Hamiltonian forms $H$
associated with a Lagrangian density $L$ is
complete if, for any solution $\ol s$ of the first order Euler-Lagrange
equations  (\ref{306a}) and (\ref{306b}), there exists a solution
$r$ of the Hamilton equations (\ref{3.11a}) and (\ref{3.11b}) for
some multimomentum Hamiltonian form $H$ from this family so that
\ben
&& r=\wh L\circ\ol s,\nonumber\\
&& \ol s =\wh H\circ r, \nonumber \\
&& \ol s= J^1(\pi_{\Pi Y}\circ r). \label{2.37}
\een
Let $L$ be a semiregular Lagrangian density. Then,
in virtue of Proposition 3.24, such a complete family of associated
multumomentum Hamiltonian forms
exists iff, for every solution $\ol s$ of the Euler-Lagrange
equations for $L$, there is a multimomentum Hamiltonian form $H$ from this
family such that Eq.(\ref{2.36}) holds.

The complete family of multimomentum Hamiltonian forms associated with a
given Lagrangian density fails to be defined uniquelly.
For instance, the multimomentum Hamiltonian forms (\ref{3.6})
constitute the complete family associated with the zero
Lagrangian density, but this family is not minimal. At the same
time, we consider only solutions of the Hamilton equations
which live on the Lagrangian
constraint space (\ref{420}). Therefore, we can restrict ourselves to
local multimomentum Hamiltonian forms and the
Hamilton equations which are defined on
some open neighborhood $\wt Q$ of the Lagrangian
constraint space $Q$. Being an
open imbedded subbundle of the Legendre bundle,  $\wt Q$ inherits the
multisymplectic structure of $\Pi$ including multisymplectic canonical
transformations which keep the constraint space $Q$.

We shall investigate existence of complete families of associated
multimomentum
Hamiltonian forms when a Lagrangian density is almost regular.
In this case, the Lagrangian constraint space $Q$, by very definition,
is an imbedded subbundle of the Legendre bundle
and the Legendre morphism
\begin{equation}
J^1Y\op\to_Y Q \label{4.29}
\end{equation}
is a fibred manifold.

\begin{proposition}{3.14} Let $L$ be an almost regular Lagrangian density.

(i) On an open neighborhood of each point $q\in Q$, there exist
local multimomentum Hamiltonian forms associated with $L$.

(ii) Let $L$ be still a semiregular Lagrangian density.
On an open neighborhood of each point $q\in
Q$, there exists a complete family of local associated multimomentum
Hamiltonian forms.
\end{proposition}

In field theories with quadratic and affine Lagrangian densities,
the complete families of the associated globally defined multimomentum
Hamiltonian forms always exist. Moreover, the following example shows
that the complete family of associated multimomentum Hamiltonian forms
may exist if even a Lagrangian density is neither semiregular nor
almost regular.

\begin{example}{15}
Let $Y$ be the bundle $\Bbb R^3\to\Bbb R^2$ coordinatized
by $(x^1,x^2,y)$. The jet manifold $J^1Y$ and the Legendre bundle
$\Pi$ over $Y$ are coordinatized by
\[
(x^1,x^2,y,y_1,y_2)
\]
and
\[
(x^1,x^2,y,p^1,p^2)
\]
respectively. Let us consider the Lagrangian density
\[
L=[\frac13(y_1)^3+\frac12(y_2)^2]\om.
\]
The associated Legendre morphism reads
\ben
&&p^1\circ\wh L=(y_1)^2, \label{204}\\
&& p^2\circ\wh L= y_2.\nonumber
\een
The corresponding constraint space $Q$ is given by the coordinate
relation $p^1\geq 0$. It is not even a submanifold of $\Pi$.
There exist two associated multimomentum Hamiltonian forms
\be
&&H_+=p^\la dy\w\om_\la -[\frac23 (p^1)^{\frac32} + \frac12(p^2)^2]\om,\\
&&H_-=p^\la dy\w\om_\la -[-\frac23 (p^1)^{\frac32} + \frac12(p^2)^2]\om
\ee
on $Q$ which correspond to different solutions
\[
y_1=\sqrt{p^1}
\]
and
\[
y_1=-\sqrt{p^1}
\]
of Eq.(\ref{204}).
One can verify that the multimomentum Hamiltonian forms $H_+$ and $H_-$
constitute the complete family.
\end{example}

 Note that, if the imbedded constraint space $Q$ is not an open
subbundle of the Legendre bundle $\Pi$, it can not be provided with the
multisymplectic structure.
Roughly speaking, not all canonical momenta $p^\la_i$ live
on such a constraint space in general.
Therefore, to consider solutions of the Hamilton
equations even on $Q$, one must set up a multimomentum Hamiltonian
form and  the corresponding  Hamilton equations at least on some open
neighborhood of the constraint space $Q$.

Another way is to construct the De Donder-Hamilton type
equations on the imbedded constraint space $Q$ itself.
Let an almost regular Lagrangian density $L$ be also semiregular. Let
$H_Q$ be the restriction of the associated multimomentum Hamiltonian forms
to the constraint space $Q$. In virtue of Proposition 16.8, it is
uniquely defined. For sections $r$
of the fibred manifold $Q\to X$, we can write the equations
\begin{equation}
r^*(u\rfloor dH_Q) =0 \label{N44}
\end{equation}
where $u$ is an arbitrary vertical vector field on $Q\to X$. Since
\[
dH_Q\neq dH\mid_Q,
\]
these equations fail to be equivalent to the Hamilton
equations restricted to the constraint space $Q$. At the same time, we have
\[
\wh\Xi_L(J^1Y) = H_Q(Q)
\]
 and so, Eqs.(\ref{N44}) are equivalent to the
De Donder-Hamilton equations (\ref{N46}).

Thus, the feature of the multimomentum Hamiltonian approach to
constraint field systems is clarified. In the framework of this
approach, field equations are set up at least on an open neighborhood
of the Lagrangian
constraint space though only their solutions living on the
constraint space are considered. Whereas in the framework of the
De Donder Hamiltonian formalism and, in fact, the Lagrangian formalism,
field equations themselves are defined only on the constraint space.

\section{}

This Section covers the Cauchy problem for the Euler-Lagrange
equations and the Hamilton equations. The key point consists in the fact
that the Hamilton equations must be modified in order to formulate the
Cauchy task. We observe that, although the first order Lagrangian
equations are equivalent to the second order ones, the Cauchy tasks for
them fail to be equivalent. At the same time, if a Lagrangian density
is regular, the Cauchy problem for the second order Euler-Lagrange
equations is equivalent to the Cauchy problem for the Hamilton equations.

The system of the first order Euler-Lagrange equations
(\ref{306a}) and (\ref{306b}) has the standard form
\begin{equation}
S^\la_{ab}(x, \phi)\dr_\la\phi^b = f_a(x,\phi) \label{307}
\end{equation}
for the Cauchy problem or, to be more precise, for the general
Cauchy problem since the coefficients $S^\la_{ab}$ depend on the
variable functions $\phi$ in general. Here, $\phi^b$ are the collective
notation of field functions $\ol s^i$ and $\ol s^i_\la$. However,
the characteristic form
\begin{equation}
\det(S^\la_{ab}c_\la), \qquad c_\la\in\Bbb R, \label{308}
\end{equation}
of the first order Euler-Lagrange equations
is singular, that is, it is not different
>from 0 because of Eqs.(\ref{306a}).

If the index $i$ takes the single value, the second order Euler-Lagrange
equations (\ref{2.29}) also have the
standard form for the Cauchy problem. In
general case, one can  reduce the Cauchy problem for these equations
to the standard form as follows.

Let us examine the extension of the Euler-Lagrange
operator (\ref{305}) to the repeated jet manifold $J^1J^1Y$ which
is given by the horizontal exterior form
\begin{equation}
\cE''_L= (\dr_i-\wh\dr_\la\dr^\la_i)\cL dy^i\w\om ,\label{309}
\end{equation}
\[
\wh\dr_\la =\dr_\la +y^i_{(\la)}\dr_i+y^i_{\m\la}\dr^\m_i,
\]
on $J^1J^1Y\to Y$.
Note that the exterior form (\ref{309}) differs from the exterior form
(\ref{304}) as follows:
\[
\La_L=\cE''_L +(y^i_{(\la)}-y^i_\la)d\pi^\la_i\w\om.
\]

We single out a local coordinate $x^1$ and consider local sections
$\ol s$ of the fibred jet manifold $J^1Y\to X$ which satisfy the
conditions
\ben
&&\dr_1\ol s^i =s^i_1, \nonumber \\
&& \dr_1\ol s^i_\la =\dr_\la\ol s^i_1, \qquad \la\neq 1. \label{310}
\een
It should be noted that the first order jet prolongations
$J^1\ol s$ of such
sections $\ol s$ do not take their values into the sesquiholonomic jet
manifold $\wh J^2Y$ in general. Let us consider the system of the first
order differential equations (\ref{310}) and the first order
differential equations
\[
\cE''_L\circ J^1\ol s =0,
\]
\[
(y^i_\la, y^i_{(\la)}, y^i_{\m\la})\circ J^1\ol s=(\ol s^i_\la,\dr_\la\ol
s^i,  \dr_\la\ol s^i_\m),
\]
for sections $\ol s$ of the fibred jet manifold $J^1Y\to X$.
The latter take the form
\begin{equation}
 \dr_i\cL-(\dr_\la+\dr_\la\ol s^j\dr_j
+\dr_\la\overline s^j_\m\dr^\m_j)\dr^\la_i\cL=0. \label{311}
\end{equation}
This system
has the standard fashion (\ref{307}) for the Cauchy
problem on the initial conditions
\ben
&&\ol s^i(x)=\phi^i(x),\nonumber\\
&&\ol s^i_\la(x)=\dr_\la\phi^i(x), \qquad \la\neq 1,\nonumber\\
&&\ol s^i_1(x)=\phi_1^i(x) \label{314}
\een
on some local hypersurface of $X$ transversal to coordinate
lines $x^1$. Its characteristic form (\ref{308}) is defined only
by the Hessian (\ref{313}). In virtue of the
well-known theorem,$^{33}$ if local functions ($\ol s^i(x)$, $\ol s^i_\la(x)$)
are solutions of the Cauchy problem for Eqs.(\ref{310}) and (\ref{311})
[in analytic functions or
functions of class $C^2$], they satisfy still Eqs.(\ref{306a}).
It follows that they take their values into
the second order jet manifold $J^2Y$ and  $\ol s^i(x)$ are
solutions of the second order Euler-Lagrange equations (\ref{2.29}).

We thus observe that, on solutions of the Cauchy problem
for the second order Euler-Lagrange equations (\ref{2.29}),
Eqs.(\ref{306a}) with $\la\neq 1$ are the
corollaries of Eqs.(\ref{310}) and the initial conditions
(\ref{314}). In other words, the system of first order Euler-Lagrange
equations is overdefined on solutions of the Cauchy problem.

\begin{lemma}{17.1} The characteristic form (\ref{308}) for the system
of Eqs.(\ref{310}) and (\ref{311}) is singular iff there
exists a collection of real numbers $c^j$ such that, whenever
$b_\m\in\Bbb R$,
\begin{equation}
\dr^\m_j\dr^\la_i\cL c^jb_\m =0. \label{315}
\end{equation}
\end{lemma}

In particular, if the determinant of the Hessian (\ref{313}) of
a Lagrangian density $L$ is different from zero,
the characteristic form (\ref{308}) for the system of Eqs.(\ref{310})
and (\ref{311}) is not singular. It means that the Cauchy problem for
these equations may be formulated.

Let us turn now to the Cauchy problem for the Hamilton equations.
The Hamilton equations (\ref{3.11a}) and (\ref{3.11b}) have the
standard form (\ref{307}) for the
Cauchy problem, but one faces the same difficulties as for
the Caushy problem of the first
order Euler-Lagrange equations (\ref{306a})
and (\ref{306b}). Their characteristic
form (\ref{308}) fails to be different
>from 0. These difficulties are overcomed in the same way as in case of
the second order Euler-Lagrange equations.

Let us single out a local coordinate $x^1$ and replace
Eqs.(\ref{3.11a}) by the equations
\ben
&&\dr_1 r^i =\dr^i_1\cH, \nonumber\\
&& \wh\dr_1\dr_\la^i\cH
=\wh\dr_\la\dr_1^i\cH, \qquad \la\neq 1, \label{4.3}\\
 && \wh\dr_\m= \dr_\m +\dr_\m r^i\dr_i +\dr_\m
r^\la_i\dr^i_\la.\nonumber
\een
The system of Eqs.(\ref{4.3}) and (\ref{3.11b}) have the
standard form for the Cauchy problem on the initial conditions
\ben
&& r^i(x')=\phi^i(x'),\nonumber\\
&& r_i^\m(x')=\phi_i^\m (x'), \nonumber\\
&& \dr_\la r^i =\dr^i_\la\cH, \qquad \la\neq 1,\label{4.4}
\een
on some local hypersurface $S$ of $X$  transversal to coordinate lines
$x^1$.

\begin{proposition}{17.2}
If $r^i$ are solutions of the Cauchy problem for
Eqs.(\ref{4.3}) and (\ref{3.11b}) [in functions of
class $C^2$] on the initial conditions (\ref{4.4}), they satisfy all
Eqs.(\ref{3.11a}).
\end{proposition}

It follows that, in order to formulate the Cauchy problem for the Hamilton
equations in the multimomentum
Hamiltonian formalism, one should single a one
of coordinates and  consider the system of equations (\ref{4.3}) and
(\ref{3.11b}).
It is readily observed that, if a Lagrangian density is regular,
the system of equations (\ref{314}) and (\ref{306b}) is equivalent
to the system of equations (\ref{4.3}) and (\ref{3.11b}) for
the Cauchy task of the second order Euler-Lagrange equations.

\newpage


\centerline{\large \bf Lecture 5. FIELD THEORY}
\bigskip\bigskip

With the multimomentum Hamiltonoan formalism, the tools are now at hand
to canonically analize constraint field systems on the covariant
finite-dimensional level. Maintaining covariance has the principal
advantages of describing field theories, for any space-time splitting
shades the covariant picture of field constraints.

The Lagrangian densities of field models are almost always semiregular
and almost regular that enables one to utilize the relations stated above
between Lagrangian and Hamiltonian formalisms. Moreover, the Lagrangian
densities of all fundamental field theories are quadratic and affine in
the derivatives of field functions.
Gauge theory exemplifies the degenerate
quadratic Lagrangian density, whereas gravity and fermion fields are
described by the affine ones. In first
Section of this Lecture, we shall spell out
Lagrangian and Hamiltonian systems in case of affine and almost regular
quadratic Lagrangian densities. The goal is the general procedure of
describing constraint field systems.

The most of contemporary field
models are concerned with gauge theory. We observe that
several attributes of gauge theory such as gauge freedom and gauge
conditions are the common attributes of systems with degenerate
quadratic and affine Lagrangian densities, without appealing to any
symmetry group.

\section{Constraint field systems}

This Section presents the complete
families of multimomentum Hamiltonian forms associated with affine and
almost regular quadratic Lagrangian densities. The key ingredient in
our consideration is splitting of the configuration space in the dynamic
sector and the gauge sector. The latter consists with the kernal of the
Legendre morphism. As an immediate
consequence of this splitting, a part of the
Hamilton equations comes to the gauge-type conditions independent of
canonical momenta. Different associated multimomentum Hamiltonian forms are
responsible for different such conditions.

Given a fibred manifold $Y\to X$,
let us consider a quadratic Lagrangian density
\ben
&&L=\cL\om, \nonumber\\
&&\cL=\frac12 a^{\la\m}_{ij}(y)y^i_\la y^j_\m +
b^\la_i(y)y^i_\la + c(y), \label{N12}
\een
where $a$, $b$ and $c$ are local functions on $Y$ with the corresponding
transformation laws. The associated Legendre morphism reads
\begin{equation}
p^\la_i\circ\wh L= a^{\la\m}_{ij} y^j_\m +b^\la_i. \label{N13}
\end{equation}

\begin{lemma}{18.1} The Lagrangian density (\ref{N12}) is semiregular.
\end{lemma}

The Legendre morphism (\ref{N13})
is an affine morphism over $Y$. It implies the corresponding linear
morphism
\be
&&\ol L: T^*X\op\otimes_YVY\to\Pi, \\
&& p^\la_i\circ\ol L=a^{\la\m}_{ij}\ol y^j_\m,
\ee
over $Y$ where $\ol y^j_\mu$ are bundle coordinates of the vector
bundle (\ref{23}).

Almost all quadratic Lagrangian
densities of field models take the particular form
\ben
&&\cL=\frac12 a^{\la\m}_{ij}\ol y^i_\la\ol y^j_\m + c, \label{N15}\\
&& \ol y^i_\m = y^i_\m -\G^i_\m, \nonumber
\een
where $\G$ is a certain connection on $Y\to X$ and
$y^i_\m \to \ol y^i_\m$ is the
corresponding covariant differential (\ref{38}).
This fashion is equivalent to condition that the Lgrangian
constraint space $Q$
defined by the Legendre morphism (\ref{N13})
contains the image of $Y$ under the canonical
zero section $\wh 0(Y)$ of the Legendre bundle $\Pi\to Y$.
Let us set
\[
\Ker\wh L= \wh L^{-1}(\wh 0(Y)).
\]
It is an affine subbundle of the jet bundle $J^1Y\to Y$
which is modelled on the vector bundle
\[
\Ker\ol L= \ol L^{-1}(\wh 0(Y)).
\]
Then, there exists a connection $\G$ on $Y\to X$
which takes its values into $\Ker\wh L$:
\ben
&&\G: Y\to \Ker\wh L, \label{N16}\\
&& a^{\la\m}_{ij}\G^j_\m + b^\la_i =0. \label{250}
\een
With this connection, the Lagrangian density
(\ref{N12}) can be brought into
the form (\ref{N15}). For instance, if the Lagrangian
density (\ref{N12}) is regular, the
connection (\ref{N16}) is the unique solution of the algebraic equations
(\ref{250}).

\begin{proposition}{18.2}
Let $L$ be an almost regular quadratic Lagrangian
density such that $\wh 0(Y)\subset Q$. Then, there exists a linear
pullback-valued horizontal 1-form
\ben
&&\si: \Pi\to T^*X\op\otimes_YVY, \label{N17}\\
&& \ol y^i_\la\circ\si =\si^{ij}_{\la\m}p^\m_j, \nonumber
\een
on the Legendre bundle $\Pi\to Y$ such that
\ben
&&\ol L\circ\si\circ i_Q= i_Q, \nonumber\\
&& a^{\la\mu}_{ij}\si^{jk}_{\mu\al}=\delta^\la_\al\delta^k_i, \label{N45}
\een
where $i_Q$ denotes the natural imbedding of $Q$ into $\Pi$.
\end{proposition}

If the Lagrangian density (\ref{N12}) is regular, the form (\ref{N17})
is determined uniquely by the algebraic equations (\ref{N45}).

The connection (\ref{N16}) and the form (\ref{N17}) play the prominent
role in our construction.

Since $\ol L$ and $\si$ are linear morphisms, their composition
$\ol L\circ\si$ is a surjective submersion of $\Pi$ onto $Q$.
It follows that
\begin{equation}
\si=\si\circ\ol L\circ\si, \label{N21}
\end{equation}
and the jet bundle $J^1Y\to Y$ has the splitting
\begin{equation}
J^1Y=\Ker\wh L\op\oplus_Y{\rm Im}\si, \label{N18}
\end{equation}
\[
y^i_\la= [y^i_\la
-\si^{ik}_{\la\al} (a^{\al\m}_{kj}y^j_\m + b^\al_k)]+
[\si^{ik}_{\la\al} (a^{\al\m}_{kj}y^j_\m + b^\al_k)].
\]
 Moreover, there exists the form $\si$ [Eq.(\ref{N17})] such that also the
Legendre bundle meets the splitting
\begin{equation}
\Pi=\Ker\si \op\oplus_Y Q. \label{N20}
\end{equation}

Given the form $\si$ [Eq.(\ref{N17})] and the connection $\G$
[Eq.(\ref{N16})], let us consider the affine momentum morphism
\ben
&&\Phi=\wh\G+\si,\nonumber\\
&& \Phi^i_\la = \G^i_\la (y) + \si^{ij}_{\la\m}p^\m_j, \label{N19}
\een
where $\wh\G$ is the pullback (\ref{104}) of $\G$ onto $\Pi$.
It is easy to see that this momentum morphism satisfies Eq.(\ref{4.26})
where $\wh L$ is the Legendre morphism (\ref{N13}). Conversely, every
affine momentum morphism satisfying the condition
(\ref{4.26}) is of the type
(\ref{N19}). Building on the condition (\ref{N21}), one can check that
the multimomentum Hamiltonian form $H_{L\Phi}$ [Eq.(\ref{4.24})]
corresponding to $\Phi$ [Eq.(\ref{N19})] is associated with the
Lagrangian density (\ref{N12}). It is given by the expression
\begin{equation}
H= p^\la_idy^i\w\om_\la - [\G^i_\la
(p^\la_i-\frac12 b^\la_i) +\frac12 \si^{ij}_{\la\m}p^\la_ip^\m_j-c]\om.
\label{N22}
\end{equation}
This expression consists with the canonical splitting (\ref{3.8}) of
$H$ where
\[
\G_H=\G.
\]

 Note that, on $\Ker\si$, the multimomentum
Hamiltonian form (\ref{N22}) becomes affine in canonical momenta.

We claim that the multimomentum Hamiltonian forms (\ref{N22})
where $\G$ are connections (\ref{N16}) constitute the complete family.

Given the multimomentum Hamiltonian form (\ref{N22}),
let us consider the Hamilton equations (\ref{3.11a}) for
sections $r$ of the fibred Legendre manifold $\Pi\to X$. They read
\begin{equation}
J^1s= (\wh\G+\si)\circ r, \qquad s=\pi_{\Pi Y}\circ r, \label{N29}
\end{equation}
or
\[
\nabla_\la r^i=\si^{ij}_{\la\mu}r^\mu_j
\]
where
\[
\nabla_\la r^i=\dr_\la r^i- (\G\circ s)^i_\la
\]
is the covariant derivative (\ref{39}).

With splitting (\ref{N18}), we have the following surjections
\be
&&{\cal S}:=\pr_1: J^1Y\to\Ker\wh L, \\
&&{\cal S}: y^i_\la\to y^i_\la
-\si^{ik}_{\la\al} (a^{\al\m}_{kj}y^j_\m + b^\al_k),
\ee
and
\be
&&\cF:=\pr_2: J^1Y\to{\rm Im}\si, \\
&& \cF=\si\circ\wh L:
y^i_\la\to \si^{ik}_{\la\al} (a^{\al\m}_{kj}y^j_\m + b^\al_k).
\ee
With respect to these surjections, the Hamilton equations (\ref{N29})
break into two parts
\ben
&&{\cal S}\circ J^1s=\G\circ s, \label{N23}\\
&&\nabla_\la r^i=
\si^{ik}_{\la\al} (a^{\al\m}_{kj}\dr_\mu r^j + b^\al_k).\nonumber
\een
and
\ben
&&\cF \circ J^1s=\si\circ r, \label{N28}\\
&&\si^{ik}_{\la\al} (a^{\al\m}_{kj}\dr_\mu r^j + b^\al_k)=
\si^{ik}_{\la\al}r^\al_k.\nonumber
\een
The Hamilton equations (\ref{N23}) are independent of canonical momenta
$r^\al_k$ and make the sense
of  gauge-type  conditions.

By analogy with gauge theory, one can think of the preimage
${\cF}^{-1}(\ol y)$ of every point $\ol y\in{\rm Im}\si$ as being a
gauge-type class. Then, Eq.(\ref{N23}) makes the sense of a
gauge-type condition. We shall say that this condition is
universal if it singles out one representative of every gauge-type class.
It is readily observed that Eq.(\ref{N23}) is the universal gauge-type
condition on sections $\g$ of the jet bundle $J^1\Pi\to\Pi$
when the algebraic Hamilton equations (\ref{3.10a}) and (\ref{3.10b})
are considered, otherwise on sections $r$ of the Legendre manifold
$\Pi\to X$.
At the same time, one can conclude that it is one or another condition
for the quantity
\[
{\cal S}\circ J^1s=\dr_\la r^i-
\si^{ik}_{\la\al} (a^{\al\m}_{kj}\dr_\mu r^j + b^\al_k)
\]
 which may supplement the underdetermined Euler-Lagrange
equations for the degenerate Lagrangian density (\ref{N12}).

Moreover, for every section $s$ of the fibred manifold $Y\to X$
(in particular, every solution of the Euler-Lagrange equations),
there exists a connection $\G$ [Eq.(\ref{N16})] such
that the gauge-type conditions (\ref{N23}) holds. Indeed, let $\G'$ be a
connection whose integral section is $s$. Set
\be
&&\G={\cal S}\circ\G',\\
&&\G={\G'}^i_\la -\si^{ik}_{\la\al} (a^{\al\mu}_{kj}{\G'}^j_\mu
+b^\al_k).
\ee
In this case, the momentum morphism (\ref{N19}) satisfies Eq.(\ref{2.36})
\[
\Phi\circ\wh L\circ J^1s=J^1s.
\]
It follows that the multimomentum Hamiltonian forms (\ref{N22}) constitute
really the complete family.

It must be noted that the multimomentum Hamiltonian forms from this
family differ from each other only in connections $\G$ [Eq.(\ref{N16})]
which imply the different gauge-type conditions (\ref{N23}).

The complete family of the multimomentum Hamiltonian forms
(\ref{N22}) is by no means minimal. It may be often minimized
by choice of a certain subset of connections (\ref{N16}).

Let us turn now to an affine Lagrangian density
\ben
&&L=\cL\om, \nonumber\\
&& \cL=b^\la_i(y)y^i_\la + c(y). \label{N24}
\een
It is almost regular and semiregular. The corresponding Legendre morphism
\begin{equation}
p^\la_i\circ\wh L = b^\la_i(y) \label{N25}
\end{equation}
is a bundle over the imbedded constraint submanifold $Q=b(Y)$ of
the Legendre manifold $\Pi$
determind by the section $b$ of the Legendre bundle $\Pi\to Y$.

Let $\G$ be a connection on the fibred manifold $Y\to X$ and
$\wh\G$ the associated momentum morphism. This momentum morphism
satisfies Eq.(\ref{4.26}) where $\wh L$ is the Legendre morphism
(\ref{N25}). Therefore, let us consider the multimomentum Hamiltonian form
(\ref{4.24})
\begin{equation}
H=H_\G+L\circ \G=p^\la_idy^i\w\om_\la - (p^\la_i -b^\la_i)\G_\la^i\om +
c\om,  \label{N26}
\end{equation}
corresponding to $\wh\G$. It is easy to see that this multimomentum
Hamiltonian form is associated with the affine Lagrangian density
(\ref{N24}).

Note that this multimomentum Hamiltonian form is affine
in canonical momenta everywhere on $\Pi$.
The corresponding momentum morphism reads
\begin{equation}
y^i_\la\circ\wh H = \G^i_\la \label{N27}
\end{equation}
and so, the Hamilton
equations (\ref{3.11a}) reduce to the gauge-type condition
\[
\dr_\la r^i=\G^i_\la.
\]
Their solution is an integral section of the connection $\G$.
Conversely, for each
section $s$ of the fibred manifold $Y\to X$, there exists
the connection $\G$ on $Y$
which has the integral section $s$. Then,
the corresponding momentum morphism
(\ref{N27}) obeys the condition
\[
\wh H\circ\wh L\circ J^1s = J^1s.
\]
It follows that the multimomentum Hamiltonian forms (\ref{N26}) constitute
the complete family.

Thus, the universal procedure is now at hand to describe constraint field
theories including gauge  theory and gravitation theory.

\section{Hamiltonian gauge theory}

Gauge theory, in conjuction with the mechanism of spontaneous symmetry
breaking, makes the adequate picture
of fundamental interactions where principal connections
model the mediators of interaction possessing a certain group of
symmetries. This Section covers the gauge theory with unbroken
symmetries.

Gauge theory of
principal connections is described by the
degenerate quadratic Lagrangian density,
and its multimomentum Hamiltonian formulation shadows the general
procedure for degenerate quadratic models from previous Section.
The feature of gauge theory
consists  in the fact that  splittings (\ref{N18}) and (\ref{N20})
of configuration and phase spaces of fields are canonical.

In the rest of this Chapter, a base manifold $X$ is proposed to be
oriented. We call it the world manifold. By $g_{\mu\nu}$ [or
$g^{\mu\nu}$] is meant a nondegenerate fibre metric in the tangent [or
cotangent] bundle of $X$. We denote
\[
g=\det (g_{\mu\nu}).
\]
Structure groups of bundles throughout are supposed to be
finite-dimensional real connected Lie groups.

Let
\[
 \pi_P :P\to X
\]
be a principal bundle with a structure Lie group $G$.
There is the 1:1 correspondence between principal connections on $P$ and
global sections of the principal connection bundle $C$ [Eq.(\ref{68})].
This bundle
is provided with  the fibred coordinates $(x^\la,k^m_\mu)$ so that
\[
(k^m_\mu\circ A)(x)=A^m_\mu(x)
\]
are coefficients of the local connection 1-form (\ref{1.32}).

In gauge theory, global sections of the principal connection bundle
(\ref{68}) are treated the gauge potentials. Their finite-dimensional
configuration space is
the first order jet manifold $J^1C$ of  $C$. It is endowed
 with the adapted coordinates
$(x^\la, k^m_\mu, k^m_{\mu\lambda})$ [Eq.(\ref{S2})].
The affine jet bundle $J^1C\to C$ is modelled on the vector bundle
\[
T^*X\op\otimes_C (C\times T^*X\otimes V^GP).
\]

On the configuration space (\ref{N31}), the conventional
Yang-Mills Lagrangian density $L_{YM}$ of gauge potentials
is given by the expression
\begin{equation}
L_{YM}=\frac{1}{4\ve^2}a^G_{mn}g^{\lambda\mu}g^{\beta\nu}\cF^m_{\lambda
\beta}\cF^n_{\mu\nu}\sqrt{\mid g\mid}\,\omega \label{5.1}
\end{equation}
where  $a^G$ is a nondegenerate $G$-invariant metric
in the Lie algebra ${\got g}_r$ and $\ve$ is a coupling constant.
This Lagrangian density is almost regular and semiregular.

\begin{remark}
Note that the Lagrangian density (\ref{5.1}) is the unique gauge
invariant quadratic Lagrangian density on $J^1C$.
In gauge theory several types of gauge transformations
are considered.
Here, we are concerned only with so-called principal morphisms.
Given a principal bundle $P\to X$,
by a principal morphism is meant its $G$-equivariant
isomorphism $\Phi_P$ over $X$ together with the first order
jet prolongations $J^1\Phi_P$. Whenever $g\in G$, we have
\[
r_g\circ\Phi_P=\Phi_P\circ r_g.
\]
Every such isomorphism $\Phi_P$ is brought into the form
\begin{equation}
\Phi_P(p)=pf_s(p),  \qquad p\in P,\label{S15}
\end{equation}
where $f_s$ is a $G$-valued equivariant function on $P$:
\[
 f_s(qg)=g^{-1}f_s(q)g, \qquad g\in G.
\]
There is the 1:1 correspondense between these functions and the global
section $s$ of the group bundle $\wt P$:
\[
s(\pi(p))p=pf_s(p).
\]
 Principal morphisms $\Phi_P$
constitute the gauge group which is isomorphic to the group of global
 sections of the $P$-associated group bundle $\wt P$.
The Sobolev completion of the
gauge group is a Banach Lie group. Its Lie algebra
in turn is the Sobolev completion of the algebra of generators of
infinitesimal principal morphsms. These generators are represented by
the corresponding vertical vector fields on a $P$-associated bundle which
carries representation of the gauge group. We call them principal
vector fields. Note that the jet lift $\ol u$ [Eq.(\ref{1.21})] of
a principal vector field $u$ on a $P$-associated bundle
is a principal vector
field on the fibred jet manifold $J^1Y\to X$.
Then, it is readily observed that a
Lagrangian density $L$  on the configuration space $J^1Y$ is gauge
invariant iff, whenever principal vector field $u$ on $Y\to X$,
\[
{\bf L}_{\ol u} L=0
\]
where ${\bf L}$ denotes the Lie derivative of exterior forms. For
instance, local principal vector fields on the principal connection
bundle $C$ read
\begin{equation}
u=(\dr_\mu\al^\mu+c^m_{nl}k^l_\mu\al^n)\dr^\mu_m \label{225}
\end{equation}
where $\al^m(x)$ are local real functions on $X$.
\end{remark}

The finite-dimensional phase space of gauge potentials is
the Legendre bundle
\be
&&\pi_{\Pi C}: \Pi\to C,\\
&&\Pi =
\op\wedge^n T^*X\op\otimes_C TX\op\otimes_C [C\times\ol C]^*,
\ee
over the principal connection bundle $C$ [Eq.(\ref{68})].
It is endowed with the canonical coordinates
\[
(x^\la,k^m_\mu,p^{\mu\lambda}_m).
\]
The Legendre bundle $\Pi$ over $C$ as like as the jet bundle $J^1C\to C$
admits the canonical splitting
\begin{equation}
\Pi=
\Pi_+\op\oplus_C\Pi_-,\label{N32}
\end{equation}
\[
 p^{\mu\lambda}_m=
p^{(\mu\lambda)}_m + p^{[\mu\lambda]}_m=\frac{1}{2}(p^{\mu\lambda}_m+
p^{\lambda\mu}_m)) + \frac{1}{2}(p^{\mu\lambda}_m-
p^{\lambda\mu}_m).
\]

The Legendre morphism associated with the Lagrangian density (\ref{5.1})
takes the form
 \bea
 &&p^{(\mu\lambda)}_m\circ\wh L_{YM}=0, \label{5.2a}\\
&&p^{[\mu\lambda]}_m\circ\wh
L_{YM}=\ve^{-2}a^G_{mn}g^{\lambda\alpha}g^{\mu\beta}
\cF^n_{\alpha\beta}\sqrt{\mid g\mid}. \label{5.2b}
\eea
A glance at this morphism shows that
\[
\Ker\wh L_{YM}=C_+
\]
 and
\[
Q=\wh L_{YM}(J^1C)=\Pi_-.
\]
 It follows that splittings (\ref{N31}) and
(\ref{N32}) are similar to the splittings (\ref{N18}) and (\ref{N20})
in general case of a degenerate Lagrangian density.

Thus, to construct the complete family of multimomentum Hamiltonian forms
associated with the Yang-Mills
Lagrangian density (\ref{5.1}), we can follow
the standard procedure for degenerate Lagrangian densities from previous
Section.

In accordance with this procedure, let us consider
connections on the principal connection bundle $C$ which
take their values into $\Ker\wh L$:
\ben
&&  S:C\to C_+, \label{69}\\
&&S^m_{\m\la}-S^m_{\la\m}-c^m_{nl}k^n_\la k^l_\m=0. \nonumber
\een
Recall that, 
given a symmetric linear connection $K^*$ [Eq.(\ref{408})]
on the cotangent bundle $T^*X$ of $X$,  every principal connection $B$ on
a principal bundle $P$ gives rise to the connection
\be
&& S_B: C\to C_+, \\
&& S_B\circ B={\cal S}\circ J^1B,
\ee
[Eq.(\ref{3.7})] on the principal connection bundle $C$. 

Given a connection (\ref{69}), one can check directly that the
multimomentum Hamiltonian form
\ben
&&H=p^{\mu\lambda}_mdk^m_\mu\wedge\omega_\lambda-
p^{\mu\lambda}_mS^m_{\mu\lambda}\omega-\wt{\cH}_{YM}\omega, \label{5.3}\\
&&\wt{\cH}_{YM}= \frac{\ve^2}{4}a^{mn}_Gg_{\mu\nu}
g_{\lambda\beta} p^{[\mu\lambda]}_m p^{[\nu\beta]}_n\mid g\mid ^{-1/2},
\nonumber
\een
is associated with the Lagrangian density $L_{YM}$.

\begin{remark} In contrast with the Lagrangian density $L_{YM}$, the
multimomentum Hamiltonian forms (\ref{5.3}) are not gauge invariant,
otherwise their restriction
\[
H_Q=p^{\mu\lambda}_m(dk^m_\mu\wedge\omega_\lambda-
\frac12 c^m_{nl}k^n_\la k^l_\mu\om) -\wt{\cH}_{YM}\omega
\]
to the constraint space $Q$. By gauge transformations in multimomentum
canonical variables are meant isomorphisms of the Legendre bundle $\Pi$
over $C$ which are induced by principal morphisms of the principal
connection bundle $C$. The corresponding principal vector fields
on $\Pi\to X$ read
\[
u=(\dr_\mu\al^\mu+c^m_{nl}k^l_\mu\al^n)\dr^\mu_m -c^l_{nm}p^{\mu\la}_l
\al^n\dr^m_{\mu\la}.
\]
\end{remark}

We can justify that the multimomentum Hamiltonian forms (\ref{5.3})
 constitute the complete family.
Moreover, to get the complete family, it suffices to take the
subset of connections (\ref{3.7}).

Let $K^*$ be a fixed symmetric linear connection (\ref{408})
on the cotangent bundle $T^*X$ of $X$. In virtue of Lemma 19.1,
every principal connection $B$ on
$P$ gives rise to the connection $S_B$ [Eq.(\ref{3.7})].
We shall denote by $H_B$ the multimomentum Hamiltonian form (\ref{5.3})
where $S$ is the connection $S_B$ [Eq.(\ref{3.7})].

Given $H_B$, the corresponding Hamilton equations
for  sections $r$ of the Legendre bundle $\Pi\to X$ are
\ben
&&\dr_\lambda r^{\mu\lambda}_m=-c^n_{lm}r^l_\nu
r^{[\mu\nu]}_n+c^n_{ml}B^l_\nu r^{(\mu\nu)}_n
-K^\mu{}_{\lambda\nu}r^{(\lambda\nu)}_m,
\label{5.5} \\
&&\dr_\lambda r^m_\mu+ \dr_\mu
r^m_\lambda=2S_B{}^m_{(\mu\lambda)}\label{5.6}
\een
and plus Eq.(\ref{5.2b}).

The Hamilton equations (\ref{5.6}) and (\ref{5.2b})
are similar to Eqs.(\ref{N23}) and (\ref{N28}) respectively. The Hamilton
equations (\ref{5.2b}) and (\ref{5.5}) restricted to the constraint space
(\ref{5.2a}) are precisely the Yang-Mills equations for a gauge potential
\[
A=\p_{\Pi C}\circ r.
\]
Whenever $H_B$, these equations are the same. They exemplify the
De Donder-Hamilton equations (\ref{N44}).

Different multimomentum
Hamiltonian forms $H_B$ lead to different Eqs.(\ref{5.6}).
The Hamilton equation (\ref{5.6}) is independent of canonical momenta
and it is just the gauge-type condition (\ref{N23})
\[
S_B{}\circ A={\cal S}\circ J^1A.
\]
A glance at this condition shows that,
given a solution $A$ of the Yang-Mills equations,
there always exists a
multimomentum Hamiltonian form $H_B$ which
obeys the condition (\ref{2.36})
 \[
\wh H_B\circ\wh L_{YM}\circ J^1A=J^1A.
\]
 For instance, this is
$H_{B=A}.$ It follows that the Hamiltonian forms $H_B$
constitute really the complete family.

It must be noted that
the gauge-type condition (\ref{5.6}) differs from the familiar gauge
condition in gauge theory. The latter singles out a representative of each
gauge coset (with accuracy to Gribov's ambiguity).
Namely, if a gauge potential $A$ is a solution of the Yang-Mills
equations, there exists a gauge conjugate potential $A'$ which also is a
solution of the same Yang-Mills equations and satisfies a given gauge
condition. At the same time, not every solution of the Yang-Mills equations
is a solution of a system of the Yang-Mills equations and one or another
gauge condition. In other words,
there are solutions of Yang-Mills equations
which are not singled out by the gauge
conditions known in gauge theory. In
this sense, the system of these gauge conditions is not complete. In gauge
theory, this lack is not essential since one think of all
gauge conjugate potentials as being physically equivalent, otherwise
in case of other constraint field theories, e.g. the Proca field.

In the framework of the multimomentum Hamiltonian description of quadratic
Lagrangian systems, there is a complete set of gauge-type conditions
in the sense that,
for any solution of the Euler-Lagrange equations, there exists a system of
Hamilton equations equivalent to these Euler-Lagrange equations and a
supplementary gauge-type condition to which this solution satisfies.

Given a principal connection $B$, the gauge-type condition (\ref{5.6})
is universal on solutions of the algebraic Hamilton equations, but it
is rather rigid on solutions of the differential Hamilton equations.
In gauge theory where gauge conjugate solutions are
treated physically equivalent, one may replace Eq.(\ref{5.6}) by one
or another condition for
\[
({\cal S}\circ J^1A)^m_{\mu\la} =\frac12 (\dr_\la A^m_\mu
+\dr_\mu A^m_\la +c^m_{nl}A^n_\la A^l_\mu)
\]
which can supplement the Yang-Mills equations. In particular,
\[
g^{\mu\la}({\cal S}\circ J^1A)^m_{\mu\la} =\al^m(x)
\]
recovers the familiar generalized Lorentz gauge condition.

\section{Electromagnetic fields}

As a test case of Hamiltonian gauge theory, let us consider
electromagnetic fields on a 4-dimensional world manifold $X^4$.

In accordance with the gauge approach, electromagnetic
potentials are identified with principal connections on a principal bundle
$P\to X^4$ with the structure group $U(1)$. In this case, the adjoint
bundle is isomorphic to the trivial linear bundle
\[
V^GP=X^4\times\Bbb R.
\]
The corresponding principal connection bundle $C$ [Eq.(\ref{68})]
is an affine bundle modelled on the cotangent bundle $T^*X$ of $X^4$.
It is coordinatized by $(x^\la,k_\mu)$.

The finite-dimensional configuration space of electromagnetic
potentials is the affine jet bundle $J^1C\to C$ modelled on the
pullback tensor bundle
\[
\ol C: =\op\otimes^2T^*X\op\times_XC\to C.
\]
Its canonical splitting (\ref{N31}) reads
\begin{equation}
J^1C=C_+\op\oplus_C (\op\wedge^2 T^*X\op\times_X C) \label{208}
\end{equation}
where $C_+\to C$ is the affine bundle modelled on the pullback symmetric
tensor bundle
\[
\ol C_+=\op\vee^2 T^*X\op\times_XC.
\]
Relative to the adapted coordinates
\[
(x^\la,k_\mu,k_{\mu,\la})
\]
of $J^1C$, the splitting (\ref{208}) is expressed as
\[
k_{\mu\la}=\frac12 ({\cal S}_{\mu\la} +\cF_{\mu\la}) =k_{(\mu\la)} +
k_{[\mu\la]}.
\]
For any section $A$ of $C$, we observe that
\[
F_{\mu\la}=\cF_{\mu\la}\circ J^1A=\dr_\la A_\m -\dr_\m A_\la
\]
is the familiar strength of an electromagnetic field.

For the sake of simplicity, we further
let $X^4$ be the Minkowski space with the Minkowski metric
\[
\eta=(+,---).
\]

On the configuration space (\ref{208}), the conventional
Lagrangian density of an electromagnetic field is written
\begin{equation}
L_E=-\frac{1}{16\pi}\eta^{\lambda\mu}\eta^{\beta\nu}\cF_{\lambda
\beta}\cF_{\mu\nu}\omega. \label{209}
\end{equation}

The finite-dimensional phase space of electromagnetic potentials is
the Legendre bundle
\[
\Pi =
(\op\wedge^4 T^*X\otimes TX\otimes TX)\op\times_X C
\]
endowed with the canonical coordinates
\[
(x^\la,k_\mu,p^{\mu\lambda}).
\]
With respect to these coordinates,
the Legendre morphism associated with the Lagrangian density (\ref{209})
is written
 \bea
 &&p^{(\mu\lambda)}\circ\wh L_E=0, \label{210a}\\
&&p^{[\mu\lambda]}\circ\wh
L_E=-\frac1{4\pi}\eta^{\lambda\alpha}\eta^{\mu\beta}
\cF_{\alpha\beta}. \label{210b}
\eea

The multimomentum Hamiltonian forms
\ben
&&H_B=p^{\mu\lambda}dk_\mu\wedge\omega_\lambda-
p^{\mu\lambda}S_{B\mu\lambda}\omega-\wt{\cH}_E\omega, \label{211}\\
&&S_{B\mu\lambda}=\frac{1}{2} (\dr_\mu B_\lambda+\dr_\lambda
B_\mu ),\nonumber\\
&&\wt{\cH}_E=-\pi \eta_{\mu\nu}
\eta_{\lambda\beta} p^{[\mu\lambda]}p^{[\nu\beta]},\nonumber
\een
parameterized by electromagnetic potentials $B$ are associated with
the Lagrangian density (\ref{209}) and constitute the complete family.

Given the multimomentum Hamiltonian form $H_B$ [Eq.(\ref{211})],
the corresponding Hamilton equations are
\ben
&&\dr_\lambda r^{\mu\lambda}=0 \label{212} \\
&&\dr_\lambda r_\mu+ \dr_\mu r_\lambda=
\dr_\lambda B_\mu+ \dr_\mu B_\lambda \label{213}
\een
and plus Eq.(\ref{210b}). On the constraint space (\ref{210a}),
Eqs.(\ref{210b}) and (\ref{212}) come to the Maxwell equations without
matter sources. At the same time, Eq.(\ref{213}) independent
on canonical momenta plays
the role of a gauge-type condition. Although its solution
\[
\pi_{\Pi C}\circ r=B.
\]
is by no means unique, this gauge-type condition, $B$ being fixed,
is rather rigid. Since gauge conjugate electromagnetic potentials are
treated physically equivalent, one may replace Eq.(\ref{213}) by one
or another condition for the quantity
\[
\dr_\la r_\mu+\dr_\mu r_\la.
\]
In particular,
the familiar generalized Lorentz gauge condition
\[
\eta^{\mu\la}\dr_\mu r_\la =\al (x)
\]
is recovered.

In gauge theory, the gauge conditions which supplement the
Euler-Lagrange equations are well-known, otherwise in general case of
constraint field systems.

\section{Proca fields}

The model of massive vector Proca fields exemplifies a
constraint field theory which shadows the electromagnetic theory, but
without gauge invariance. Note that at
present, the Proca field model is not
actual, for vector mesons appeared to be composite particles and the
Higgs mechanism of generating a mass is dominant in contemporary
particle physics.

The vector
Proca fields are represented by sections of the cotangent bundle
$T^*X\to X^4$, whereas electromagnetic potentials are
sections of the affine bundle modelled on $T^*X$. The finite-dimensional
configuration space of Proca fields
is the affine jet bundle
\[
J^1T^*X\to T^*X
\]
modelled on the pullback tensor bundle
\begin{equation}
\op\otimes^2T^*X\times TX\to TX. \label{219}
\end{equation}
It is provided with the adapted coordinates
\[
(x^\la,k_\mu,k_{\mu,\la})
\]
where $k_\mu=\dot x_\mu$ are the familiar induced coordinates of $T^*X$.

Let $X^4$ be the Minkowski space.
The Lagrangian density of Proca fields is viewed as the electromagnetic
Lagrangian density (\ref{209}) minus the massive-type term:
\begin{equation}
L_P=L_E -\frac1{8\pi} m^2\eta^{\mu\la}k_\mu k_\la\omega. \label{214}
\end{equation}
It is semiregular and almost regular.

The finite-dimensional phase space of Proca fields is
the Legendre bundle
\[
\Pi =\op\wedge^4 T^*X\otimes TX\otimes TX\times T^*X
\]
endowed with the canonical coordinates
\[
(x^\la,k_\mu,p^{\mu\lambda}).
\]

With respect to these coordinates,
the Legendre morphism associated with the Lagrangian density (\ref{214})
takes the form
 \bea
 &&p^{(\mu\lambda)}\circ\wh L_P=0, \label{215a}\\
&&p^{[\mu\lambda]}\circ\wh
L_P=-\frac1{4\pi}\eta^{\lambda\alpha}\eta^{\mu\beta}
\cF_{\alpha\beta}. \label{215b}
\eea
It is viewed exactly as the Legendre morphism $\wh L_E$. We have
\[
\Ker\wh L_P=\op\vee^2T^*X\times T^*X
\]
and
\[
Q=\op\w^4T^*X\op\otimes_{T^*X} (\op\w^2TX),
\]
\[
p^{(\mu\la)}=0.
\]
Following the general procedure of describing constraint field systems
>from Section 18, let us consider the form $\si$ [Eq.(\ref{N17})]:
\[
\ol k_{\mu\la}\circ\si = -2\pi\eta_{\mu\nu}\eta_{\la\beta}p^{[\nu\beta]}
\]
where $\ol k_{\mu\la}$ are the bundle coordinates of the bundle
(\ref{219}). Since
\[
{\rm Im}\si =\op\w^2T^*X\times T^*X
\]
and
\[
\Ker\si =\op\w^4T^*X\otimes (\op\vee^2TX)\times T^*X,
\]
one can perform the corresponding splitting (\ref{N18}) of the
configuration space
\ben
&&J^1T^*X=\op\vee^2T^*X\op\oplus_{T^*X} \op\w^2T^*X,\label{218}\\
&&k_{\mu\la}=\frac12 ({\cal S}_{\mu\la} +\cF_{\mu\la}) =k_{(\mu\la)} +
k_{[\mu\la]}\nonumber
\een
and the splitting (\ref{N20}) of the phase space
\be
&&\Pi =[\op\wedge^4 T^*X\op\otimes_{T^*X}
(\op\vee^2TX)]\op\oplus_{T^*X}Q,\\
&&p^{\mu\la}=p^{(\mu\la)}+p^{[\mu\la]}.
\ee

Following Section 18, let us consider connections on the cotangent
bundle $T^*X$ taking
their values into $\Ker\wh L_P$. Every such a connection is expressed as
\[
S=\phi_{\mu\la}dx^\mu\otimes\dr^\la
\]
where $\phi$ is a symmetric soldering form on $T^*X$, for $K=0$ in the
Minkowski space. By analogy with case of the electormagnetic field,
it suffies to take the connections
\[
S_B=dx^\la\otimes [\dr_\la +\frac{1}{2} (\dr_\mu B_\lambda+\dr_\lambda
B_\mu )\dr_\mu]
\]
where $B$ is a section of $T^*X$. Then one can justify that
the multimomentum Hamiltonian forms
\be
&&H_B=p^{\mu\lambda}dk_\mu\wedge\omega_\lambda-
p^{\mu\lambda}S_{B\mu\lambda}\omega-\wt{\cH}_P\omega,\\
&&\wt{\cH}_P=\wt{\cH}_E +\frac1{8\pi}m^2\eta^{\mu\nu}k_\mu k_\nu,
\ee
are associated with
the Lagrangian density $L_P$ [Eq.(\ref{209})]
and constitute the complete family.

Given the multimomentum Hamiltonian form $H_B$,
the corresponding Hamilton equations are
\ben
&&\dr_\lambda r^{\mu\lambda}=-\frac1{4\pi}m^2\eta^{\mu\nu}r_\nu,
\label{216} \\
&&\dr_\lambda r_\mu+ \dr_\mu r_\lambda=
\dr_\lambda B_\mu+ \dr_\mu B_\lambda \label{217}
\een
and plus Eq.(\ref{215b}). On the constraint space (\ref{215a}),
Eqs.(\ref{215b}) and (\ref{216}) come to the Euler-Lagrange equations
which are supplemented by the gauge-type condition (\ref{217}).
In view of the splitting (\ref{218}), this supplementary condition
is universal on solutions of the algebraic Hamilton equations for $H_B$,
but is rather rigid on solutions of the differential Hamilton equations.
At the same time, one may replace Eq.(\ref{217}) by one
or another condition for the quantity
\[
\dr_\la r_\mu+\dr_\mu r_\la,
\]
e.g., the generalized Lorentz gauge condition
\[
\eta^{\mu\la}\dr_\mu r_\la =\al (x).
\]
But, in contrast with case of the electromagnetic field, no such condition
exausts all physically non-equivalent solutions of the Euler-Lagrange
equations for Proca fields. For instance, the Lorentz gauge condition
\[
\eta^{\mu\la}\dr_\mu r_\la =0
\]
is compatible with the wave solutions.

Note that, in another model, massive vector fields
represented by sections of the cotangent
bundle $T^*X$ may be described also by the hyperregular Lagrangian
density
\[
L=\frac{1}{2}\eta^{\mu\nu}(\eta^{\la\al}k_{\mu\la}
k_{\nu\al}-m^2k_\mu k_\nu)\om
\]
on the configuration space $J^1T^*X$. In this case, they make the
sense of matter fields.

\section{Matter fields}

 In gauge theory, matter fields possessing only internal symmetries,
i.e., scalar fields are described by sections of a vector bundle
 associated with a principal bundle $P$. One calls it the matter bundle.

Let $Y\to X$ be a bundle associated with a principal bundle $P\to X$.
The structure group $G$ of $P$ acts freely on the standard fibre $V$ of
$Y$ on the left. The total space of the $P$-associated bundle $Y$,
by definition, is the quotient
\[
Y=(P\times V)/G
\]
of the product $P\times V$
by identification of its elements $(qg\times gv)$
for all $g\in G$. The $P$-associated bundle $Y$ is provided with
an atlas
\[
\Psi=\{U_\xi, \psi_\xi\}
\]
associated with an atlas
\[
\Psi^P=\{U_\xi, z_\xi\}
\]
of the principal bundle $P$ as follows:
\[
\psi^{-1}_\xi (x\times V)= [z_\xi (x)]_V (V), \qquad x\in U_\xi,
\]
where by $[p]_V$ is meant the restriction of the canonical map
\[
[P]_V:P\times V\to Y
\]
to the subset $p\times V$.

\begin{remark} For each $P$ associated
bundle $Y$, there exists the fibre-preserving representation morphism
\[
\wt P\times Y\ni (\wt p,y)\mapsto \wt py\in Y
\]
where $\wt P$ is the $P$-associated group bundle. Building on this
representation morphism, one can induce principal morphisms of $Y$:
\[
\Phi_s: Y\ni y\mapsto (s\circ \pi)(y)y\in Y
\]
where $s$ is a global section of $\wt P$. The corresponding
principal vertical vector fields on the $P$-associated vector bundle
$Y\to X$ read
\[
u=\al^m(x)I_m{}^i{}_jy^j\dr_i
\]
where $I_m$ are generators of the structure group $G$ acting on $V$
and $\al^m(x)$ are arbitrary local functions on $X$.
\end{remark}

Every principal connection $A$ on a principal bundle $P$ yields the
associated connection $\G$ [Eq.(\ref{S4})] on a $P$-associated 
bundle $Y\to X$.
Relative to the associated atlases $\Psi$ of $Y$ and $\Psi^P$ of $P$,
this connection is written
\begin{equation}
\G=dx^\la\otimes [\dr_\la +A^m_\mu (x)I_m{}^i{}_jy^j\dr_i] \label{G4}
\end{equation}
where $A^m_\mu (x)$ are coefficients of the local connection 1-form
(\ref{1.32}). The curvature (\ref{13})
of the connection (\ref{G4}) reads
\[
R^i_{\la\mu}= F^m_{\la\mu}I_m{}^i{}_jy^i.
\]

Turning now to the model of scalar matter fields, we let $Y\to X$ be
a $P$-associated matter bundle. It is assumed to be
provided with a $G$-invariant
fibre metric $a^Y$. Because of the canonical
vertical splitting (\ref{1.10}), the metric $a^Y$  is also a
fibre metric in the vertical tangent bundle $VY\to X$.
A linear connection $\Gamma$ on a matter bundle $Y$ is supposed
to be an associated principal connection (\ref{G4}).

The finite-dimensional configuration space of matter fields is the jet
manifold $J^1Y$ provided with the adapted coordinates
$(x^\la, y^i, y^i_\la)$. Relative to these coordinates,
the Lagrangian density of scalar matter fields in
the presence of a background gauge field, i.e., a
connection  $\Gamma$ [Eq.(\ref{G4})] on $Y$  reads
\begin{equation}
L_M=\frac{1}{2}a^Y_{ij}[g^{\mu\nu}(y^i_\mu-\Gamma^i_\mu)
(y^j_\nu-\Gamma^j_\nu)-m^2y^iy^j]\sqrt{\mid g\mid}\omega.\label{5.12}
\end{equation}
It is hyperregular.

The finite-dimensional phase space of matter fields is the Legendre bundle
\[
\Pi=(\op\w^nT^*X\otimes TX\otimes Y^*)\times Y
\]
where by $Y^*$ is meant the vector bundle dual to $Y\to X$. It is provided
with the canonical coordinates $(x^\la, y^i, p^\la_i)$. The unique
multimomentum Hamiltonan form on $\Pi$ associated with the Lagrangian
density (\ref{5.12}) reads
\ben
&& H_M=p^\lambda_idy^i\wedge\omega_\lambda-p^\lambda_i
A^m_\lambda I_m{}^i{}_jy^j\omega-\wt{\cH}_M, \label{220}\\
&& \wt{\cH}_M=\frac12(a^{ij}_Yg_{\mu\nu}p^\mu_ip^\nu_j\mid g\mid^{-1}
 + m^2a^Y_{ij}y^iy^j)\sqrt{\mid g\mid}, \nonumber
\een
where $a_Y$ is the fibre metric in $Y^*$ dual to $a^Y$. The corresponding
Hamilton equations for sections $r$ of the Legendre manifold $\Pi\to X$
take the form
\be
&& \dr_\la r^i = A^m_\lambda I_m{}^i{}_jr^j
+ a^{ij}_Yg_{\la\mu}r^\mu_j\mid g\mid^{-1/2},\\
&& \dr_\la r^\la_i= -
A^m_\lambda I_m{}^i{}_jr^\la_j - m^2a^Y_{ij}r^j\sqrt{\mid g\mid}.
\ee
They are equivalent to the Euler-Lagrange equations corresponding
to the Lagrangian density (\ref{5.12}).

In case
of unbroken symmetries, the total configuration space of
gauge potentials and matter fields  is the direct product
\begin{equation}
J^1Y\op\times_X J^1C. \label{N52}
\end{equation}
Accordingly, the total Lagrangian density describing matter fields
in the presence of dynamic gauge potentials is the sum of the
Lagrangian density (\ref{5.1}) and the Lagrangian
density (\ref{5.12}) where we must set up
\begin{equation}
\Gamma^i_\lambda =k^m_\lambda I_m{}^i{}_jy^j.\label{5.14}
\end{equation}
The associated multimomentum Hamiltonian forms are the sum of the
multimomentum Hamiltonian forms (\ref{5.3}) where $S=S_B$
[Eq.(\ref{3.7})] and the multimomentum Hamiltonian form (\ref{220})
where the connection $\G$ is given by the expression (\ref{5.14}). In
this case, the Hamilton equation (\ref{5.5}) contains the familiar
N\"oether current
\[
J^\la_m= p^\la_iI_m{}^i{}_jy^j
\]
as a matter source of gauge potentials.

\section{Hamilton equations of General Relativity}

 The contemporary concept of  gravitation interactions is based on
the gauge gravitation theory with  two types of gravitational
fields. These are tetrad  gravitational fields and
Lorentz gauge potentials.
They correspond to different matter sources: the
energy-momentum tensor and the spin current of matter. At
present, all Lagrangian densities of
 classical and quantum gravitation
theories are expressed in these variables. They are of the first order
with respect to these fields.
Only General Relativity without spin
matter sources utilizes traditionally   the Hilbert-Einstein Lagrangian
density $L_{HE}$ which is of the second order  with respect
to a pseudo-Riemannian metric. One can reduce its order by means of
the Palatini variables when the Levi-Civita
connection is regarded on the same footing as a pseudo-Riemannian
metric.

This Section is devoted to the so-called
affine-metric gravitation theory when
gravitational variables are both pseudo-Riemannian metrics $g$ on a world
manifold $X^4$ and linear connections $K$ [Eq.(\ref{408})]
on the tangent bundle of $X^4$. We
call them a world metric and a world connection respectively.
Given a world metric, every
world connection meets the well-known decomposition in the Cristoffel
symbols, contorsion and the nonmetricity term.
We here are not concerned with
the matter sources of a general linear connection, for they, except
fermion fields, are non-Lagrangian and hypothetical as a rule.

In this Section, $X$ is an oriented 4-dimensional
world manifold which obeys the
well-known topological conditions in order that a gravitational field
exists on $X^4$.

Let $LX\to X^4$ be the principal bundle of linear frames in the
tangent spaces to $X^4$. The structure group of $LX$ is the group
\[
GL_4=GL^+(4,\Bbb R)
\]
of general linear transfromations of the standard fibre $\Bbb R^4$ of
the tangent bundle $TX$.
The world connections are associated with  principal connections on the
principal bundle $LX\to X^4$.
 Hence, there is the 1:1 correspondence between the
world connections and the global sections of the principal connection
bundle
\begin{equation}
C=J^1LX/GL_4. \label{251}
\end{equation}
 Therefore, we can apply the standard scheme of
gauge theory in order to describe the configuration and phase spaces of
world connections.

 There is the 1:1 correspondence between the world metrics $g$ on
$X^4$  and the global sections of the  bundle $\Sigma_g$ of
pseudo-Riemannian bilinear
forms in tangent spaces to $X^4$. This bundle is
associated with the $GL_4$-principal bundle $LX$.
The 2-fold covering of the bundle $\Si_g$ is the quotient bundle
\[
\Sigma=LX/SO(3,1)
\]
where by $SO(3,1)$ is meant the connected Lorentz group.

Thus, the total configuration space of the
affine-metric gravitational variables is represented by the
 product of the corresponding jet manifolds:
\begin{equation}
J^1C\op\times_{X^4}J^1\Sigma_g. \label{N33}
\end{equation}
Given a
holonomic bundle atlas of $LX$ associated with  induced coordinates of
$TX$ and $T^*X$, this
 configuration space is provided with the adapted coordinates
\[
(x^\mu, g^{\alpha\beta}, k^\alpha{}_{\beta\mu}, g^{\alpha\beta}{}_\lambda,
k^\alpha{}_{\beta\mu\lambda}).
\]

Also the total phase space $\Pi$ of the affine-metric gravity
is the product of the Legendre bundles
over the above-mentioned bundles $C$ and $\Si_g$.
It is coordinatized by the corresponding canonical coordinates
\[
(x^\mu, g^{\alpha\beta}, k^\alpha{}_{\beta\mu},
p_{\alpha\beta}{}^\lambda, p_\alpha{}^{\beta\mu\lambda}).
\]

On the configuration space (\ref{N33}), the  Hilbert-Einstein  Lagrangian
density of General Relativity reads
\begin{equation}
L_{HE}=-\frac{1}{2\kappa}g^{\beta\lambda}\cF^\alpha{}_{\beta\alpha\lambda}
\sqrt{-g}\omega,\label{5.17}
\end{equation}
 \[
\cF^\alpha{}_{\beta\nu\lambda}=k^\alpha{}_{\beta\lambda\nu}-
k^\alpha{}_{\beta\nu\lambda}+k^\alpha{}_{\varepsilon\nu}
k^\varepsilon{}_{\beta\lambda}-k^\alpha{}_{\varepsilon\lambda}
k^\varepsilon{}_{\beta\nu}.
\]
It is affine in connection velocities $k^\al{}_{\beta\mu\la}$ and,
moreover, it is independent of metric velocities $g^{\al\beta}{}_\la$
at all. Therefore, one can follow the general procedure
of analyzing constraint field theories from Section 18.
The corresponding Legendre morphism is given by the expressions
\ben
&& p_{\alpha\beta}{}^\lambda\circ \wh L_{HE}=0,\nonumber
\\ &&   p_\alpha{}^{\beta\nu\lambda}\circ \wh L_{HE}
=\pi_\alpha{}^{\beta\nu\lambda} =\frac{1}{2\kappa}(\delta^\nu_\alpha
g^{\beta\lambda}-\delta^\lambda_\alpha g^{\beta\nu})\sqrt{-g}.\label{5.18}
\een
These relations define the constraint space of General Relativity  in
multimomentum canonical variables.

Building on all the set of connections on the bundle $C\times\Si_g$,
one can construct the complete family of  multimomentum Hamiltonian
forms (\ref{N26}) associated with the Lagrangian density (\ref{5.17}). To
minimize it this complete family,
we consider the following subset of these connections.

Let $K$ be a world connection and
\be
&& S_K{}^\alpha{}_{\beta\nu\lambda}=\frac12
[k^\alpha{}_{\varepsilon\nu}
k^\varepsilon{}_{\beta\lambda}-k^\alpha{}_{\varepsilon\lambda}
k^\varepsilon{}_{\beta\nu} +\dr_\lambda K^\alpha{}_{\beta\nu}
+\dr_\nu K^\alpha{}_{\beta\lambda}\\
&& \qquad
-2K^\varepsilon{}_{(\nu\lambda)}(K^\alpha{}_{\beta\varepsilon}
-k^\alpha{}_{\beta\varepsilon}) +
K^\varepsilon{}_{\beta\lambda}k^\alpha{}_{\varepsilon\nu}
+K^\varepsilon{}_{\beta\nu}k^\alpha{}_{\varepsilon\lambda}\\
&& \qquad -
K^\alpha{}_{\varepsilon\lambda}k^\varepsilon{}_{\beta\nu}
-K^\alpha{}_{\varepsilon\nu}k^\varepsilon{}_{\beta\lambda}]
\ee
the corresponding connection (\ref{3.7}) on the bundle $C$
[Eq.(\ref{251})]. Let $K'$ be
another symmetric world connection which induces an associated
principal connection on the bundle $\Si_g$.  On
the bundle $C\times\Si_g$, we consider the following connection
\ben
&& \G^{\al\bt}{}_\la =
-{K'}^\alpha{}_{\varepsilon\lambda}
g^{\varepsilon\beta} -
{K'}^\beta{}_{\varepsilon\lambda} g^{\alpha\varepsilon}, \nonumber
\\
 &&
\Gamma^\alpha{}_{\beta\nu\lambda} = S_K{}^\alpha{}_{\beta\nu\lambda}
-R^\alpha{}_{\beta\nu\lambda} \label{N34}
 \een
where $R$ is the curvature of the connection $K$. The corresponding
multimomentum Hamiltonian form (\ref{N26}) is given by the expression
 \ben
&& H_{HE}=(p_{\alpha\beta}{}^\lambda dg^{\alpha\beta} +
p_\alpha{}^{\beta\nu\lambda}dk^\alpha{}_{\beta\nu})\wedge\omega_\lambda
-\cH_{HE}\omega, \nonumber \\
&& \cH_{HE}=-p_{\alpha\beta}{}^\lambda({K'}^\alpha{}_{\varepsilon\lambda}
g^{\varepsilon\beta} +{K'}^\beta{}_{\varepsilon\lambda}
g^{\alpha\varepsilon})
+p_\alpha{}^{\beta\nu\lambda}\Gamma^\alpha{}_{\beta\nu\lambda}\nonumber \\
 && \qquad -R^\alpha{}_{\beta\nu\lambda}
(p_\alpha{}^{\beta\nu\lambda}-\pi_\alpha{}^{\beta\nu\lambda}). \label{5.19}
\een
It is associated with the Lagrangian density $L_{HE}$. We shall justify
that the
multimomentum Hamiltonian forms (\ref{5.19}) parameterized by all the world
connections $K$ and $K'$ constitute the complete family.

Given the multimomentum Hamiltonian form $H_{HE}$ [Eq.(\ref{5.19})],
the corresponding covariant Hamilton equations for General Relativity read
\bea
&&\dr_\lambda
g^{\alpha\beta} +{K'}^\alpha{}_{\varepsilon\lambda}g^{\varepsilon\beta}
+{K'}^\beta{}_{\varepsilon\lambda}g^{\alpha\varepsilon}=0, \label{5.20a}\\
&&\dr_\lambda k^\alpha{}_{\beta\nu}=
\Gamma^\alpha{}_{\beta\nu\lambda}
-R^\alpha{}_{\beta\nu\lambda}, \label{5.20b} \\
&&\dr_\lambda p_{\alpha\beta}{}^\lambda =p_{\varepsilon\beta}{}^\sigma
{K'}^\varepsilon{}_{\alpha\sigma} +
p_{\varepsilon\alpha}{}^\sigma {K'}^\varepsilon{}_{\beta\sigma} \nonumber\\
&& \qquad
+\frac{1}{\kappa}(R_{\alpha\beta} -\frac12g_{\alpha\beta}R)\sqrt{-g},
\label{5.20c} \\
&& \dr_\lambda p_\alpha{}^{\beta\nu\lambda}
= -p_\alpha{}^{\varepsilon[\nu\gamma]}
k^\beta{}_{\varepsilon\gamma}
 +p_\varepsilon{}^{\beta[\nu\gamma]}
k^\varepsilon{}_{\alpha\gamma} \nonumber \\
&& \qquad -
 p_\alpha{}^{\beta\varepsilon\gamma}
K^\nu{}_{(\varepsilon\gamma)} -p_\alpha{}^{\varepsilon(\nu\gamma)}
K^\beta{}_{\varepsilon\gamma}
 +p_\varepsilon{}^{\beta(\nu\gamma)}
K^\varepsilon{}_{\alpha\gamma}. \label{5.20d}
\eea
The Hamilton equations (\ref{5.20a}) and (\ref{5.20b}) are independent
of canonical momenta and so, reduce to the gauge-type condition
(\ref{N27}). In accordance with the canonical splitting (\ref{N31}) of
$J^1C$, the gauge-type
condition (\ref{5.20b}) breaks into two parts
\ben
&&\cF^\alpha{}_{\beta\nu\lambda}=R^\alpha{}_{\beta\nu\lambda},\label{5.21}\\
&& \dr_\nu(K^\alpha{}_{\beta\lambda} -k^\alpha{}_{\beta\lambda})
+\dr_\lambda(K^\alpha{}_{\beta\nu} -k^\alpha{}_{\beta\nu}) \nonumber\\
&&\qquad  -2K^\varepsilon{}_{(\nu\lambda)}
(K^\alpha{}_{\beta\varepsilon} -k^\alpha{}_{\beta\varepsilon}) +
K^\varepsilon{}_{\beta\lambda}k^\alpha{}_{\varepsilon\nu}
+K^\varepsilon{}_{\beta\nu}k^\alpha{}_{\varepsilon\lambda} \nonumber\\
&& \qquad   -K^\alpha{}_{\varepsilon\lambda}k^\varepsilon{}_{\beta\nu}
-K^\alpha{}_{\varepsilon\nu}k^\varepsilon{}_{\beta\lambda}=0.
\label{5.22}
\een
It is readily observed that,
for a given world metric $g$ and a world connection $k$, there
always exist the world connections $K'$ and $K$ such that the
gauge-type conditions (\ref{5.20a}), (\ref{5.21}) and (\ref{5.22}) hold
(e.g. $K'$ is the Levi-Civita connection of $g$ and $K=k$).
It follows that the multimomentum Hamiltonian forms (\ref{5.19}) consitute
really the complete family. At the same time, one can think of
connections $K$ and $K'$
as being variable parameters expressed into dynamical variables
$g$ and $k$ and so, can substitute
them into other equations.

 Being restricted to the constraint space (\ref{5.18}), the Hamilton
equations (\ref{5.20c}) and (\ref{5.20d}) comes to
\ben
&& \frac{1}{\kappa}(R_{\alpha\beta}
-\frac12 g_{\alpha\beta}R)\sqrt{-g} =0, \label{5.23} \\
&& D_\al(\sqrt{-g}g^{\nu\bt}) - \delta^\nu_\al
D_\la(\sqrt{-g}g^{\la\bt}) +\sqrt{-g}[g^{\nu\bt}(k^\la{}_{\al\la} -
k^\la{}_{\la\al}) \nonumber\\
&& \qquad + g^{\la\bt}(k^\nu{}_{\la\al}-k^\nu{}_{\al\la})+ \delta^\nu_\al
g^{\la\bt} (k^\m{}_{\m\la} - k^\m{}_{\la\m})] =0 \label{5.24}
\een
where
\[
D_\la g^{\al\bt}= \dr_\la g^{\al\bt} + k^\al{}_{\m\la}g^{\m\bt} +
k^\bt{}_{\m\la}g^{\al\m}.
\]
Substituting Eq.(\ref{5.21}) into Eq.(\ref{5.23}),
we obtain the Einstein equations
\begin{equation}
\cF_{\alpha\beta}-\frac12 g_{\alpha\beta}\cF= 0.\label{5.25}
\end{equation}
It is easy to see that Eqs.(\ref{5.24}) and (\ref{5.25}) are the
familiar equations of gravitation theory phrased in terms of
the nonsymmetric Palatini variables. In particular,
the former is the equation for
the torsion and the nonmetricity term of the connection $k^\al{}_{\bt\nu}$.
In the absence of matter sources, it admits the well-known solution
\be
&&k^\al{}_{\bt\nu} =\{^\al{}_{\bt\nu}\} - \frac12\delta^\al_\nu V_\bt,\\
&& D_\al g^{\bt\g}= V_\al g^{\bt\g},
\ee
where $V_\al$ is an arbitrary covector field corresponding to the
well-known projective freedom.

\section{Conservation laws}

Unless $n=1$, the physical meaning of a multimomentum Hamiltonian form
is not evident. To make it more understood, let us explore the
covariany Hamiltonian formulation of the energy-momentum conservation
law.

In the framerwork of the multimomentum Hamiltonian formalism, we get the
fundamental identity whose restriction to the Lagrangian constraint
space recovers the familiar energy-momentum conservation law, without
appealing to any symmetry condition.

Let $H$ be a Hamiltonian form on the Legendre bundle $\Pi$
(\ref{00}) over a fibred manifold $Y\to X$. Let $r$ be a section of
of the fibred Legendre manifold $\Pi\to X$ and $(y^i(x), p^\la_i(x))$ its
local components. Given a connection $\G$ on $Y\to X$, we consider the 
following $T^*X$-valued $(n-1)$-form on $X$:
\ben &&T_\G(r)=-(\G\rfloor H)\circ r,\nonumber\\
&&T_\G (r)=[p^\la_i(y^i_\mu -\G^i_\mu)-\dl^\la_\mu(p^\al_i(y^i_\al-\G^i_\al)
-\wt{\cH}_\G)] dx^\mu\otimes\om_\la, \label{5.8}\een
where $\wt{\cH}_\G$ is the Hamiltonian density in the splitting
(\ref{4.7}) of $H$ with respect to the connection $\G$.

Let
\[\tau=\tau^\la\dr_\la\]
be a vector field on $X$. Given a connection $\G$ on $Y\to X$, it
gives rise to the vector field
\[\wt\tau_\G= \tau^\la\dr_\la + \tau^\la\G^i_\la\dr_i +
(-\tau^\m p^\la_j\dr_i\G^j_\m
-p^\la_i\dr_\m\tau^\m + p^\m_i\dr_\m\tau^\la) \dr^i_\la\]
on the Legendre bundle $\Pi$. Let us calculate the Lie derivative
${\bf L}_{\wt\tau_\G}\wt H_\G$ on a section $r$. We have
\begin{equation}
({\bf L}_{\wt\tau_\G}\wt H_\G)\circ r=p^\la_iR^i_{\la\m}+d[\tau^\m
T_\G{}^\la{}_\m (r)\om_\la]-(\wt\tau_{\G V}\rfloor\cE_H)\circ r\label{221}
\end{equation}
where 
\be &&R =\frac12 R^i_{\la\m} dx^\la\wedge dx^\m\otimes\dr_i=\\
&&\quad \frac12 (\dr_\la\G^i_\m -\dr_\m\G^i_\la +\G^j_\la\dr_j\G^i_\m
-\G^j_\m\dr_j\G^i_\la) dx^\la\wedge dx^\m\otimes\dr_i; \ee
of  the connection $\G$,  $\cE_H$ is the Hamilton
operator (\ref{3.9}) and 
\[\wt\tau_{\G V}=\tau^\la(\G^i_\la-y^i_\la)\dr_i +
(-\tau^\m p^\la_j\dr_i\G^j_\m
-p^\la_i\dr_\m\tau^\m+p^\m_i\dr_\m\tau^\la-\tau^\mu p^{\la\mu}_i) \dr^i_\la\]
is the vertical part of the canonical
horizontal splitting of the vector field $\wt\tau_V$ on $\Pi$
over $J^1\Pi$. If $r$ is a
solution of the Hamilton equations, the equality (\ref{221}) comes
to the identity
\begin{equation}
(\dr_\mu+\G^i_\mu\dr_i-\dr_i\G^j_\mu
p^\la_j\dr^i_ \la)\wt{\cH}_\G-\frac{d}{dx^\la}
T_\G{}^\la{}_\mu (r)= p^\la_iR^i_{\la\mu}. \label{5.27}
\end{equation}
On solutions of the Hamilton equations, the form (\ref{5.8}) reads
\begin{equation}
T_\G(r)=[p^\la_i\dr^i_\mu\wt{\cH}_\G-
\dl^\la_\mu(p^\al_i\dr^i_\al\wt{\cH}_\G-\wt{\cH}_\G)]
dx^\mu\otimes\om_\la.\label{5.26}
\end{equation}
One can verify that the identity (\ref{5.27}) does not depend upon choice
of the connection $\G$.

For instance,
if $X={\bf R}$ and $\G$ is the trivial connection, then
\[
T_\G(r)=\wt{\cH}_0dt
\]
where $\wt{\cH}_0$ is a Hamiltonian and the identity
(\ref{5.27}) consists with the familiar energy conservation law
\[
\frac{d\wt{\cH}_0}{dt}=\frac{\dr\wt{\cH}_0}{\dr t}
\]
in analytical mechanics.

Unless $n=1$, the identity (\ref{5.27}) can not be regarded directly
as the energy-momentum conservation law. To clarify its physical meaning,
we turn to the Lagrangian formalism.
Let a multimomentum Hamiltonian form $H$ be associated with a
semiregular Lagrangian density $L$. Let $r$ be a solution
of the Hamilton equations for $H$ which lives on the Lagrangian
constraint space $Q$ and $\ol s$ the associated solution  of the
first order
Euler-Lagrange equations for $L$ so that they satisfy the conditions
(\ref{2.37}). Then, we have
\[ 
T_\G (r)=T_\G(\ol s)
\]
where is the Lagrangian canonical energy-momentum tensor (\ref{S14}).
It follows that the form (\ref{5.26}) may be treated as a Hamiltonian
canonical energy-momentum tensor with respect to a background connection
$\G$ on the fibred manifold $Y\to X$.
At the same time, the examples below will show
that, in field models, the identity (\ref{5.27}) is precisely the
energy-momentum conservation law
for the metric energy-momentum tensor, not the canonical one.

In the Lagrangian formalism, the metric 
energy-momentum tensor is defined to be
\[ \sqrt{-g} t_{\al\bt}=2\frac{\dr\cL}{\dr g^{\al\bt}}.\]
In case of a background world metric $g$, this object is well-behaved.
In the framework of the multimomentum Hamiltonian formalism, 
one can introduce the similar tensor
\begin{equation}
\sqrt{-g}t_H{}^{\al\bt}=2\frac{\dr\cH}{\dr g_{\al\bt}}.\label{5.28}
\end{equation}

If a multimomentum Hamiltonian form $H$ is associated
with a semiregular Lagrangian density $L$, we have the equalities
\be &&t^{\al\bt}(x^\la,y^i,\dr_\la^i\cH(q))
=g^{\al\mu}g^{\bt\nu}t_{\mu\nu}(x^\la,y^i,\dr_\la^i\cH(q))
 =t_H{}^{\al\bt}(q), \\
&& t^{\al\bt}(z)=g^{\al\mu}g^{\bt\nu}t_{\mu\nu}(z)
=t_H{}^{\al\bt}(x^\la,y^i,\pi^\la_i(z))\ee
where $q\in Q$ and
\[ \wh H\circ\wh L(z)=z.\]
In view of these equalities, we can think of the tensor (\ref{5.28})
restricted to the Lagrangian constraint space $Q$ as being the
Hamiltonian metric energy-momentum tensor.
On $Q$, the tensor (\ref{5.28}) does not depend upon choice of
a Hamiltonian form $H$ associated with $L$. Therefore, we shall
denote it by the common symbol $t$.

In the presence of a background world metric
$g$, the identity (\ref{5.27}) takes the form
\begin{equation}
t^\la{}_\al\{^\al{}_{\la\mu}\}\sqrt{-g}
+(\G^i_\mu\dr_i-\dr_i\G^j_\mu p^\la_j\dr^i_
\la)\wt{\cH}_\G=\frac{d}{dx^\la} T_\G{}^\la{}_\mu +p^\la_iR^i_{\la\mu}
\label{5.29} \end{equation}
where
\[\frac{d}{dx^\la} = \dr_\la +\dr_\la y^i\dr_i +\dr_\la p^\m_i\dr_\m^i\]
and by $\{^\al{}_{\la\mu}\}$ are meant the Cristoffel symbols of the
world metric $g$.

For instance, let us examine  matter fields in the presence of a
background gauge potential $A$
which are described by the Hamiltonian form (\ref{220}).
In this case, we have the equality
\be && t^\la{}_\mu\sqrt{-g}=T^\la{}_\mu=
[a^{ij}_Yg_{\mu\nu}p^\la_ip^\nu_j(-g)^{-1} \\
&& \qquad -\dl^\la_\mu
\frac12 (a^{ij}_Yg_{\al\nu}p^\al_ip^\nu_j(-g)^{-1}+
m^2a^Y_{ij}y^iy^j)]\sqrt{-g}\ee
and the gauge invariance condition
\[I_m{}^j{}_ip_j^\la\dr_\la^i\wt{\cH}=0.\]
The identity (\ref{5.29}) then takes the form of the energy-momentum
conservation law
\[\sqrt{-g}\nabla_\la t^\la{}_\mu=-p^\la_iF^m_{\la\mu}I_m{}^i{}_jy^j\]
where $\nabla_\la$ is the covariant derivative relative to the
Levi-Civita connection and $F$ is the strength of the background gauge
potential $A$.

Let us consider gauge potentials $A$ described by the complete family
of the multimomentum Hamiltonian forms (\ref{5.3}) where $S=S_B$. On the
solution $A=B,$ the curvature of the connection $S_B$ is reduced to
\[ R^m_{\la\al\mu}=\frac12(\dr_\la\cF^m_{\al\mu}-
c^m_{qn}k^q_\la\cF^n_{\al\mu}-\{^\bt{}_{\mu\la}\}\cF^m_{\bt\al}-
\{^\bt{}_{\al\la}\}\cF^m_{\mu\bt}).\]
Set
\[S^\la{}_\mu=p_m^{[\al\la]}\dr^m_{\al\mu}\wt{\cH}_{YM}=
\frac{\ve^2}{2\sqrt{-g}}a^{mn}_Gg_{\mu\nu}
g_{\al\bt} p^{[\al\la]}_m p^{[\bt\nu]}_n.\]
In virtue of Eqs.(\ref{5.2a}), (\ref{5.2b}) and (\ref{5.5}), we
obtain the relations
\[p_m^{[\la\al]}R^m_{\la\al\mu}=\dr_\la S^\la{}_\mu
-\{^\bt{}_{\mu\la}\}S^\la{}_\bt,\]
\[\dr^\bt_n\G^m_{\al\mu}p^{\al\la}_m\dr^n_{\bt\la}
\wt{\cH}_{YM}=\{^\bt{}_{\al\mu}\}S^\al{}_\bt,\]
\[t^\la{}_\mu\sqrt{-g}=2S^\la{}_\mu-\dl^\la_\mu\wt{\cH}_{YM}\]
and
\[t^\la{}_\mu\sqrt{-g}=T^\la{}_\mu+S^\la{}_\mu.\]
The identity (\ref{5.29}) then takes the form of the energy-momentum
conservation law
\[\nabla_\la t^\la{}_\mu=0.\]

In particular, the familiar energy-momentum conservation law for
electromagnetic field is reproduced.

\end{document}